\documentclass[12pt]{JINST}

\usepackage{epsfig,wrapfig} 

\def\bit{\begin{itemize}}
\def\eit{\end{itemize}}
\def\benu{\begin{enumerate}}
\def\eenu{\end{enumerate}}
\def\beq{\begin{equation}}
\def\eeq{\end{equation}}
\def\bec{\begin{center}}
\def\eec{\end{center}}
\def\btable{\begin{tabular}}
\def\etable{\end{tabular}}
\def\beqr{\begin{eqnarray}}
\def\eeqr{\end{eqnarray}}

\def\lm{\lambda}
\def\eps{\epsilon}

\def\alp{\alpha}
\def\bt{\beta}

\def\Dl{\Delta}
\def\sg{\sigma}

\def\btab{\begin{tabbing}}
\def\etab{\end{tabbing}}
\def\beqrs{\begin{eqnarray*}}
\def\eeqrs{\end{eqnarray*}}

\def\noi{\noindent}

 \fontfamily{ptm}\selectfont
\usepackage{mathptmx}           

\title{Longitudinal dynamics and tomography in the Tevatron}

\author{J.~Stogin$^a$, T.~Sen$^b$ \thanks{Corresponding author} , 
R.S.~Moore$^b$ \\
\llap{$^a$}Princeton University,  Princeton, NJ 08544, USA \\
\llap{$^b$}Fermilab, PO Box 500, Batavia, IL 60510, USA \\
Email: \email{tsen@fnal.gov}}

\abstract{Motivated by the desire to understand the longitudinal
effects of beam-beam forces, we study the longitudinal dynamics of 
protons and anti-protons at injection and top energy in the Tevatron.
Multi-turn data of the longitudinal profiles are captured to
reveal information about frequencies of oscillation,
and changes in the bunch distributions. Tomographic reconstruction
is used to create phase space maps which are subsequently used
to find the momentum distributions. Changes in these
distributions for both proton and anti-proton beams are also followed 
through the 
operational cycle. We report on the details of interesting dynamics
and some unexpected findings.}


\begin{document}

\section{Introduction}

Tomographic techniques are used to reconstruct the phase space based on
one dimensional profiles measured by profile monitors. Transverse and 
longitudinal tomography have been used to study the development of beam
tails and the onset of instabilities in several accelerators. Here we 
will focus on longitudinal dynamics of beams in the Tevatron as revealed by the  changes in longitudinal distributions and by  tomographic reconstruction using data from a wall current monitor. 

Bunches are coalesced in the Main Injector and fed into the Tevatron at an energy of
150 GeV. First protons are injected, then electrostatic separators are turned
on to place them on a helical orbit which is designed to separate them 
from the anti-protons that are subsequently injected onto a 
separate helical orbit. Anti-proton bunches are injected
four bunches at a time into gaps between the three proton bunch trains.
After each group of 3 anti-proton transfers, the gaps are cleared for the
subsequent set of transfers by "cogging" the antiprotons - changing the
antiproton RF cavity frequency to let them slip longitudinally relative
to the protons. After acceleration to 980 GeV, the 36 proton and
36 anti-proton bunches are brought into head-on collisions at two high energy 
physics detectors CDF and D0. 
In addition to the head-on collisions, each bunch experiences 70 long-range interactions with bunches
of the other beam around the ring. As a consequence of these beam-beam 
interactions, there is beam loss and emittance growth. The effects on
the transverse dynamics has been well studied and documented 
\cite{Sen, Shiltsev, Sen_ICFA}. Here we will study the impact of these beam-beam
interactions on the longitudinal dynamics via longitudinal tomography.
The effects of the long-range interactions may be seen by comparing the 
longitudinal phase space and momentum distributions of protons before and 
after the anti-protons are injected. Effects of the 
head-on collisions may be found by investigating the phase space and
momentum distributions of both
protons and anti-protons before and after collisions are initiated.
Table \ref{table: tev_param} shows some of the main parameters of the
Tevatron beams.
\begin{table}
\bec
\btable{|c|c|}\hline
Parameter & Value \\ \hline
Top energy [GeV] & 980 \\
Number of bunches per beam & 36 \\
Proton bunch intensity & 2.9$\times 10^{11}$ \\
Anti-proton bunch intensity & 0.9$\times 10^{11}$ \\
Proton transverse normalized 95 \% emittance [mm-mrad] & 18 \\
Anti-proton transverse normalized 95 \% emittance [mm-mrad] & 8 \\
$\bt^*$ at IP [m] & 0.28 \\
Proton rms bunch length at 980 GeV [nsec] &  1.7 \\
Anti-proton rms bunch length at 980 GeV [nsec] & 1.5  \\
\hline
\etable
\eec
\caption{Parameters of the Tevatron}
\label{table: tev_param}
\end{table}

The beam-beam force has a longitudinal component due to the coupling of
the transverse motion to the electric field. This has been 
estimated to lead to an energy change for a test particle per interaction
\cite{Danilov, Hogan} to be
\beq
\Dl E_{bb} = - \frac{N e^2 \alp^*}{\bt^*} = -\frac{1}{4\pi\eps_0}
\frac{N e \alp^*}{\bt^*} \; [{\rm e V}]
\label{eq: dE_bb}
\eeq
where $N$ is the bunch intensity of the opposing bunch, $\alp^*,\bt^*$ are
the Twiss parameters at the IP.
Substituting values for the parameters in the Tevatron, 
$N_p = 2.9\times 10^{11}$,$N_{\bar{p}} = 0.9\times 10^{11}$, 
$\bt^* = 0.28$m, we have
\[ 
{\rm Anti-protons :} \;\; \Dl E_{bb} = -1.5\alp^* {\rm [keV]}, \;\;\;\;
{\rm Protons:} \;\; \Dl E_{bb} = -0.5\alp^* {\rm [keV]}
\]
$\alp^*$ can be in the range $\pm (0.1 - 1)$, so the energy change per
particle per kick is of the order of keV. The relative energy change
at 980 GeV is of the order of $10^{-6}$ per kick, this is small when
compared to the energy spread in the beam, around $10^{-4}$. This is a 
very simple estimate and there are other sources of energy change from
beam-beam interactions including beams crossing at an angle and the 
70 long-range interactions. It is possible that these effects are 
cumulative and may have an observable effect on the bunch properties.
In this paper we will test this hypothesis by observing the longitudinal
dynamics subsequent to the beam-beam interactions.

\section{Measurements of longitudinal profiles} \label{sec: measure}

The longitudinal beam profiles are obtained via a resistive wall monitor
in the Tevatron. This device consists of a short ceramic vacuum pipe
with eighty 120 $\Omega$ resistors across it. A copper casing enclosing
the ceramic break is filled with ferrite to provide a low impedance
bypass for DC currents while forcing AC currents to flow through the
resistors. These signals are summed into a single intensity signal which
is input to a LeCroy DL7200 digital oscilloscope for data acquisition.
The longitudinal beam profile was sampled at a rate of 1 GHz with
periodic triggers synchronized to the RF system provided by custom
electronics.  These triggers were set to integer multiples of "turns"
(beam revolutions) where 1 turn $\approx 21 \mu$s in the Tevatron.  The
depth of the oscilloscope's internal data buffer limited the number of
recorded triggers and the length of each digitized waveform.  For our
data samples, we recorded either 128 triggers of 10 $\mu$s duration
(enough for 12 bunches of both beams) or 1024 triggers of 1 $\mu$s
duration (enough for 5 bunches total).  The digitized waveform data was
read from the oscilloscope in ASCII format via an ethernet interface.

Data was taken on two days: store 7949 on July 14 2010 and store 
8146 on October 6, 2010. The data was captured at different stages
during injection and at top energy. The intent was to observe the effects
of beam injection, acceleration and bringing the beams into collision
on the longitudinal dynamics. In store 7949, data was captured at the
following stages: 1) All 36 proton bunches injected and circulating on 
their helical orbit and four anti-proton bunches in each train injected 
and cogged. In the first train, anti-proton bunches A1-A4 are circulating.
2) After eight anti-proton bunches in each train (bunches A1-A8
in the first train) have been injected and cogged or moved towards the
head of the proton train.
3) After all 12 anti-proton bunches in each train have been injected and cogged. 
This is the stage just before acceleration.
4) After acceleration to 980 GeV and final cogging and before the beta
squeeze
5) After the beta squeeze and during initial collisions.
6) Beginning of data taking for high energy physics. Between stages 5 and 6,
the transverse beam halo is removed using movable collimators.
7) 1 hour and 47 minutes after stage 6. 
At each stage, the profiles were recorded from all 12 bunches of the first train in each beam over 128 turns with some
number of delays between each captured turn. 

In store 8146 data was captured at the following stages: 
1) Just after injecting all 36 proton bunches and opening the helix.
Only protons are circulating. 
2) After injecting first  set of anti-protons and first cogging. There are 12 
proton bunches and 4 anti-proton bunches in each train.
3) After the second set of anti-protons are injected and second cogging.
4) After the third set of anti-protons and third cogging and before acceleration.
5) After the acceleration to 980 GeV and final cogging and before beta 
squeeze.
6) Just after the beta squeeze and before collisions are initiated.
7) Just after initiating collisions.
8) 2 hours and 12 mins after start of collisions. 
At each stage the profiles were captured for 1024 turns, again with some
number of delays between captured turns. The increased amount of data
and the limited buffer size in the oscilloscope allowed storing data
only from the first three proton bunches P1-P3 and two anti-proton
bunches A2 and A3 in the first train.

\begin{figure}[h]
\centering
\includegraphics[scale=0.8]{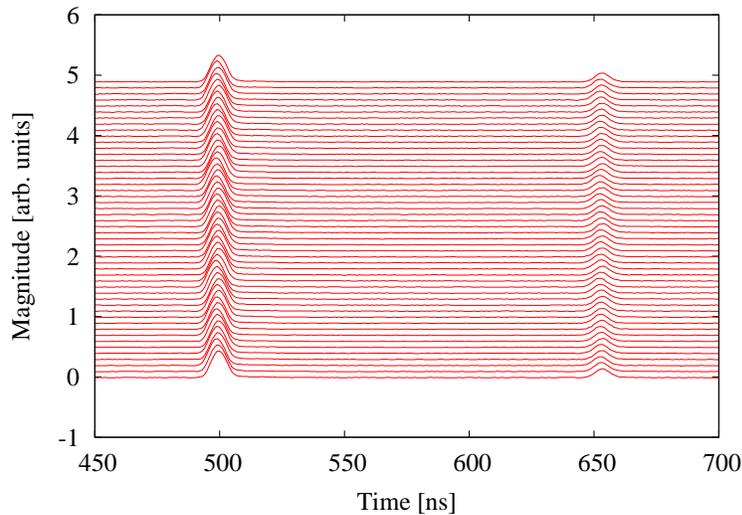}
\caption{Part of the data from Store 7949 captured from the wall current 
monitor at energy 150 GeV just before acceleration. On the left is a 
proton bunch and on the right is an anti-proton
bunch. The entire data set contains 12 proton and 12 anti-proton bunches.
The initial profiles are at the bottom while the last profiles are at the
top.}
\label{fig:data_7949}
\end{figure}

Figure \ref{fig:data_7949} shows a mountain range view of the profiles
of a proton bunch and an anti-proton bunch in store 7949. The profiles
do not show any evidence of low frequency bunch oscillations.
Earlier in  Run II, such oscillations or ``dancing''  were
observed with uncoalesced bunches\cite{Moore}. Longitudinal dampers 
reduced the centroid oscillations 
at the synchrotron frequency.

\begin{figure}[h]
\centering
\includegraphics[scale=0.5]{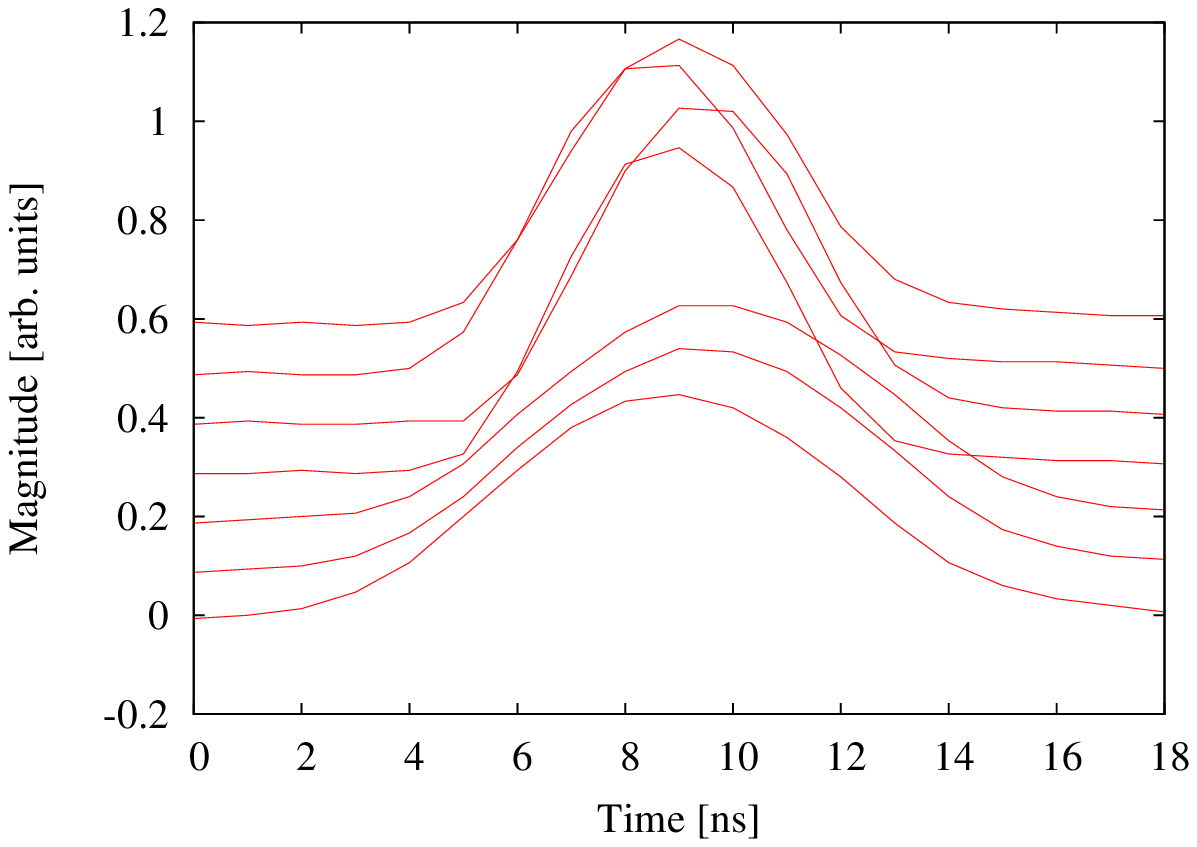}
\includegraphics[scale=0.5]{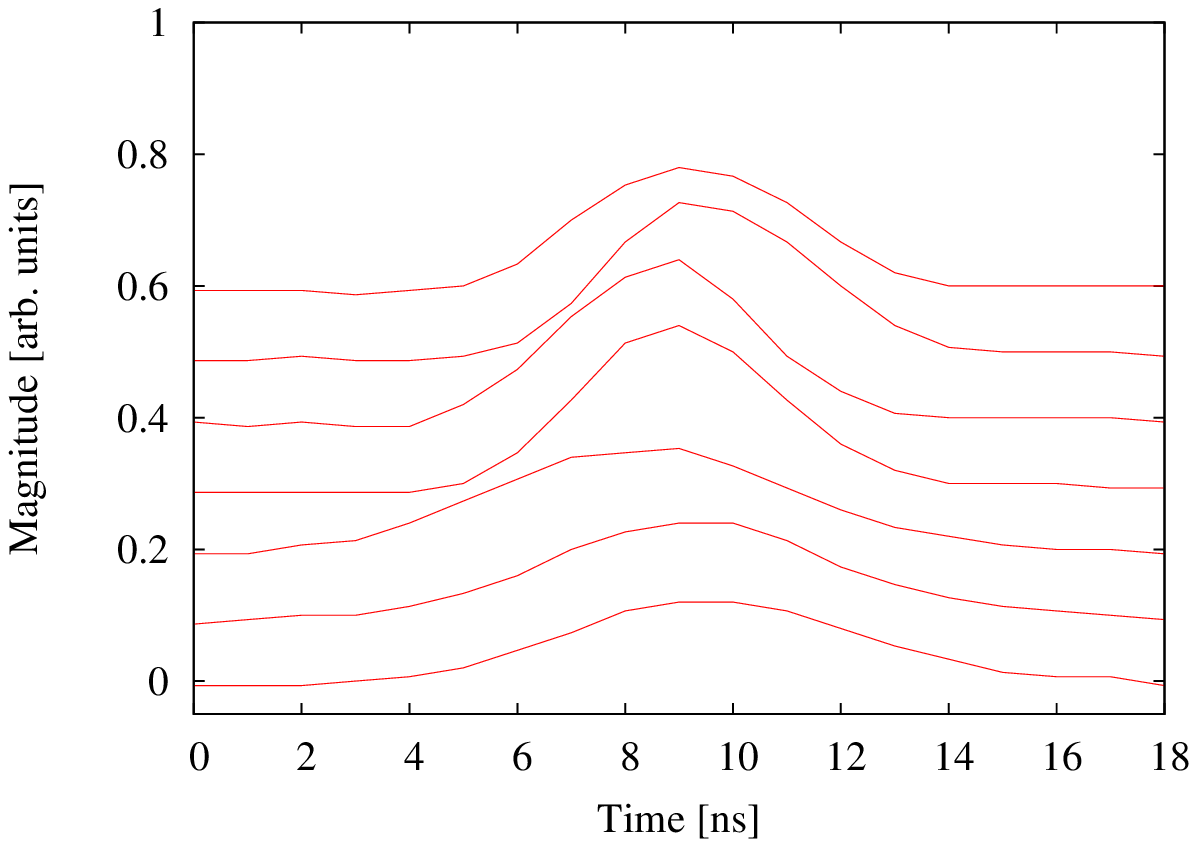}
\caption{Store 7949: Longitudinal profile of proton bunch 1 (left) and anti-
proton bunch 11 (right) at different stages from 150 GeV to 980 GeV.}
\label{fig: prot1_pbar11_7949}
\end{figure}
Figure \ref{fig: prot1_pbar11_7949} shows the profiles of proton bunch
1 and anti-proton bunch 11 at different stages in store 7949.

The distributions can be characterized by their root mean square (rms) 
length and the excess kurtosis which is defined as
\beq
k = \frac{\sg_4}{\sg_2^2} - 3
\eeq
where $\sg_4$ is the fourth moment and $\sg_2$ is the second moment. The
kurtosis measures the length of the tails relative to the core. This 
kurtosis is zero for a Gaussian, so a positive value indicates that the tails
are longer than in a Gaussian distribution. 
\begin{figure}[h]
\centering
\includegraphics[scale=0.5]{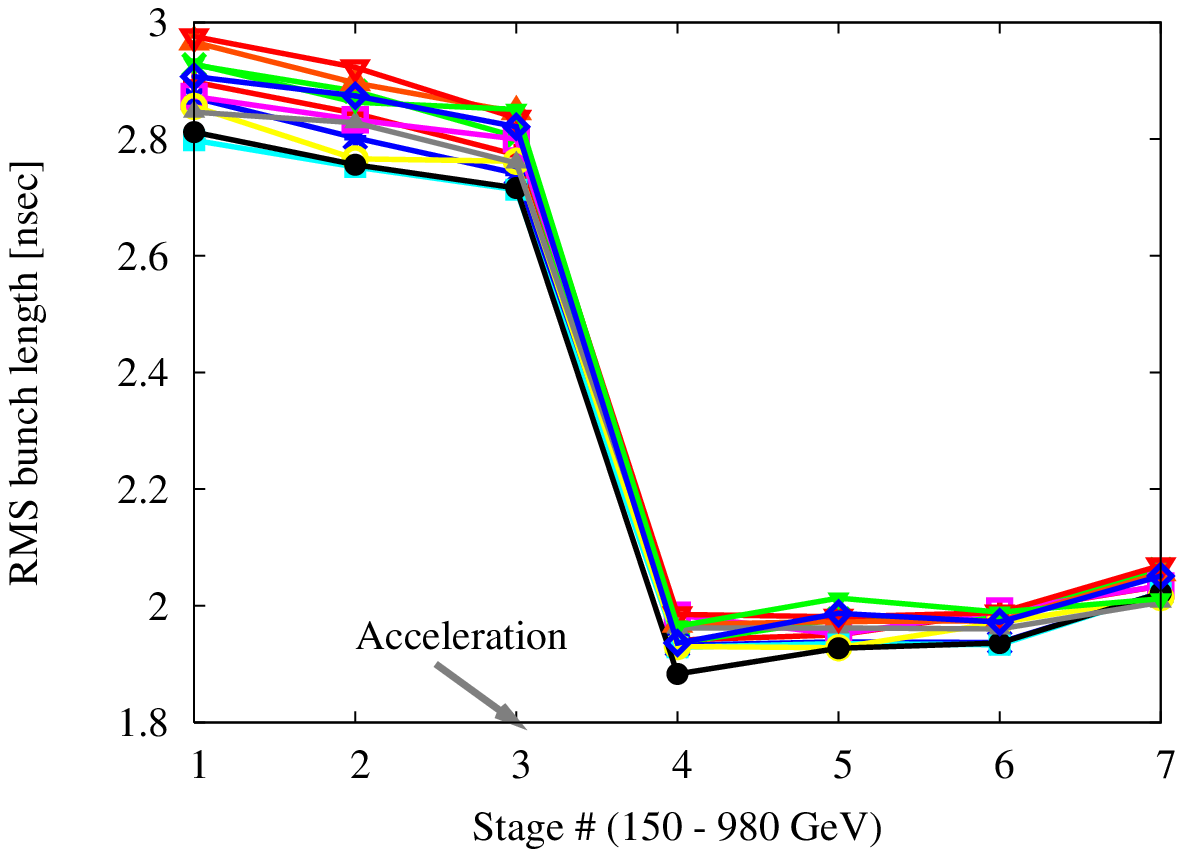}
\includegraphics[scale=0.5]{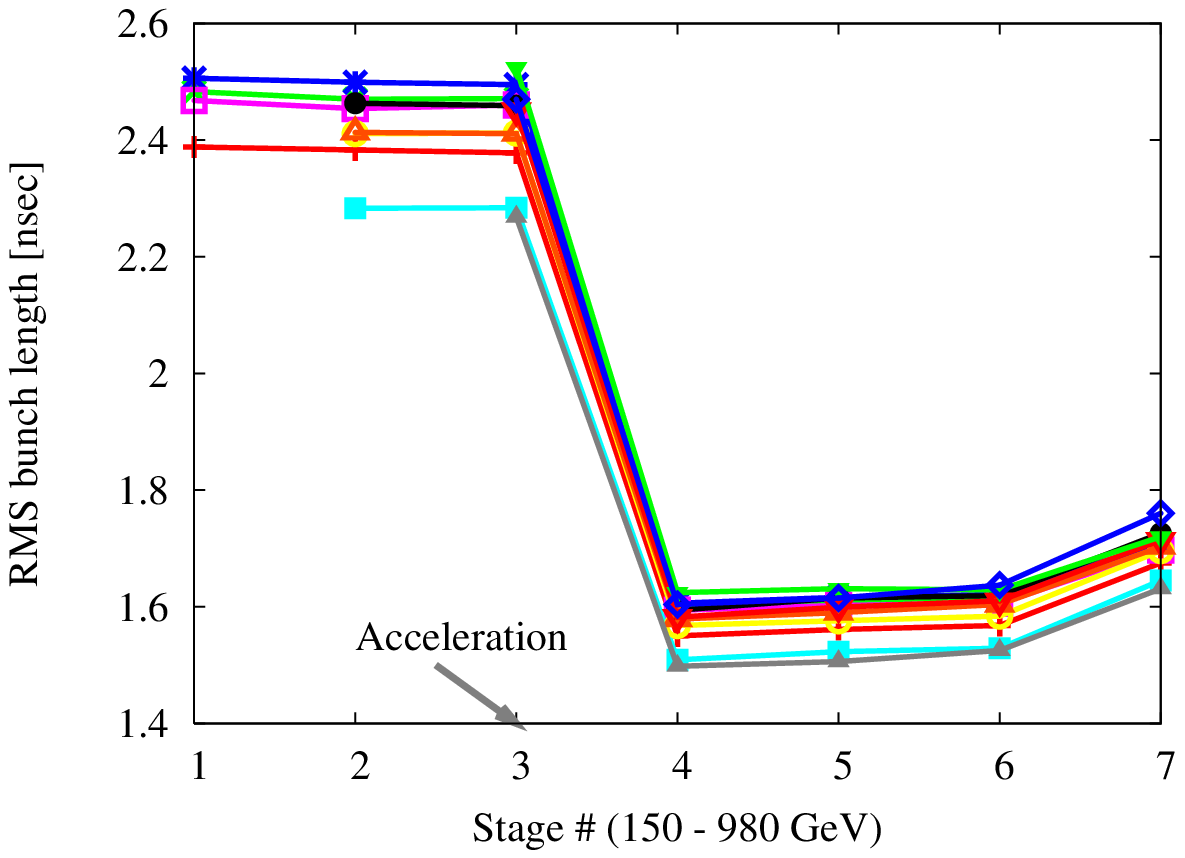}
\caption{Evolution of the bunch length of protons (left) and anti-protons
(right) for the 12 bunches during Store 7949. At the end of stage 3,
beams are accelerated from 150 GeV to 980 GeV.}
\label{fig: sigt_7949}
\includegraphics[scale=0.5]{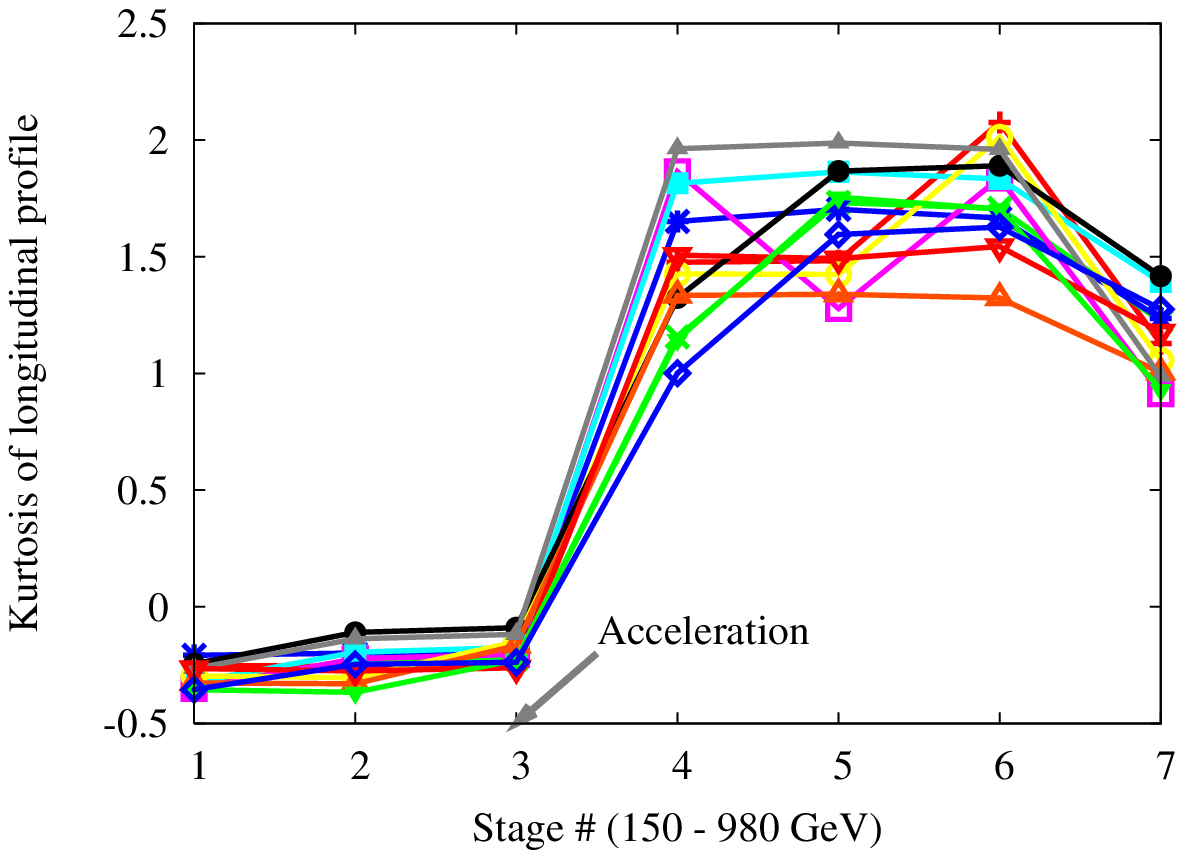}
\includegraphics[scale=0.5]{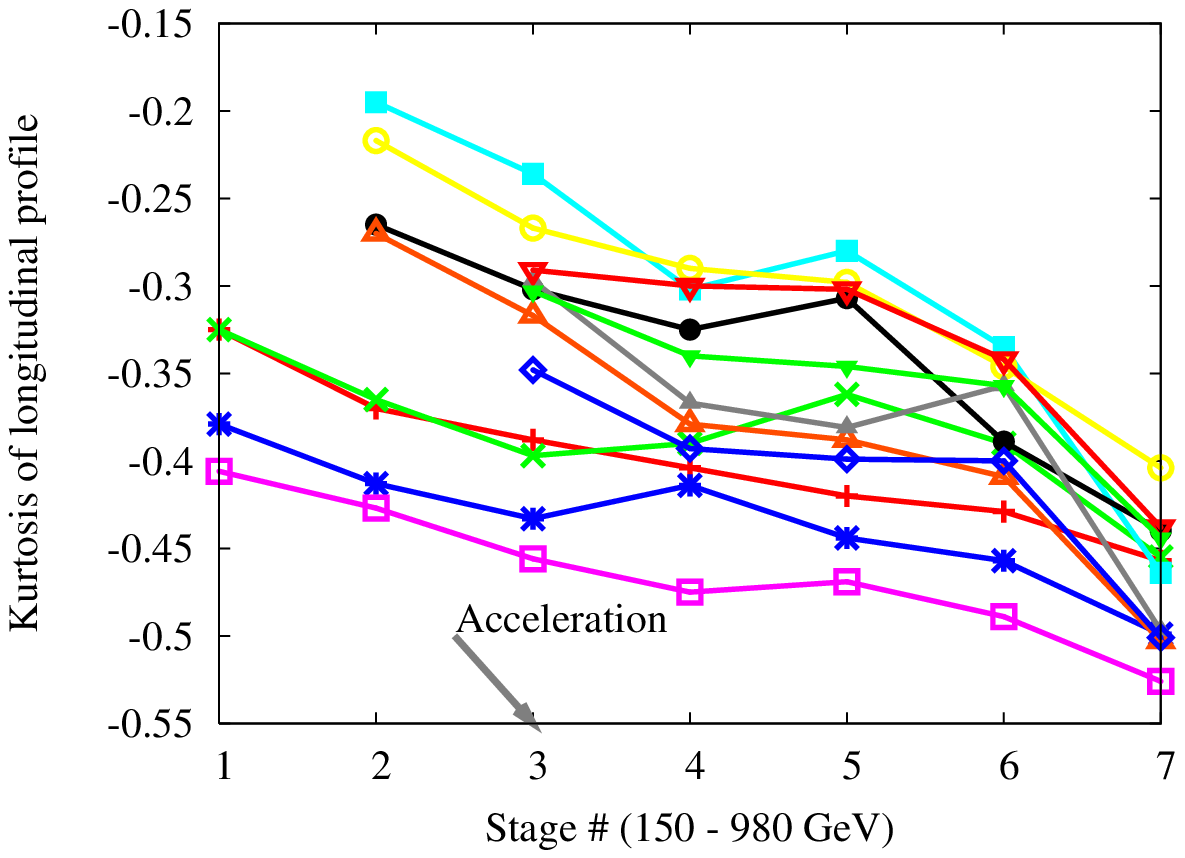}
\caption{Evolution of the kurtosis of protons (left) and of anti-protons
(right) for the bunches during Store 7949.}
\label{fig: kurt_7949}
\includegraphics[scale=0.5]{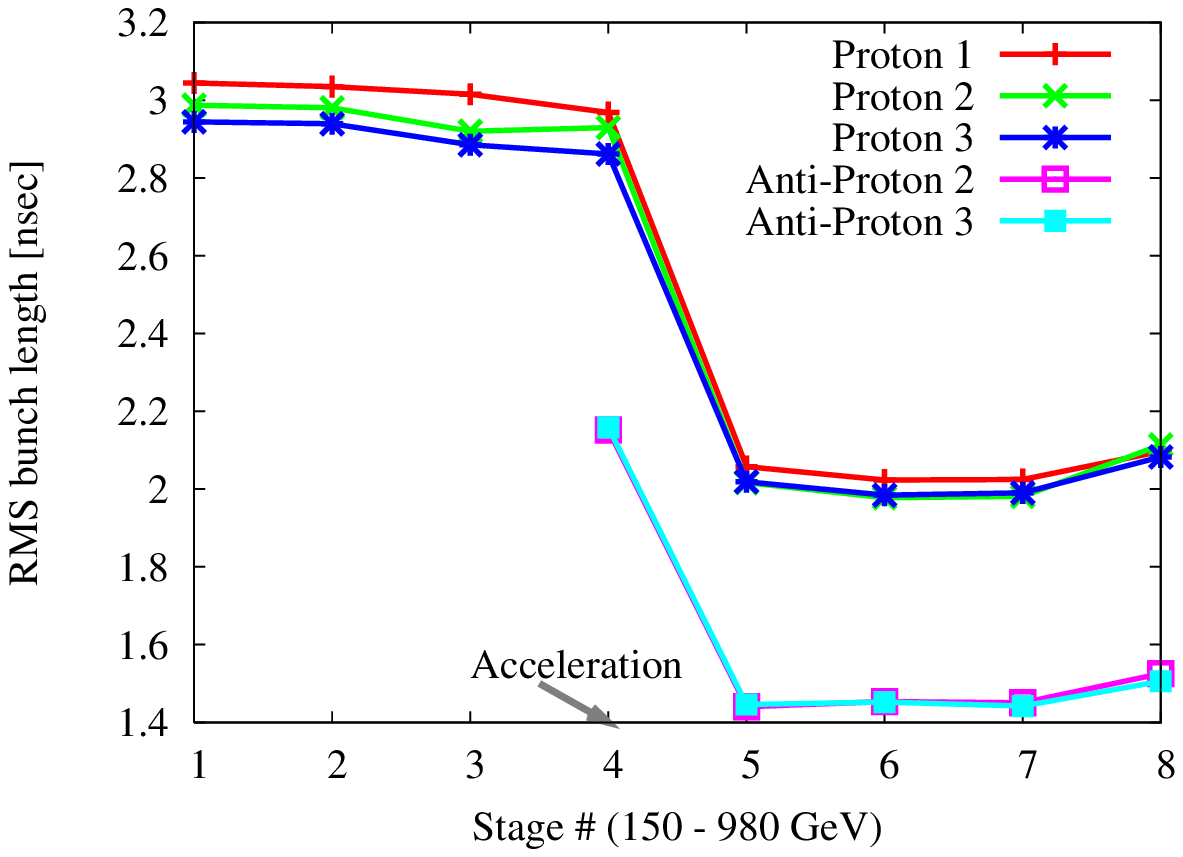}
\includegraphics[scale=0.5]{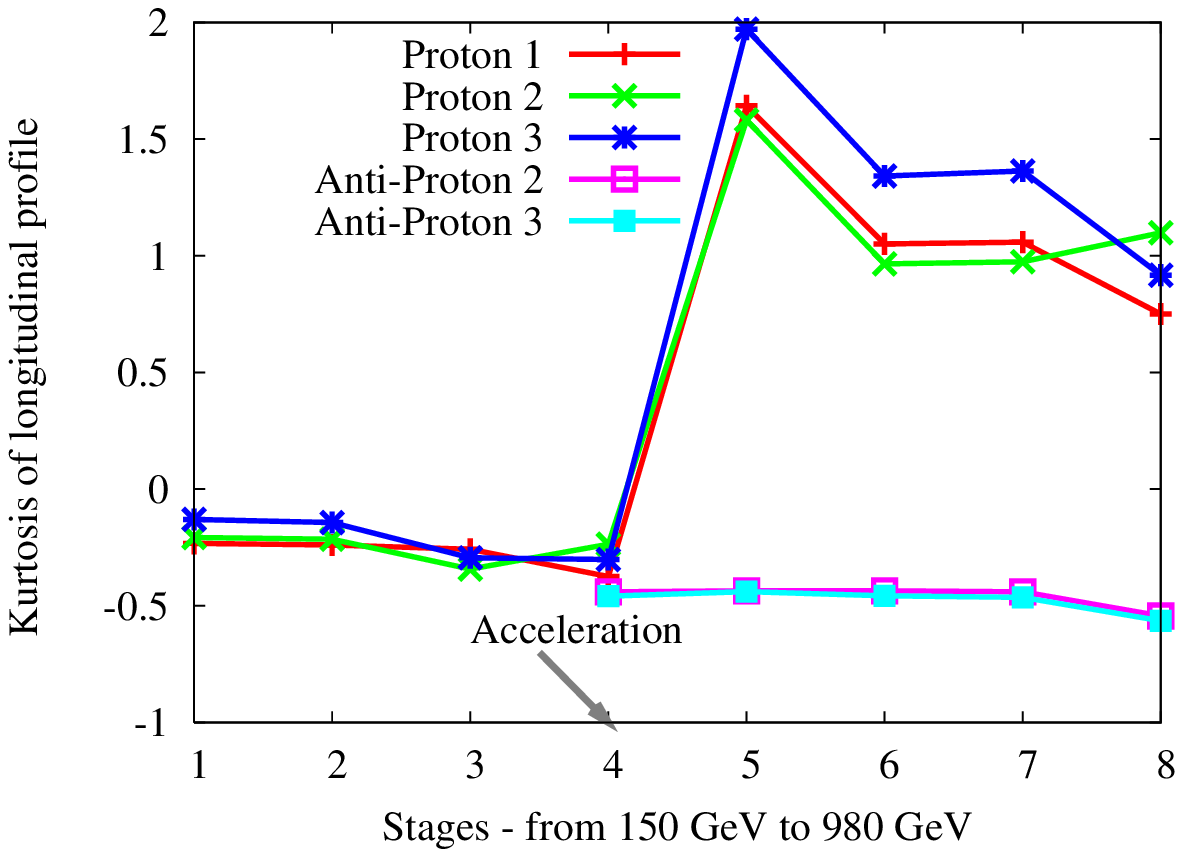}
\caption{Evolution of the bunch length (left) and kurtosis (right) for the 
5 bunches during Store 8146. In this store, acceleration occurs after
stage 4.}
\label{fig: sigt_kurt_8146}
\end{figure}
Figure \ref{fig: sigt_7949} shows the evolution of the bunch length of
protons and anti-protons in store 7949. The proton bunches are more
intense, longer and fuller in the bucket and thus more likely to be lost from the bucket. At injection energy, we see clear
evidence that the bunch length of protons shrinks while the anti-proton
bunch length stays nearly constant. Following the acceleration to top
energy and decrease in size due to adiabatic damping, the bunch length of both species subsequently
increases, primarily due to intra-beam scattering. 
The increase is the largest during the 2 hours between stages 6 and 7. 
Figure \ref{fig: kurt_7949} shows the excess kurtosis of both species.
At injection energy, the protons have negative kurtosis implying that 
their tails are shorter than for a Gaussian but the kurtosis gradually
approaches that of a Gaussian. Following acceleration there is
a rapid increase in the kurtosis even though the bunch itself is shorter,
suggesting that the beam tails have grown at the expense of the core.
At flat top, the kurtosis stays nearly constant for most bunches
(except for a couple of bunches) until stage 6, the start of collisions,
after which the kurtosis drops with time. The kurtosis of the anti-protons
has a very different behaviour - they start at negative values after
injection and they keep decreasing. There is no sharp rise in the
kurtosis during acceleration and the decrease in kurtosis continues
with time suggesting that the core is broadening at the expense of the
tails. 

Figure \ref{fig: sigt_kurt_8146} shows the evolution of the rms bunch length
of protons and anti-protons in store 8146. The statistical errors are
smaller here since the data was averaged over 1024 samples compared
to 128 samples for store 7949. Much the same conclusions
on the bunch length and kurtosis apply to the behaviour in this store
as well. The bunch length of protons decreases at injection energy and
grows at flat top during the store for both species. Again,
acceleration to top energy increases  the kurtosis sharply
for protons but not for the anti-protons. At flat top during the store, 
the kurtosis drops for both species but more quickly for protons.

During store 8130 we obtained data for the same five bunches at one stage
about 12 hrs into the store. We find that the kurtosis of the 3 proton bunches
had dropped to a range between 0.1 - 0.2 while the kurtosis of the anti-protons
had fallen to -0.6.  The proton distributions appear to approach a Gaussian distribution
over long times while the distribution of the anti-protons continues to
become more non-Gaussian over time with a larger core relative to the tails.

\section{FFT of centroid motion} \label{sec: fft}

The longitudinal profiles will be used for reconstruction of the phase space
as well to analyze the motion of the centroid. Good resolution of 
the phase space structure requires faster sampling for as many turns with as 
little delay as possible between turns, preferably turn by turn. On the other hand the 
frequency resolution of the FFT of the centroid motion is given by 
\beq
\Dl f = \frac{1}{N \Dl t} = 2 \frac{f_{max}}{N}
\eeq
where $N$ is the number of samples, $\Dl t$ is the time delay between samples
and $f_{max} = 1/(2 \Dl t)$ is the maximum frequency that can be measured. If the number of samples is fixed, a larger delay $\Delta t$ improves the frequency resolution but worsens the phase space resolution. 

In Store 7949 the data 
was captured at a rate of 1 GHz for 128 turns. At 150 GeV, 
the time delay between each of the 128 samples was 6 turns while
at 980 GeV, the delay between samples was 11 turns.
In store 8146 data was captured at the same sampling rate but over 1024 
turns with a 
delay between samples of 6 turns at 150 GeV and 11 turns at 980 GeV.
Table \ref{table:FFT_spec} shows the frequency resolution and the maximum
frequency that could be measured and the number of synchrotron periods
sampled in the two stores.
\begin{table}
\bec
\btable{|c|c|c|c|c|c|c|} \hline
Store & Energy [GeV] & No. of samples &	Delay between samples  &  $\Dl f$ & $f_{max}$ 
& $N_{synch}$ \\ \hline 
7949 & 150 & 128 &  6 turns & 62.1  Hz & 3976 Hz & 1.5 \\
     & 980 & 128 & 11 turns & 33.9 Hz & 2169 Hz & 1.1 \\ \hline
8146 & 150 & 1024 &  6 turns & 7.8 Hz & 3976 Hz & 11.9 \\
     & 980 & 1024 & 11 turns & 4.2 Hz & 2169 Hz & 8.6 \\
\hline
\etable
\eec
\caption{FFT frequency resolution $\Dl f$, maximum frequency $f_{max}$
and the number of synchrotron periods sampled $N_{synch}$ in the two stores.}
\label{table:FFT_spec}
\end{table}

\begin{figure}[h]
\centering
\includegraphics[scale=0.5]{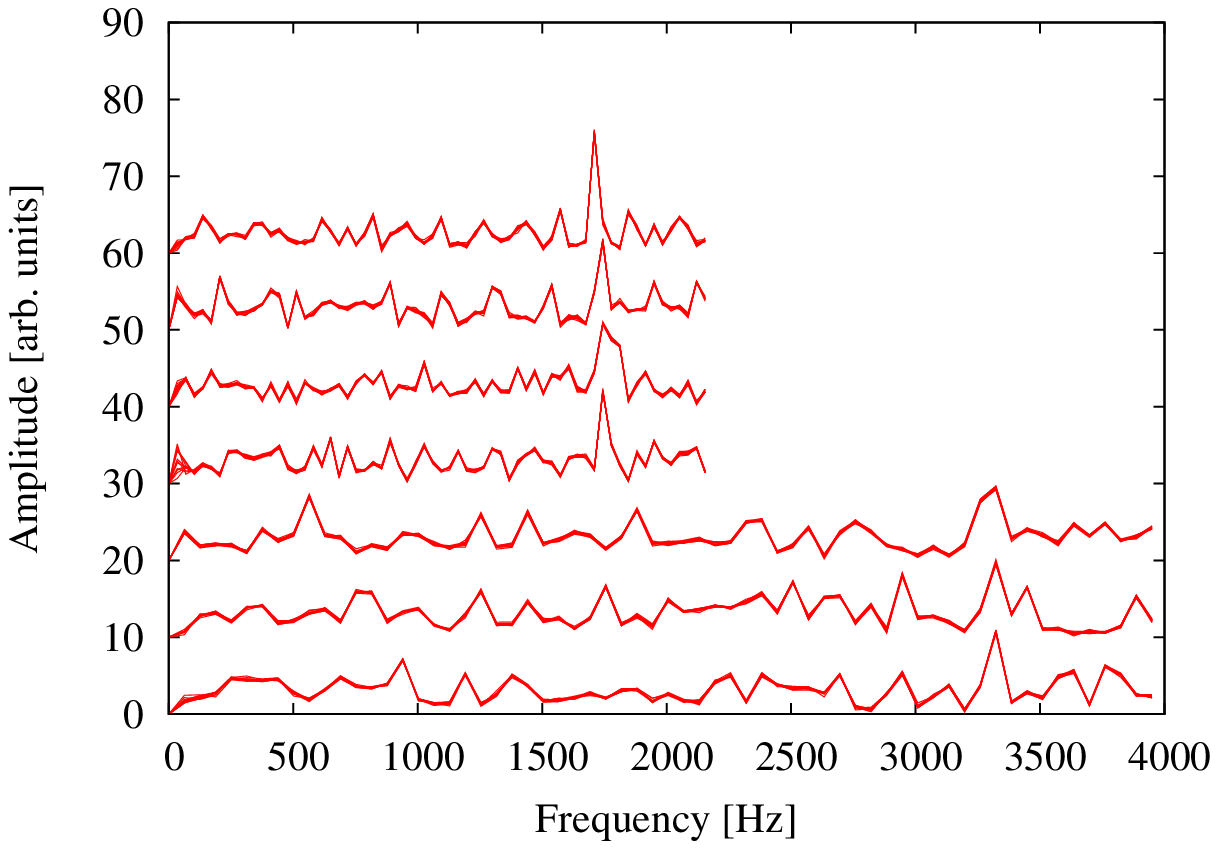}
\includegraphics[scale=0.5]{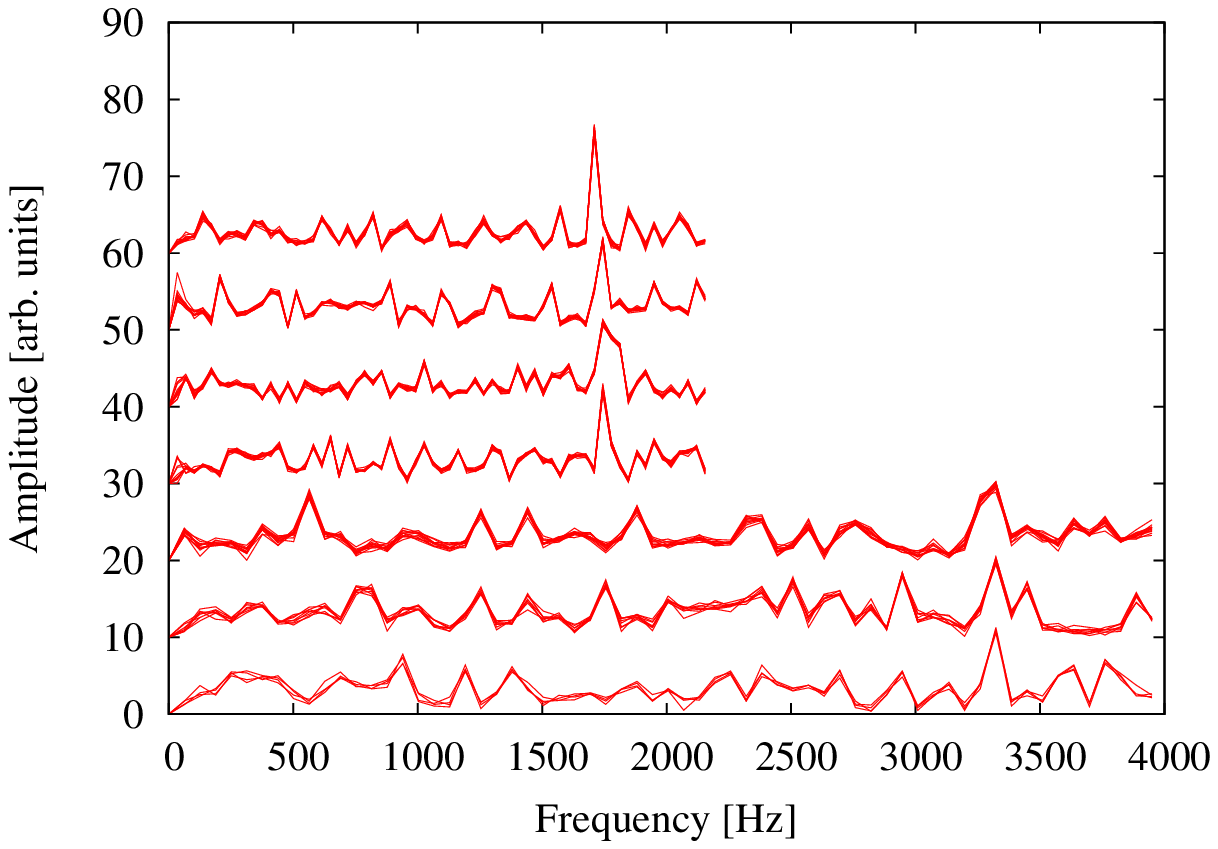}
\caption{The Fourier spectra of all 12 proton bunches (left) and 12 
anti-proton bunches (right) 
in store 7949 over all seven stages, starting with the first stage 
shown at the bottom and progressing upwards. } 
\label{fig:1st_moment}
\centering
\includegraphics[scale=0.5]{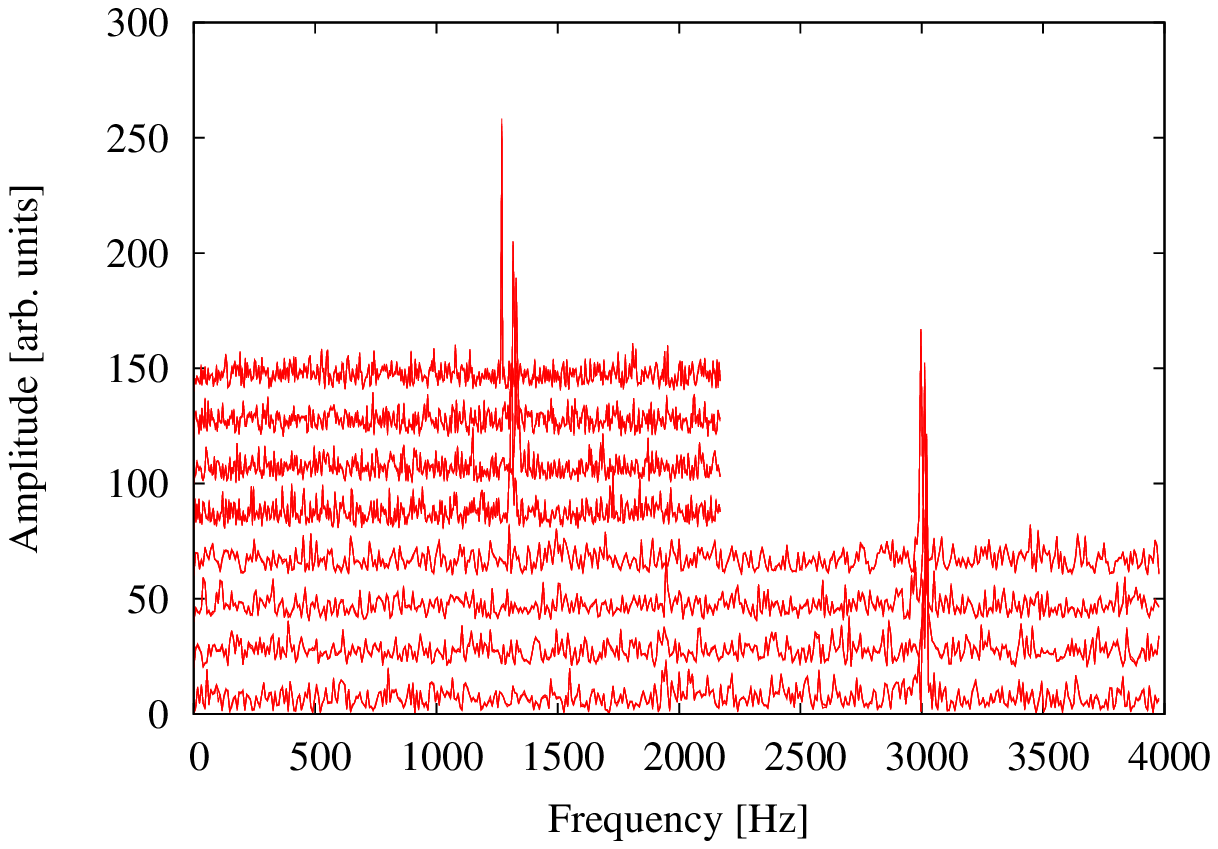}
\includegraphics[scale=0.5]{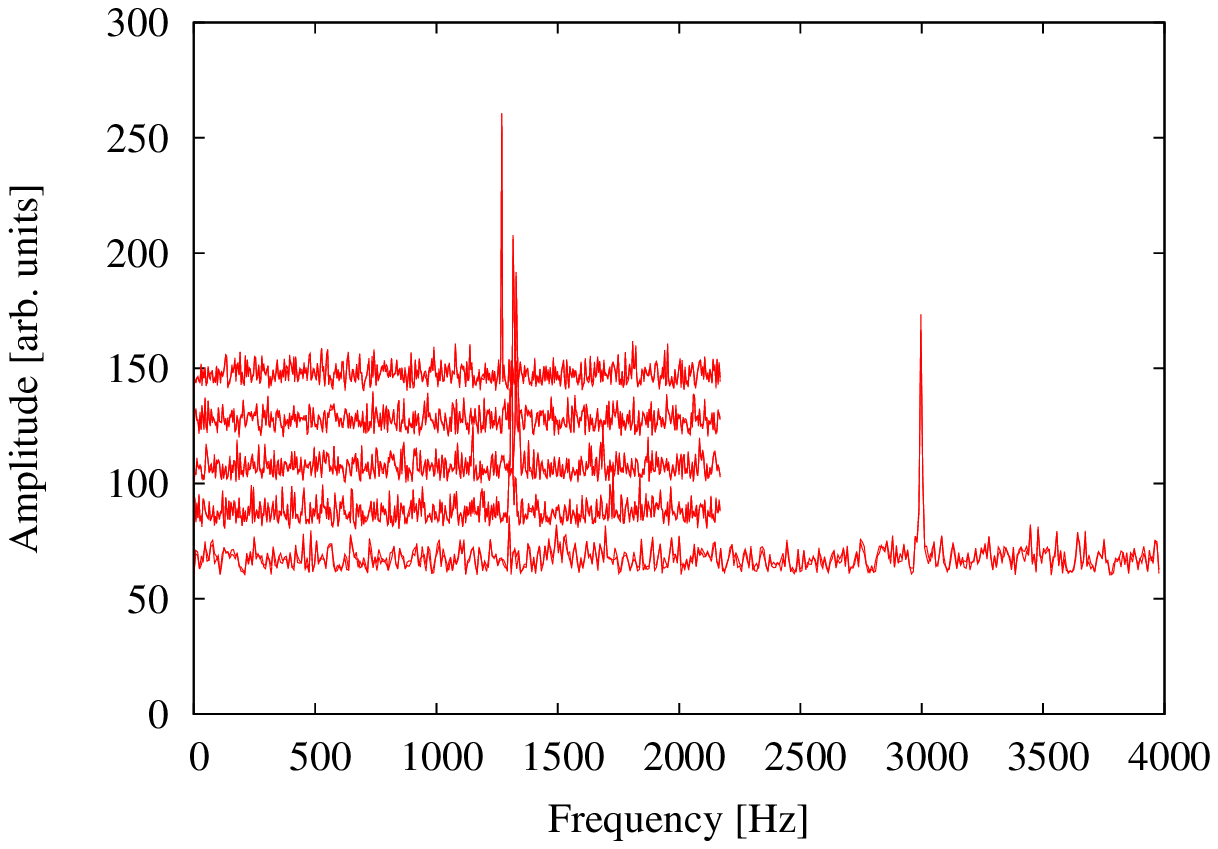}
\caption{FFT of the centroid motion of the three proton bunches (left) and 
two anti-proton bunches (right) in store 8146. }
\label{fig: fft_8146}
\end{figure}

Analysis shows that the Fourier spectrum of all 12 proton bunches
is almost the same. There is a little more 
variation amongst the anti-proton bunches. These observations are valid
at each of the seven stages. Between stages, the spectrum does change.
The stage by stage evolution is shown in Figure \ref{fig:1st_moment}.
The surprising feature is the presence of a high frequency line in each
spectrum. For every bunch, the dominant frequency 
at 150 GeV is $3322$ Hz and for the first three stages at 980 GeV is 
1744 Hz. Between the third and fourth stages (corresponding to an elapsed time of roughly 1 hr 47 mins) at 980 GeV, the dominant frequency drops by one resolution unit to $1710$ Hz. It is also worth noting that the spike at the dominant frequency seems to widen at the start of initial collisions.

Figure \ref{fig: fft_8146} shows the spectra of the centroid motion of 
bunches from store 8146. While the number of bunches is less, we again
observe the high frequency spike and that the centroid of each bunch has the 
same frequency content at each stage. All bunches show a downward shift in
the centroid frequency after 2 hours into colliding mode.

It is interesting to compare the observed frequencies with some of the natural frequencies of a beam. 
The incoherent small amplitude synchrotron frequency is given by
\beq
f_s = f_0 \sqrt{\frac{h\eta e V_{rf}}{2\pi \bt E_0}}
\eeq
where $f_0$ is the revolution frequency, $h$ is the harmonic number,
$\eta$ is the slip factor, $V_{rf}$ is the peak rf voltage,
$\bt$ is the kinematic factor and $E_0$ is the synchronous energy.
At 150 GeV, the small amplitude synchrotron frequency is 92.7 Hz
and at 980 GeV it is 36.5 Hz. More relevant is the frequency of the coherent 
small amplitude dipole oscillations which is the frequency when the bunch
oscillates as a rigid unit. It is given by \cite{Sen_coh}
\beq
f_c = f_0 \left[ \frac{e h \left|\eta\right|}{2\pi \bt^2 E_0 } 
\int_{\phi_1}^{\phi_2} \lm'(\phi) [V(\phi) - V(\phi_s)] d\tau \right]^{1/2}
\label{eq: omegac_all}
\eeq
for a bunch with a longitudinal density distribution $\lm(\phi)$, 
normalized to unity $\int \lm(\phi)d\phi = 1$, $V(\phi)$ is the rf voltage
as a function of phase and $\phi_1, \phi_2$ are the endpoints of the  bunch.
For a Gaussian bunch and for Tevatron parameters, this frequency at 980 GeV is
32.7 Hz which is close to the incoherent frequency of 36.5 Hz.

\begin{table}
\bec
\btable{|c|c|c|c|c|} \hline
Store & Energy [GeV] & Stage & Frequency [Hz] & Frequency/$f_{c}$ \\ \hline
7949 & 150 & 1 & 3322 & 101.7 \\
     & 150 & 2 & 3322 & 101.7 \\
     & 150 & 3 & 3322 & 101.7 \\
     & 980 & 4 & 1744 &  53.4 \\
     & 980 & 5 & 1744 &  53.4 \\
     & 980 & 6 & 1744 &  53.4 \\
     & 980 & 7 & 1710 &  52.4 \\
\hline
8146 & 150 & 1 & 3020  & 92.5 \\
     & 150 & 2 & 3020  &  92.5 \\
     & 150 & 3 & 3012  & 92.5  \\
     & 150 & 4 & 2996  & 91.7  \\ 
     & 980 & 5 & 1312  & 40.2 \\
     & 980 & 6 & 1329  & 40.7 \\
     & 980 & 7 & 1316  & 40.1 \\
     & 980 & 8 & 1269  & 38.9 \\
\hline
\etable
\eec
\caption{Dominant frequency in the FFT spectrum observed in the
three proton bunches and two anti-proton bunches. All bunches
have the same frequency at each stage.}
\label{table: FFTvalues}
\end{table}
Table \ref{table: FFTvalues} shows the dominant frequency observed from the
FFTs of the bunch centroid motion in the two stores and the ratio of these frequencies
to the coherent frequency. The existence of this high frequency line is
quite unexpected and the source of these frequencies is not clear. 
Since the observed high frequency drops with increased energy, it suggests 
that it is associated with synchrotron motion. We expect that the frequency 
should scale as $\propto 1/\sqrt{E}$. Using this scaling, a frequency of 
3322 Hz observed in the first two stages at 150 GeV in store 7949 should scale 
to 1300 Hz at 980 GeV. This is lower than the frequencies observed during the
four stages at 980 GeV in store 7949. Similarly scaling 2996 Hz results in
a frequency of 1172 Hz, also lower than the observed value in store 8146.
Assuming that the frequency does scale with a power of the energy, data
from store 7949 imply a scaling $E^{-0.34}$ while the data from store 8146 
imply a scaling $E^{-0.44}$. Given that there is greater uncertainty with
the data from the first store due to the lower resolution, perhaps these
scalings are not entirely inconsistent. 

The changes in the dominant frequency at the same energy in a store may 
reflect changes in the beam distribution and hence changes in the coherent 
frequencies. It is possible that the same harmonic of the coherent dipole 
frequency is excited at each stage but the coherent dipole frequency changes 
with the distribution.

\begin{figure}
\centering
\includegraphics[scale=0.7]{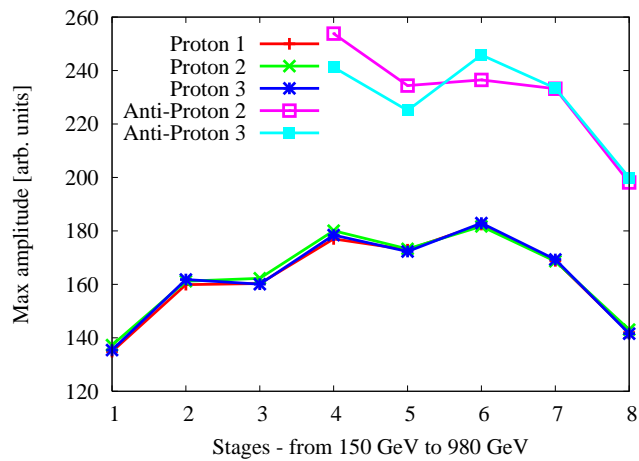}
\caption{Amplitude of the dominant frequency in the spectrum of 
each bunch as it evolves during the different stages in store 8146. }
\label{fig: fft_maxamp}
\end{figure}

The amplitude of oscillation in the proton bunches increases with time
at 150 GeV, stays nearly constant after acceleration and then gradually
decreases at 980 GeV. The amplitude is larger in the anti-proton 
bunches but it shows a similar behaviour with time at 980 GeV.
Figure \ref{fig: fft_maxamp} shows the evolution of the amplitude
of the dominant frequency for each bunch in store 8146. 
The origin of these high frequencies is unknown. 
Since all bunches in both beams have nearly the same frequencies
and these lines persist for long periods of time (e.g several hours at
980 GeV) an external source associated with the rf 
is probably driving both beams. The fact that these high frequencies do not
resonate with any of the low harmonics of the synchrotron frequency
or the coherent frequency likely explains why the bunches are not
perturbed by these driving frequencies.

\section{Phase space reconstruction}

We use the phase reconstruction method discussed in reference 
\cite{Hancock}. It uses a hybrid method combining algebraic
reconstruction techniques (ART) and particle tracking using the
difference synchrotron equations of motion. The tracking generates the 
coefficients in each phase space cell required during back projection
from the profiles to phase space.

\begin{figure}[h]
  \centering
\includegraphics[scale=0.2]{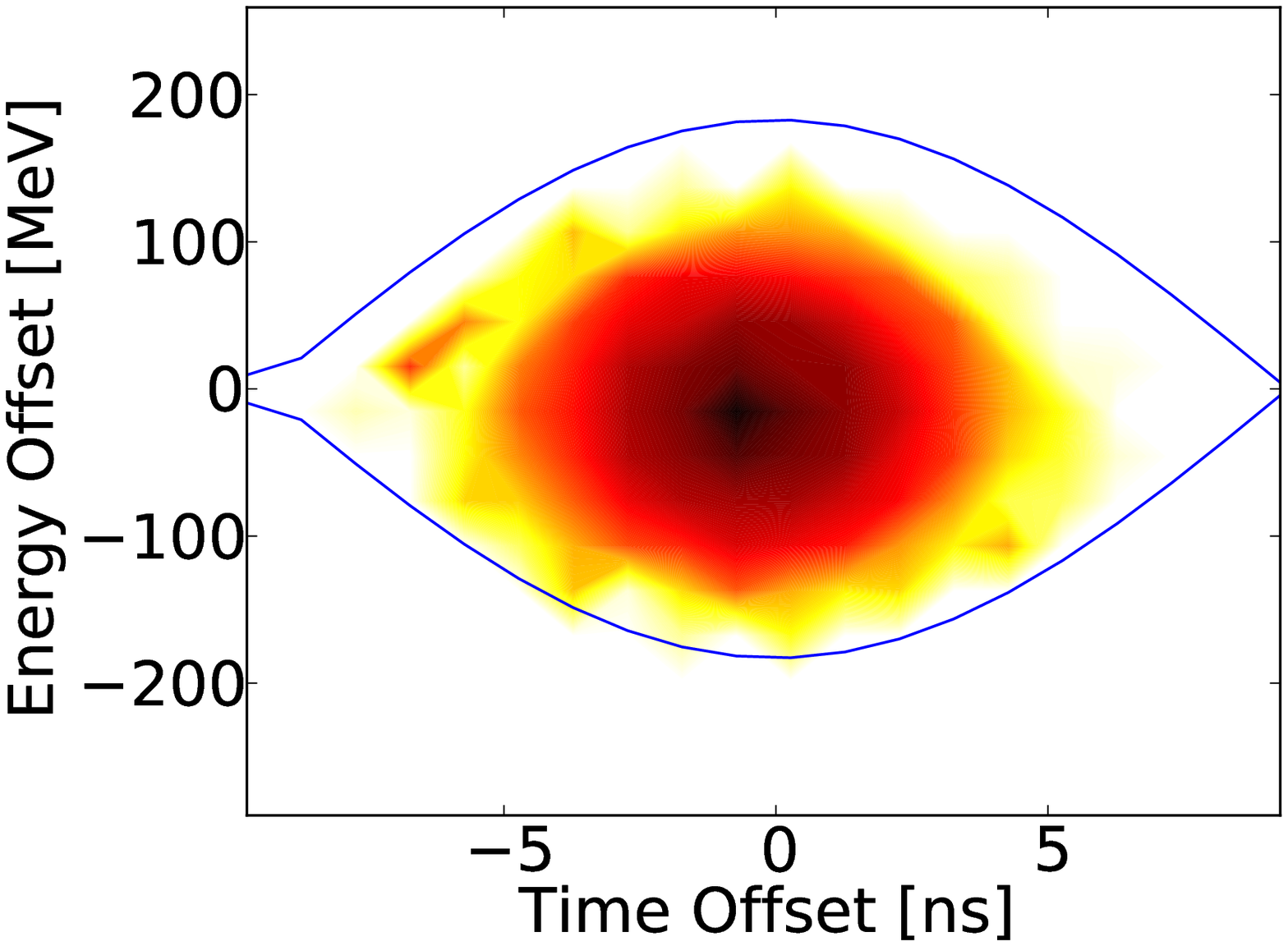}
  \includegraphics[scale=0.2]{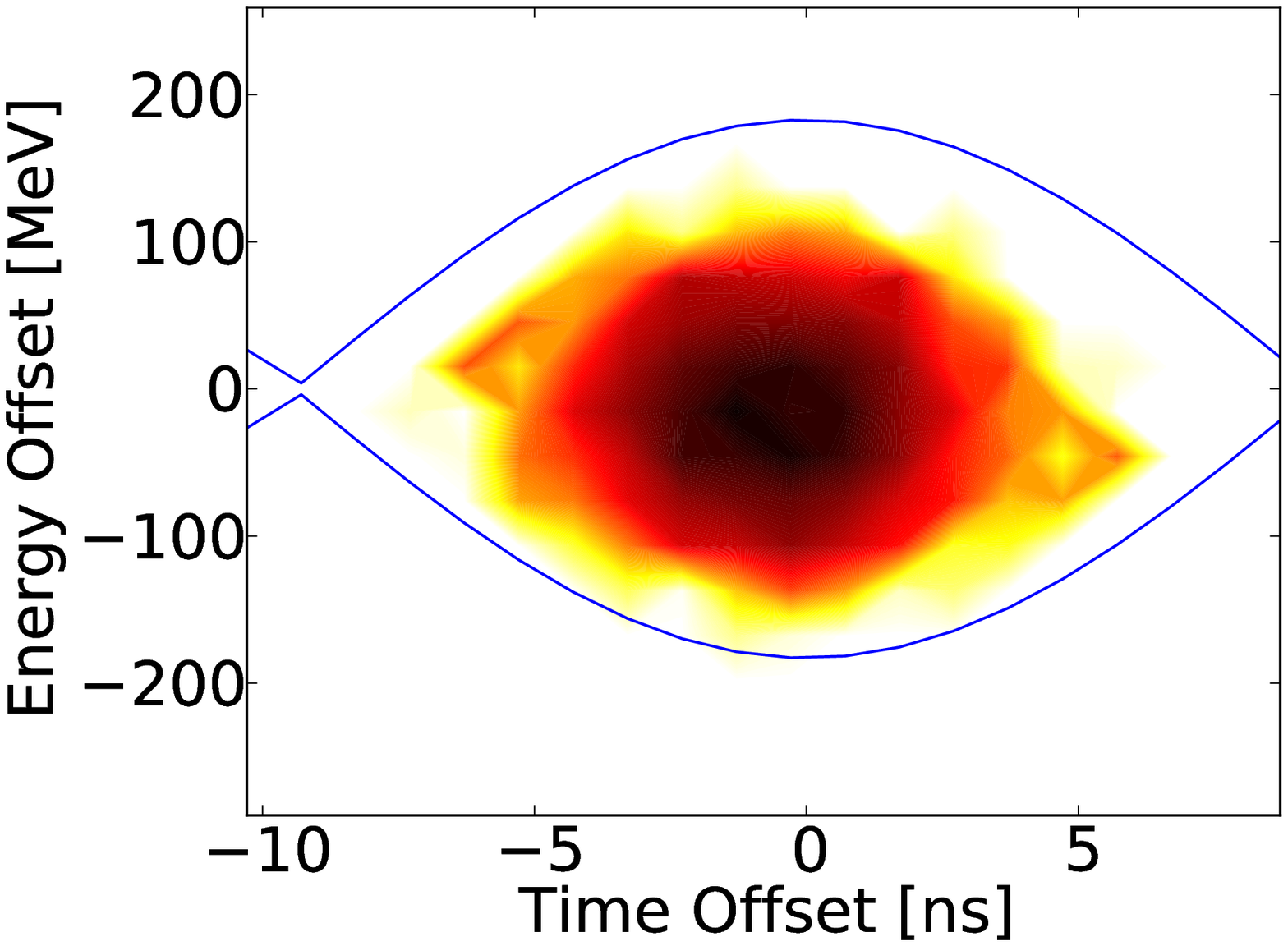}
  \includegraphics[scale=0.2]{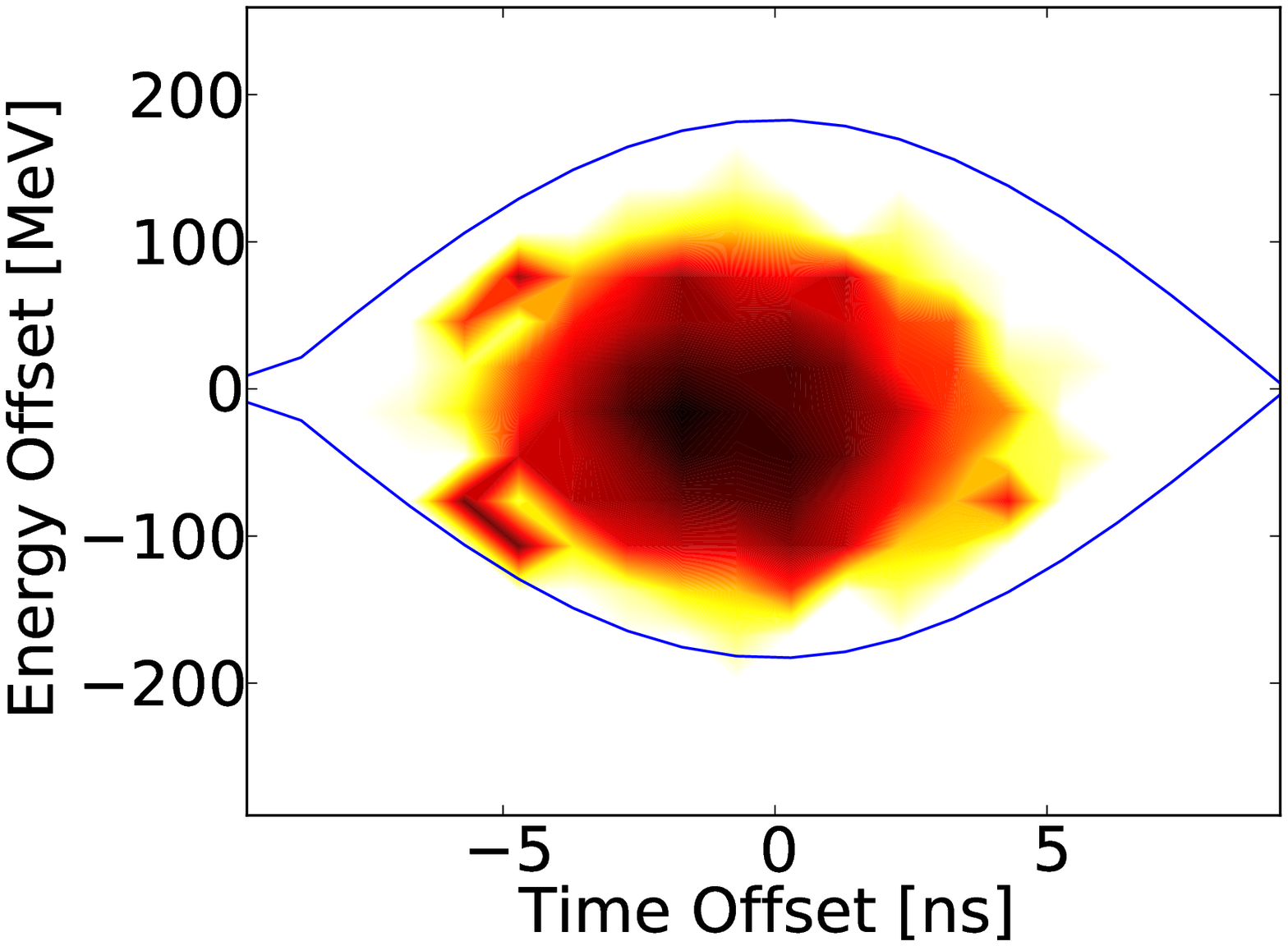}
  \newline
  \includegraphics[scale=0.15]{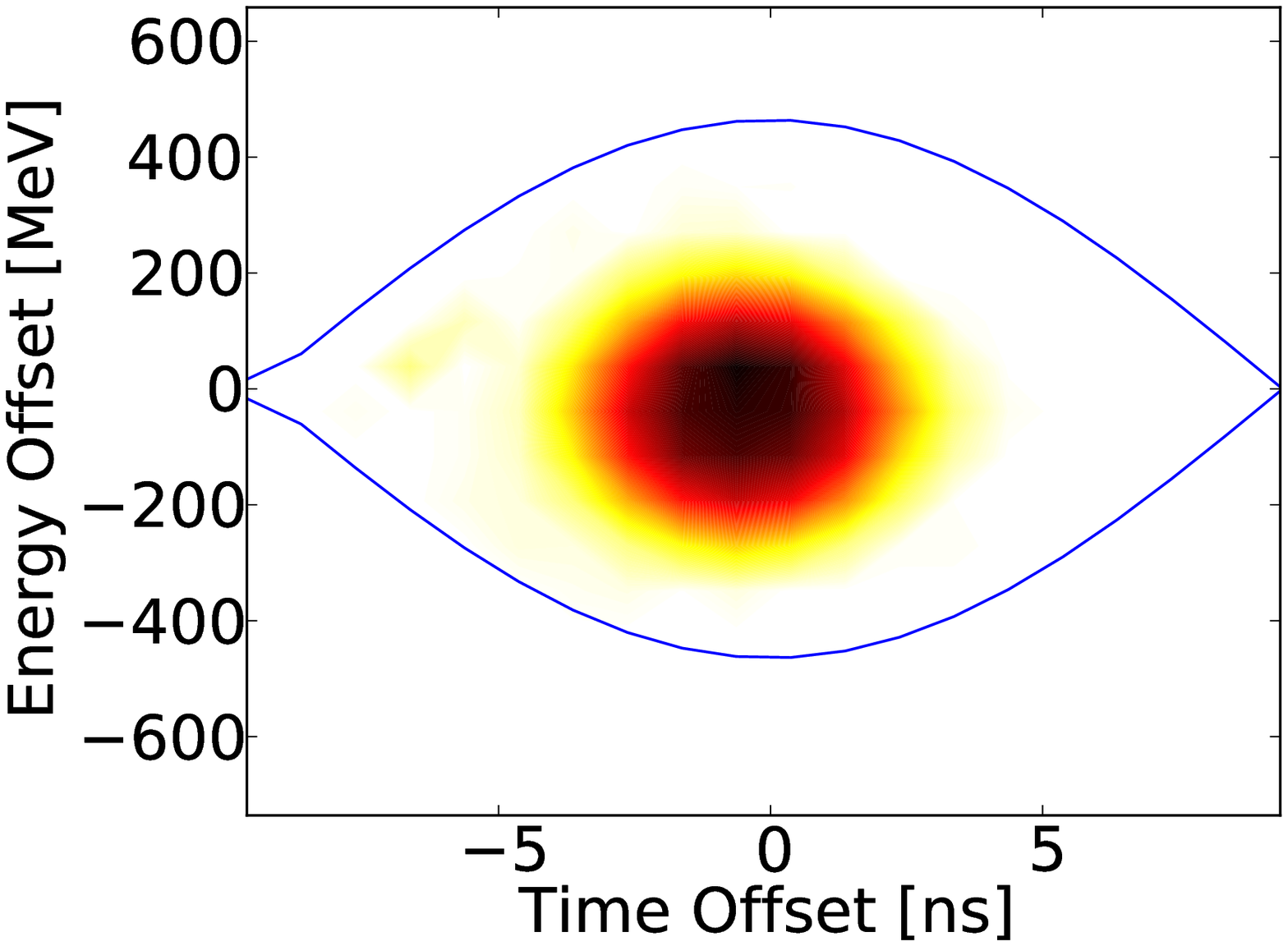}
  \includegraphics[scale=0.15]{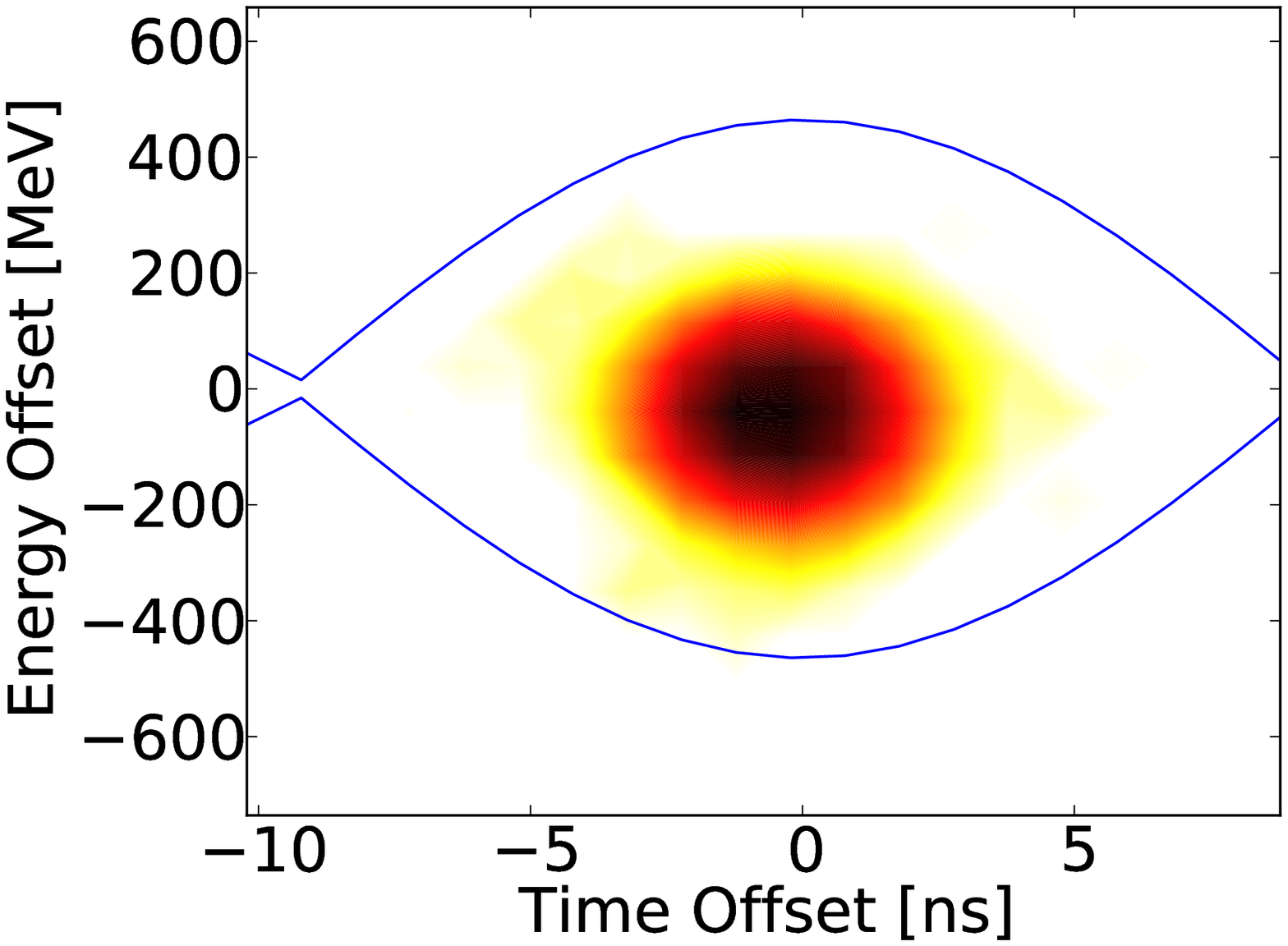}
  \includegraphics[scale=0.15]{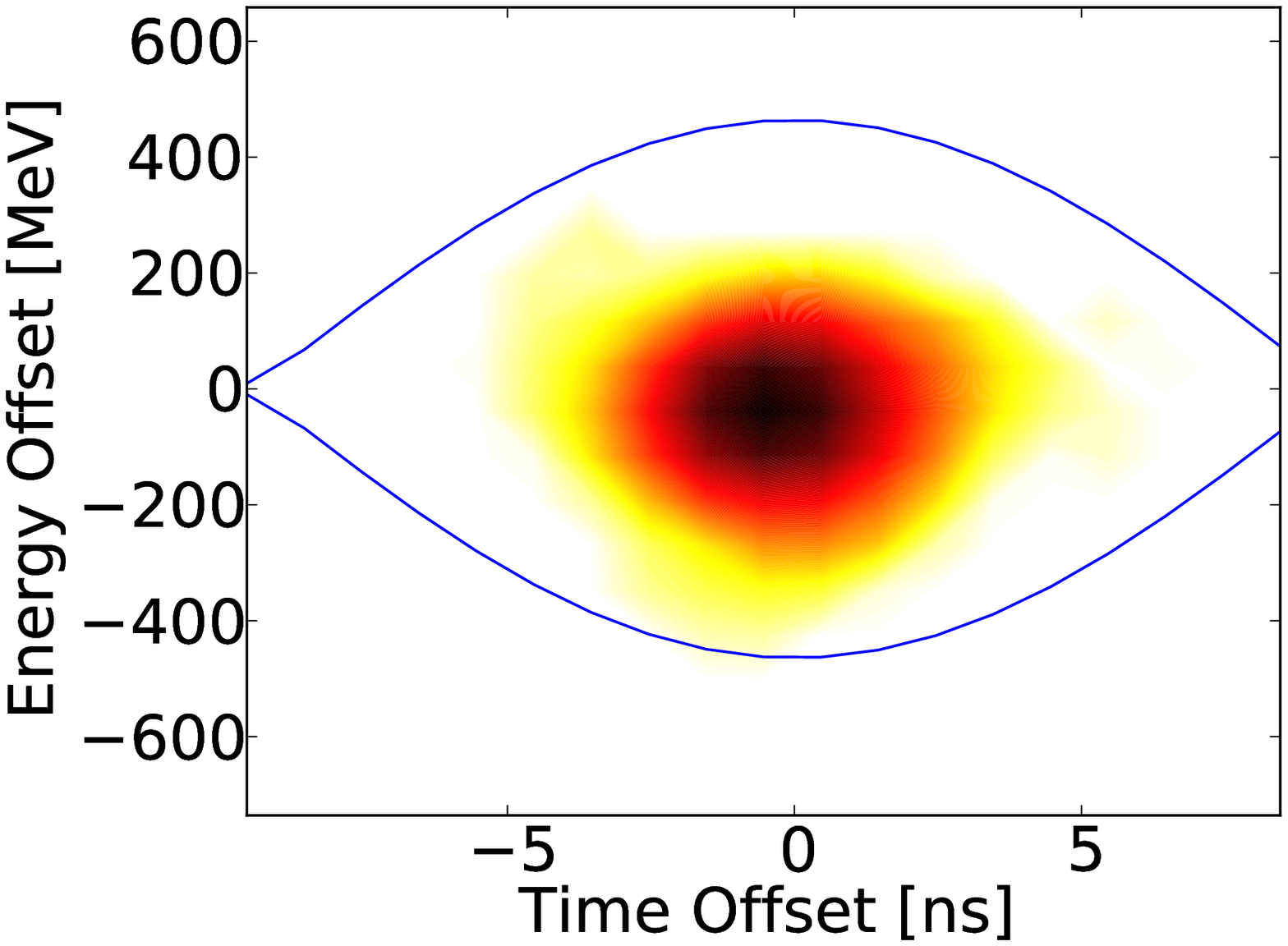}
  \includegraphics[scale=0.15]{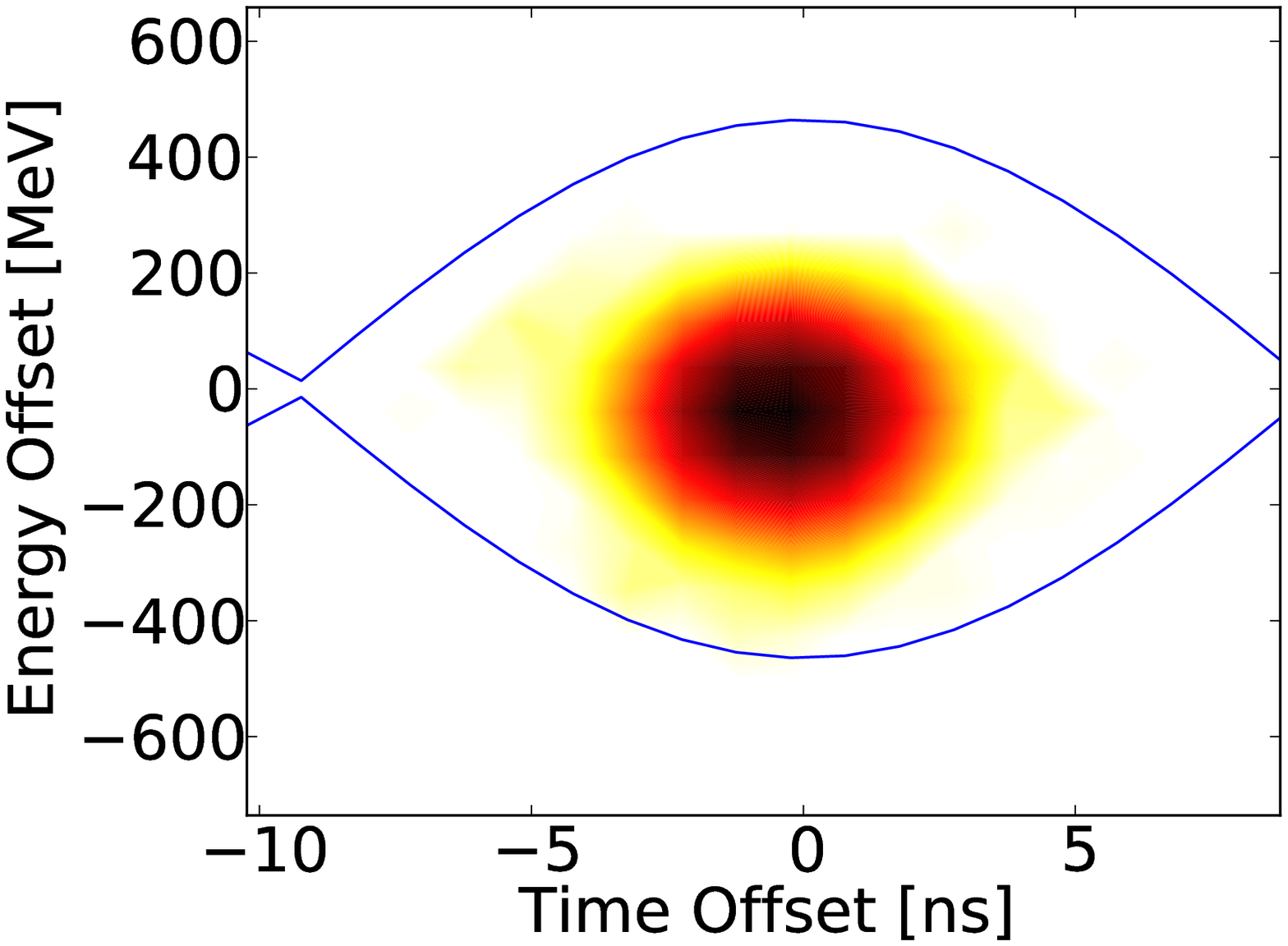}
\caption{Proton bunch 1 in Store 7949. First three images are during
injection and the last four are at top energy. }
  \label{fig:prot_1_psp_7949}
%
  \centering
  \includegraphics[scale=0.2]{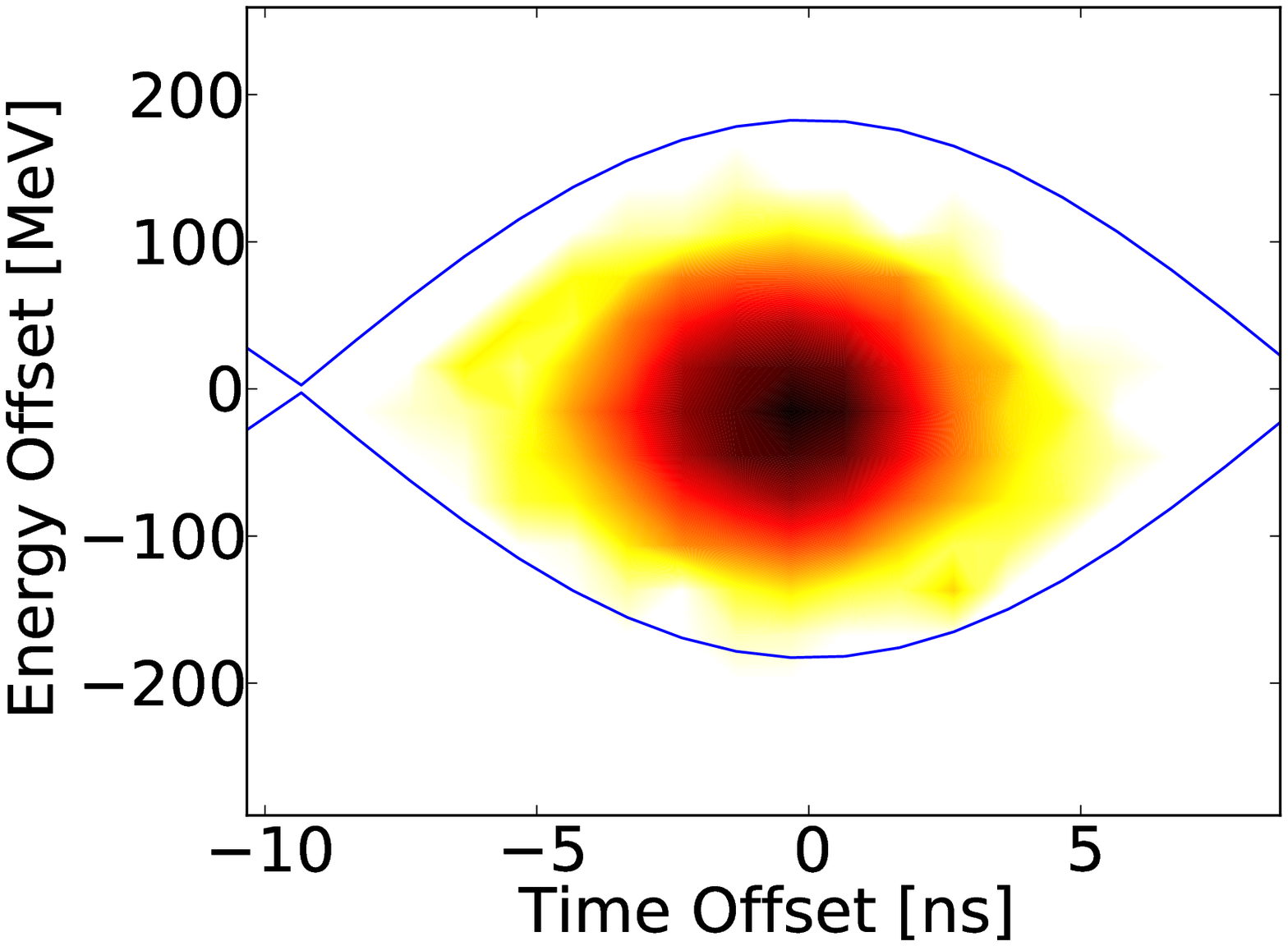}
  \includegraphics[scale=0.2]{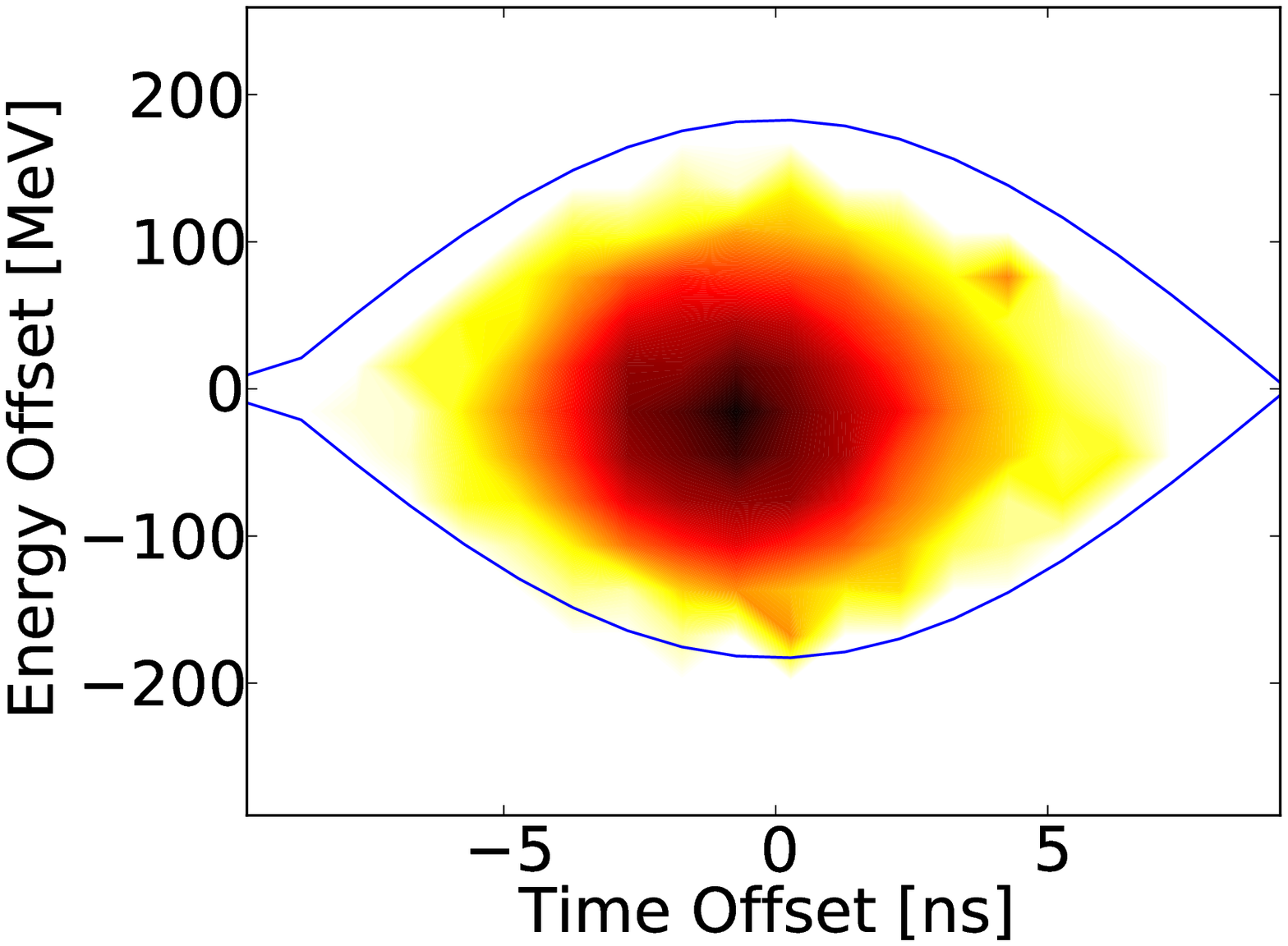}
  \includegraphics[scale=0.2]{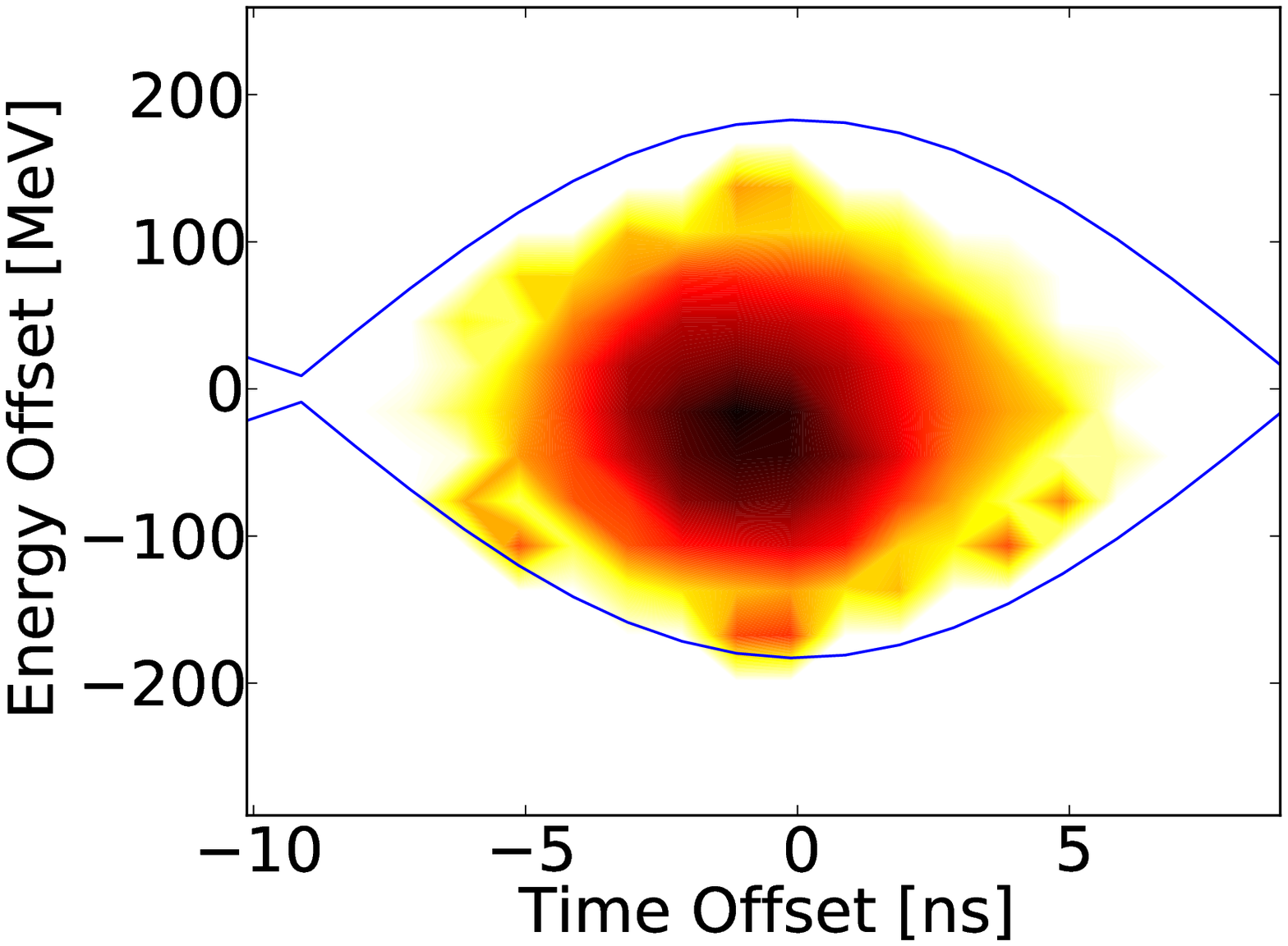}
  \newline
  \includegraphics[scale=0.15]{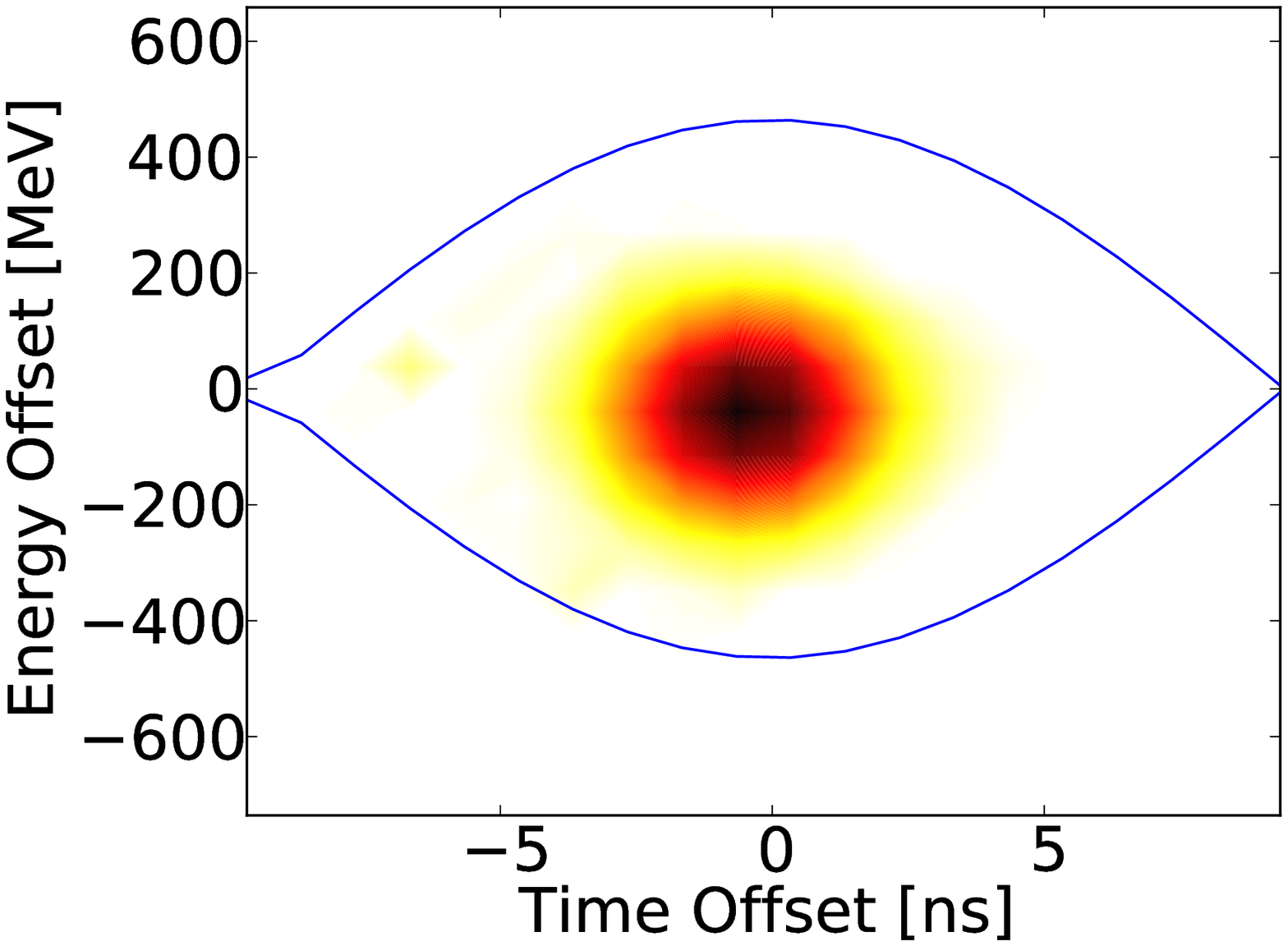}
  \includegraphics[scale=0.15]{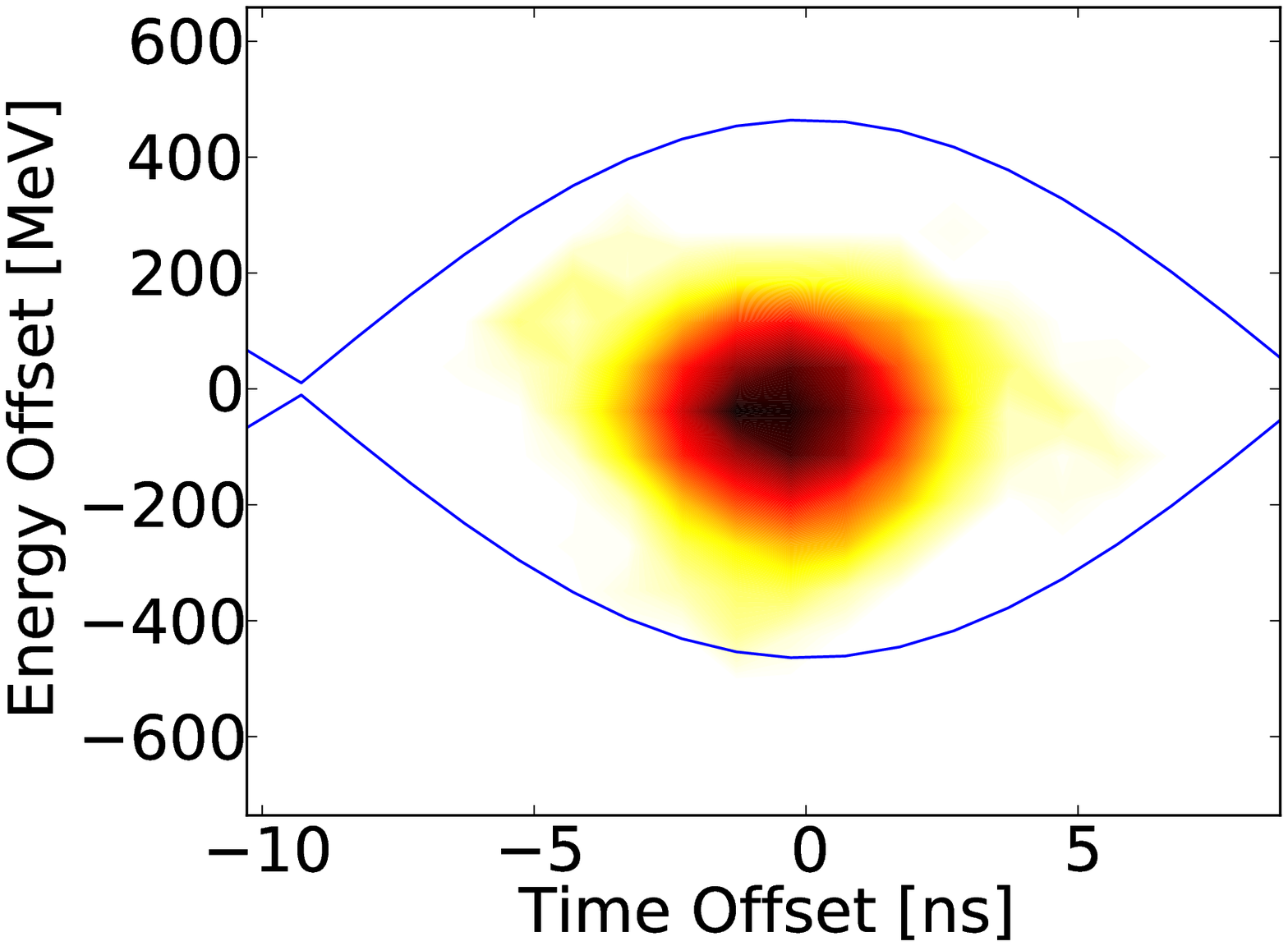}
  \includegraphics[scale=0.15]{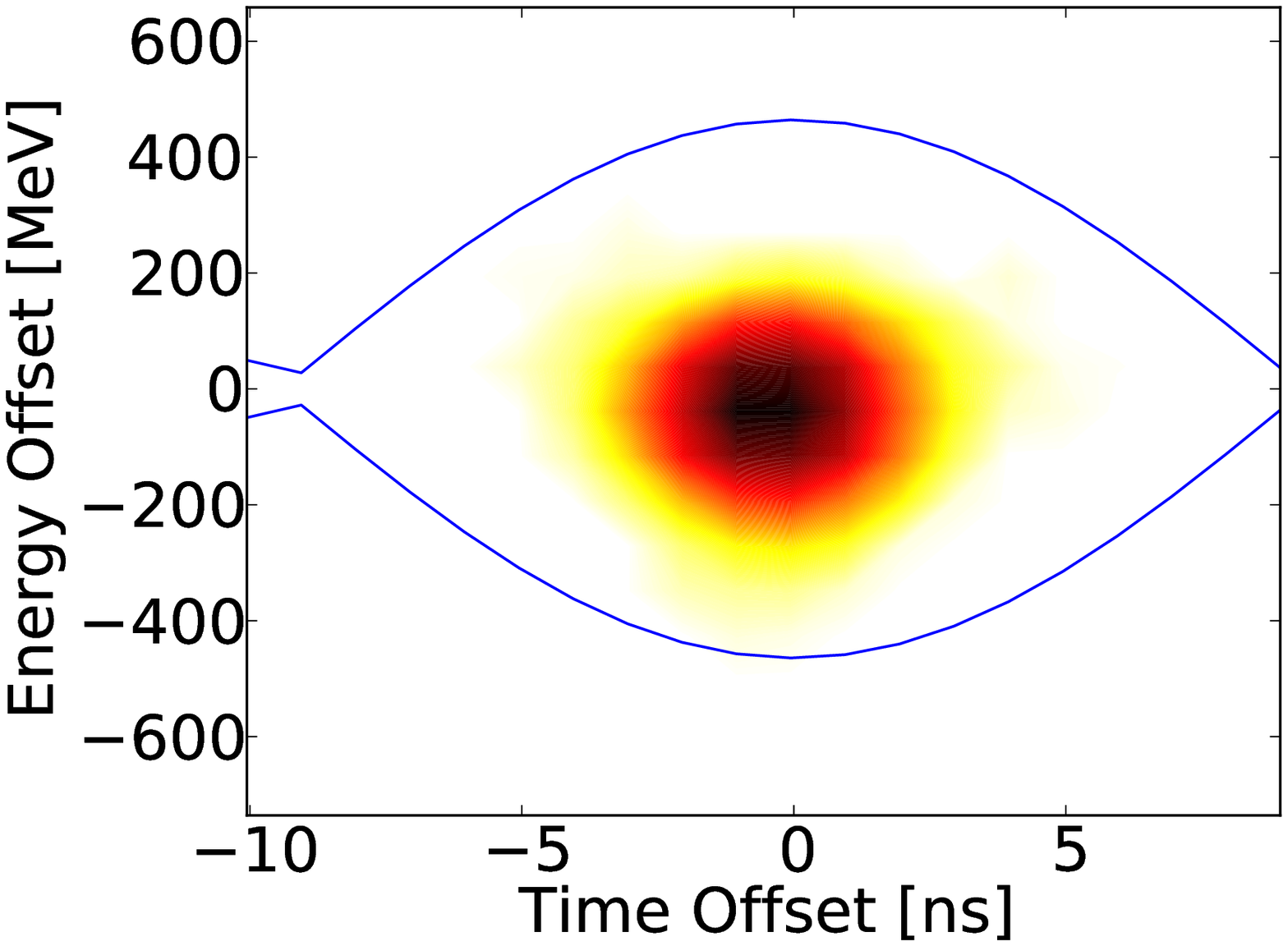}
  \includegraphics[scale=0.15]{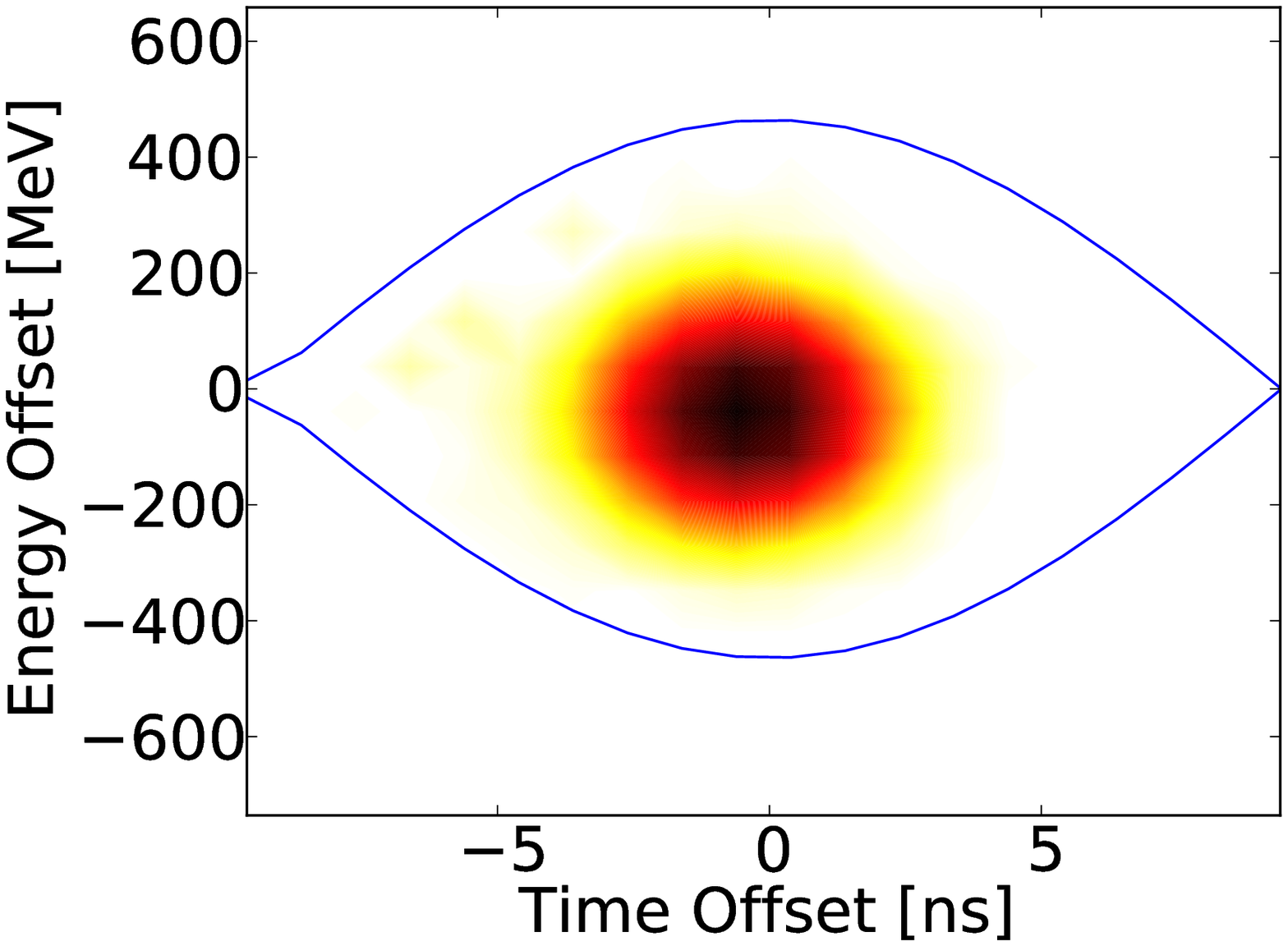}
\caption{Anti-proton bunch 11 in Store 7949: at 150 GeV(top) 
and 980 GeV (bottom). The stages are the same as in Figure 9.}
  \label{fig:pbar_11_psp_7949}
\end{figure}

Representative phase space plots for store 7949 are shown in Figures 
\ref{fig:prot_1_psp_7949} to \ref{fig:pbar_11_psp_7949}.
These show the phase space of proton bunch 1 and anti-proton bunch 11
in store 7949. 
\begin{figure}
\centering
\includegraphics[scale=0.15]{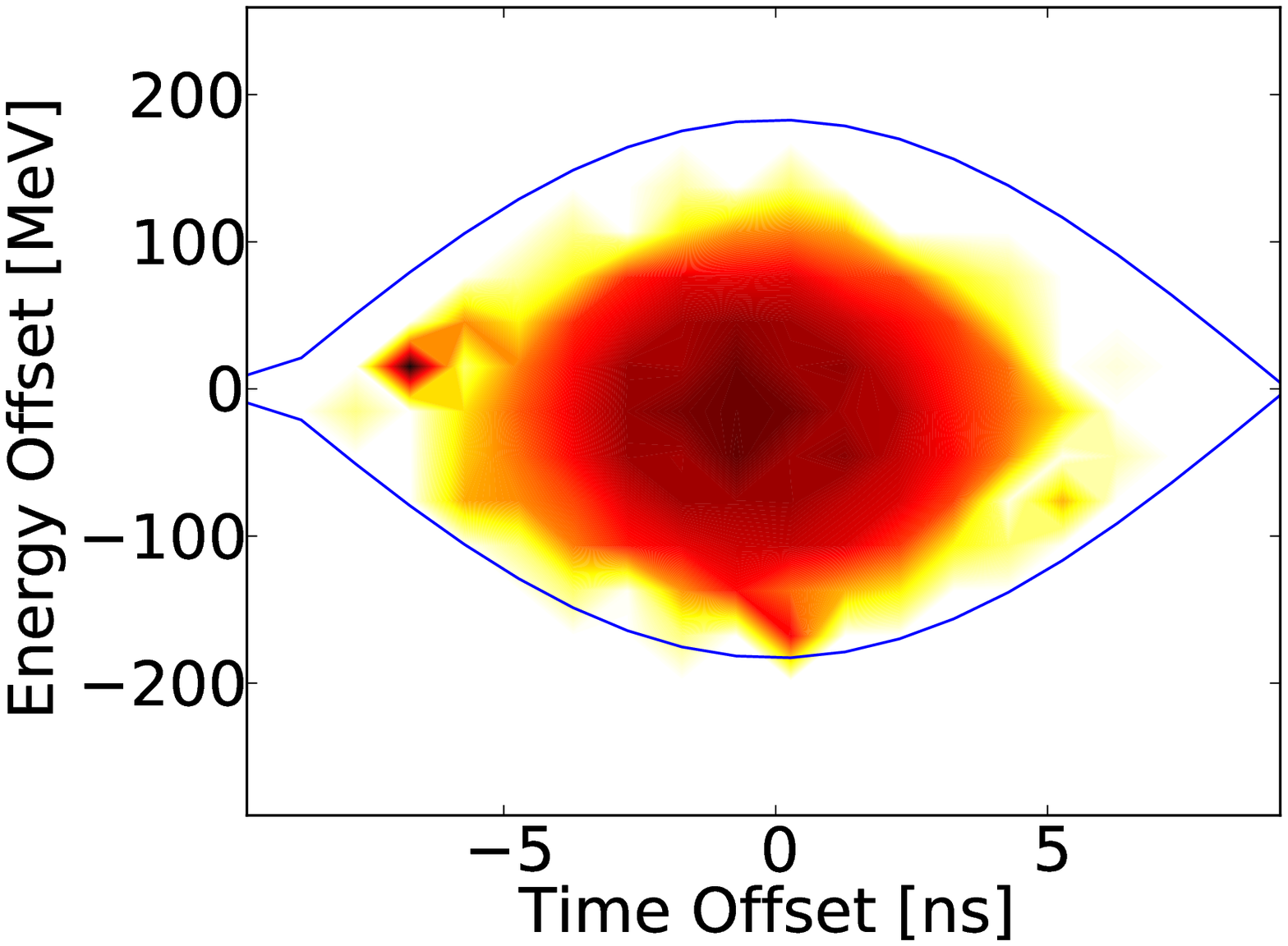}
\includegraphics[scale=0.15]{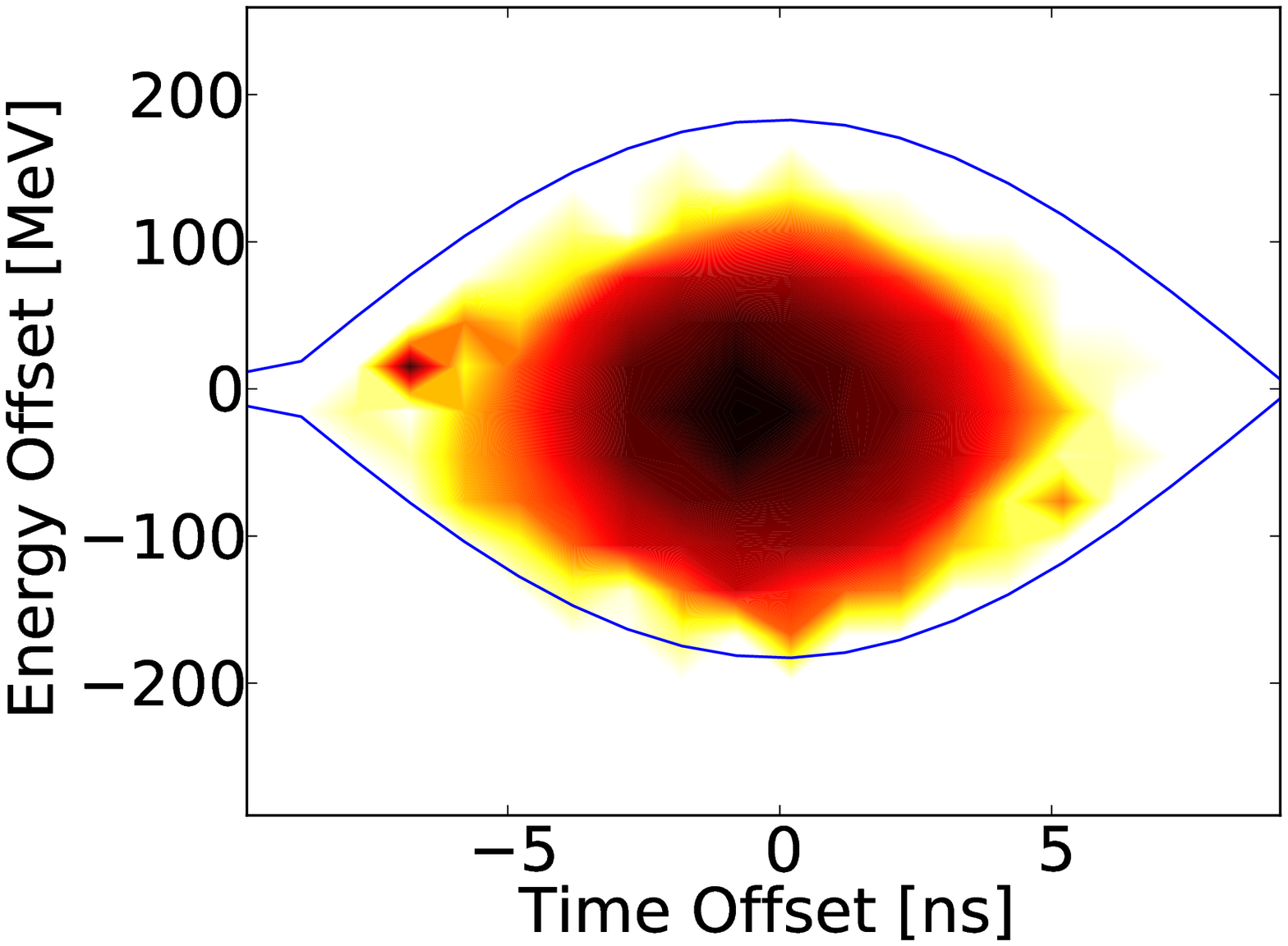}
\includegraphics[scale=0.15]{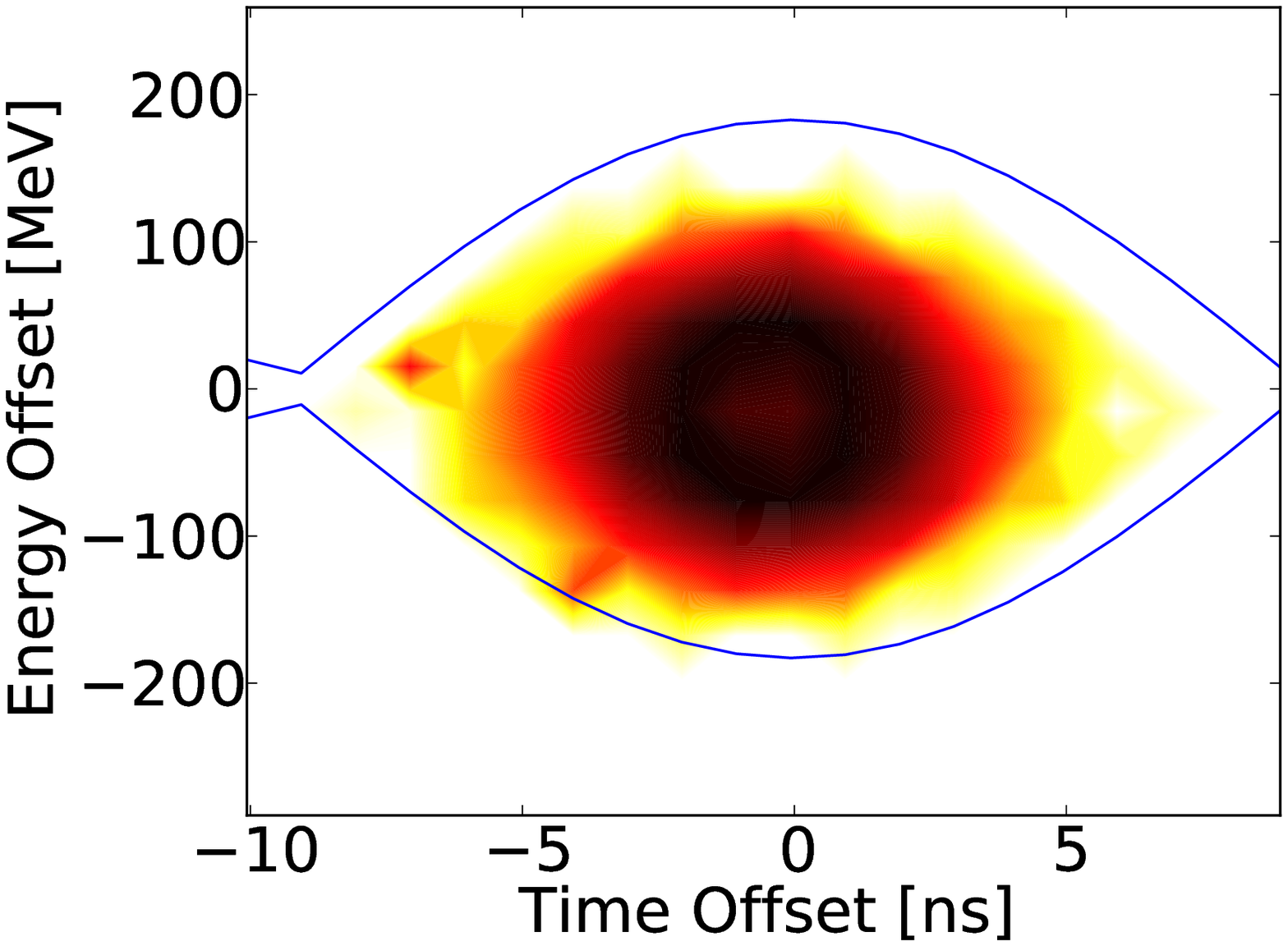}
\includegraphics[scale=0.15]{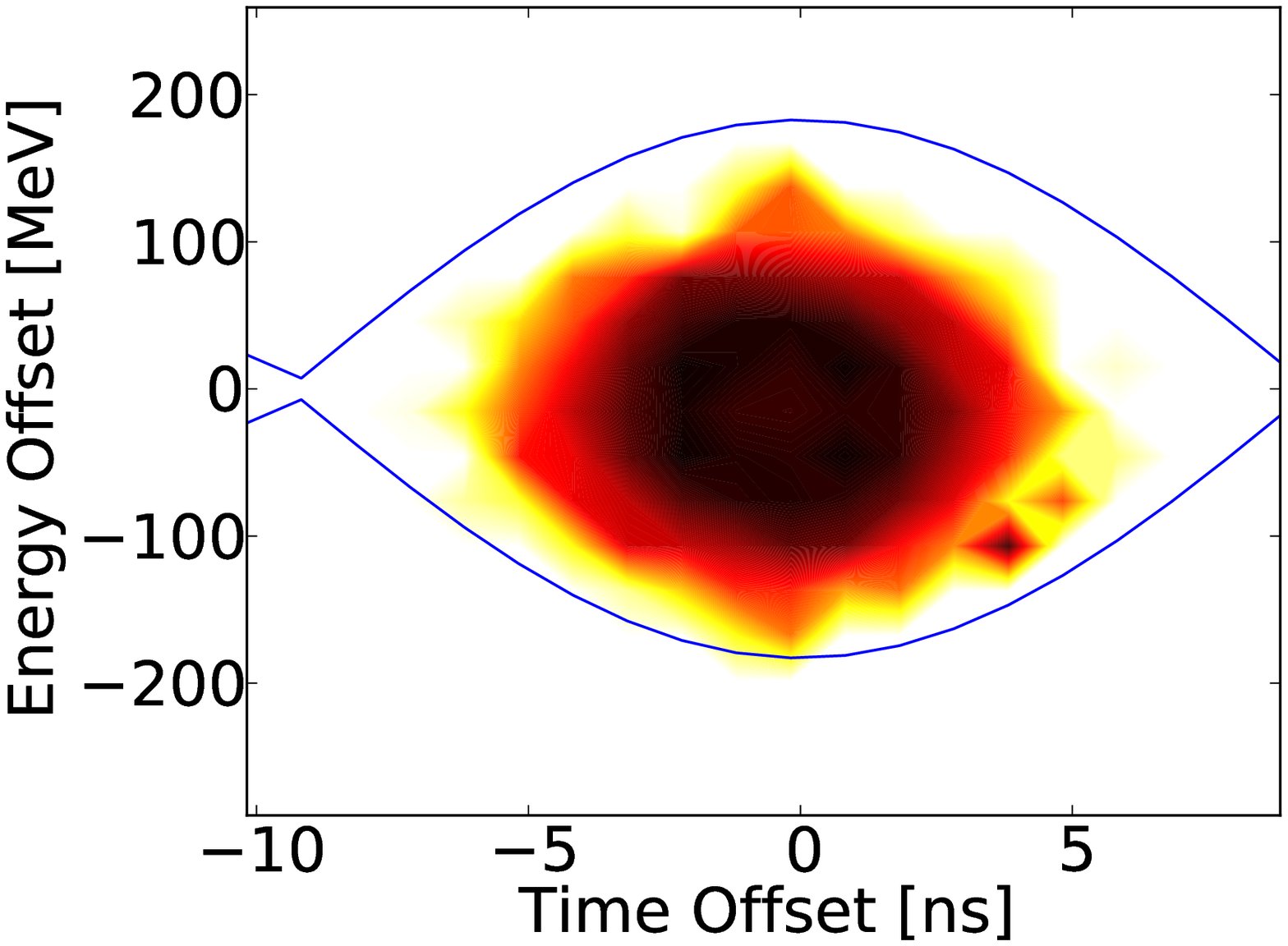}
\newline
\includegraphics[scale=0.15]{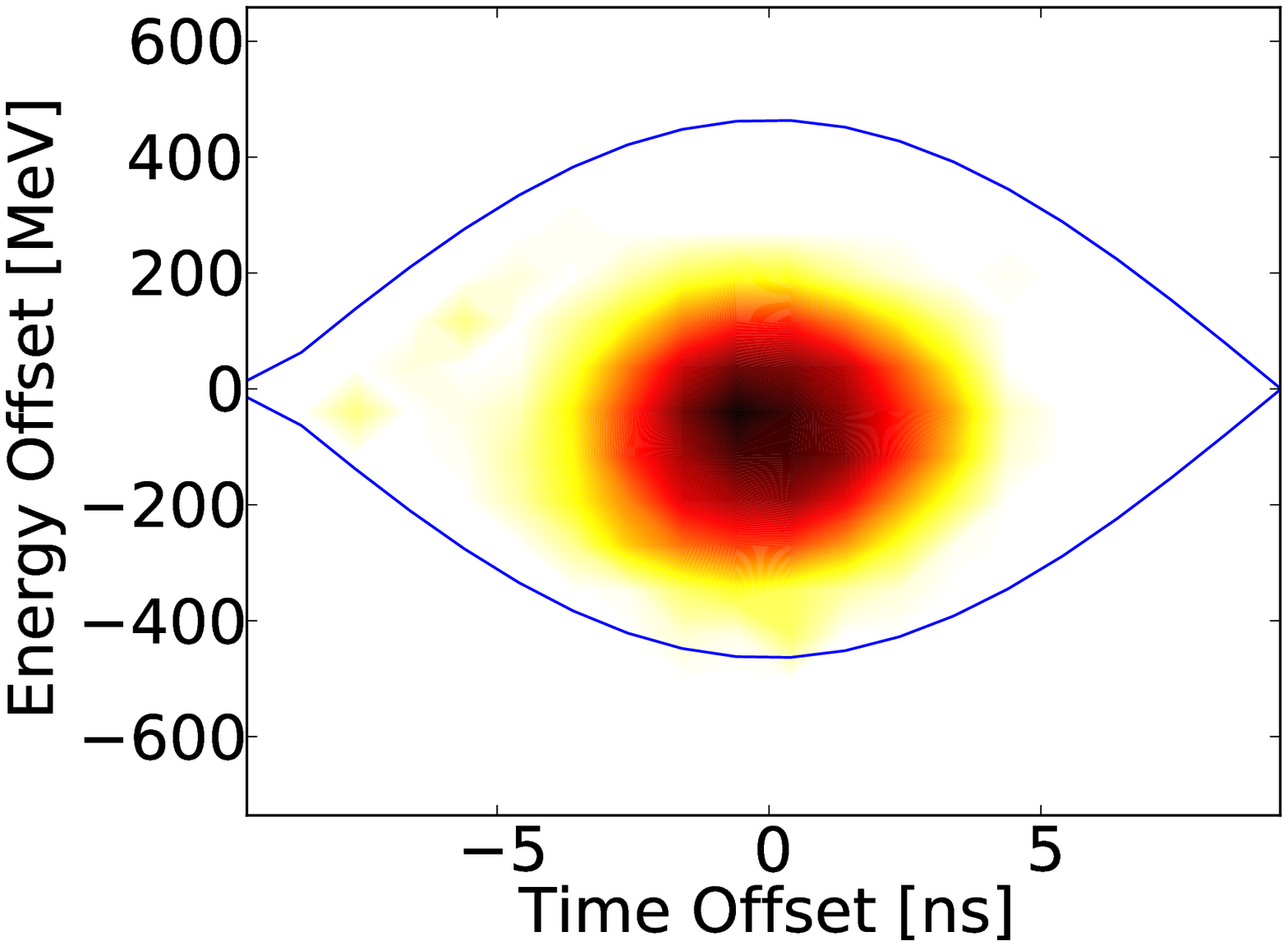}
\includegraphics[scale=0.15]{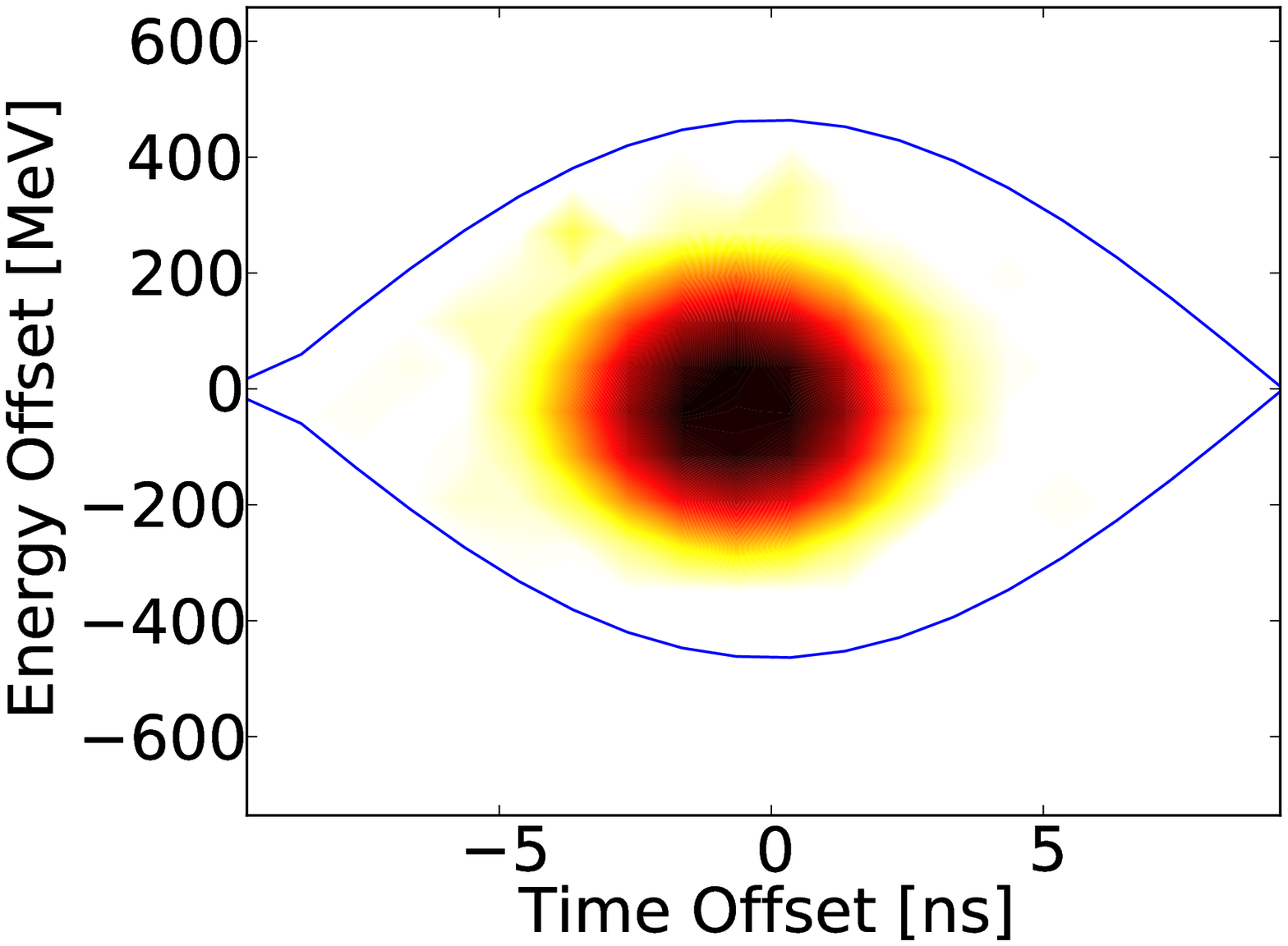}
\includegraphics[scale=0.15]{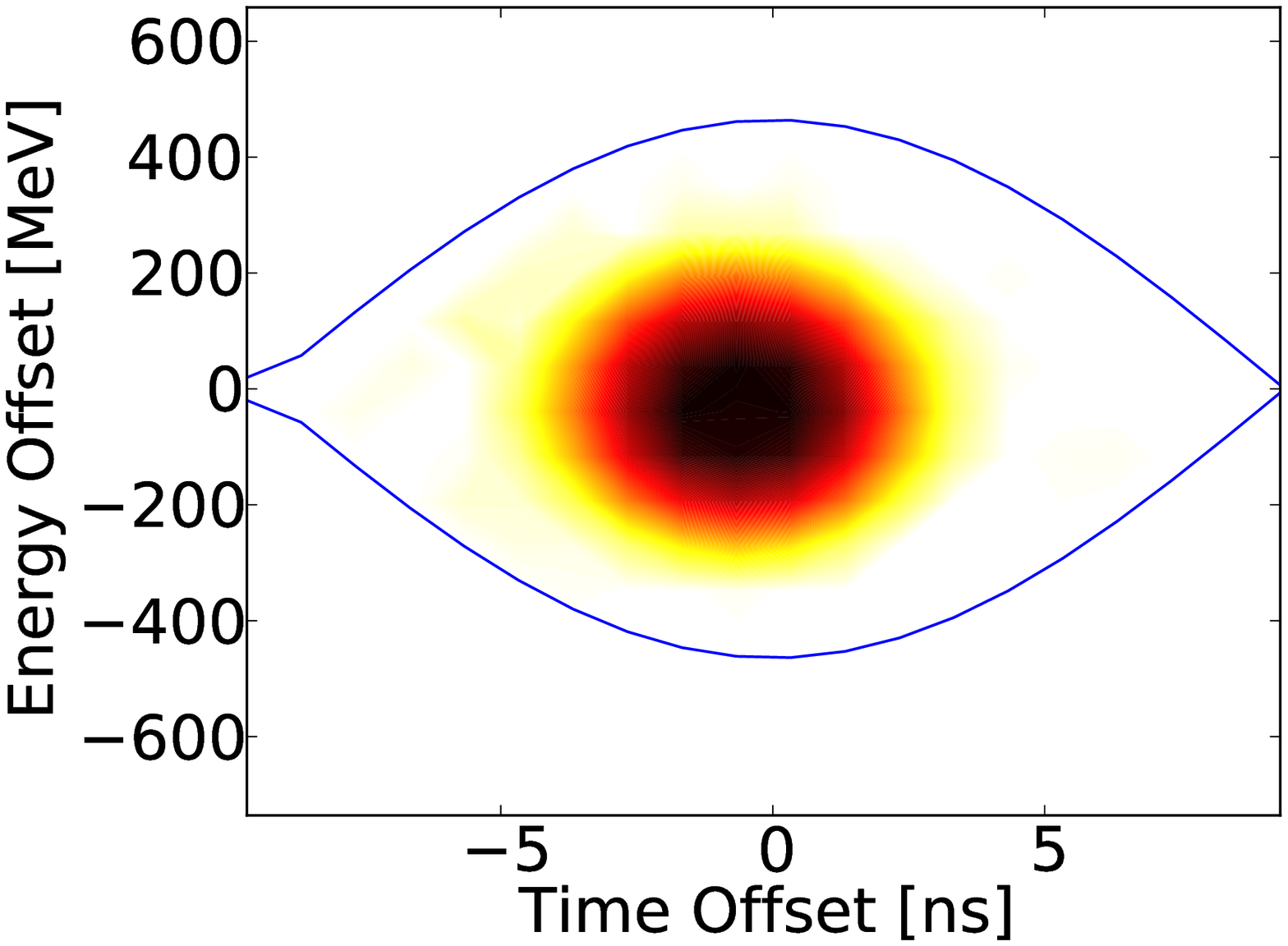}
\includegraphics[scale=0.15]{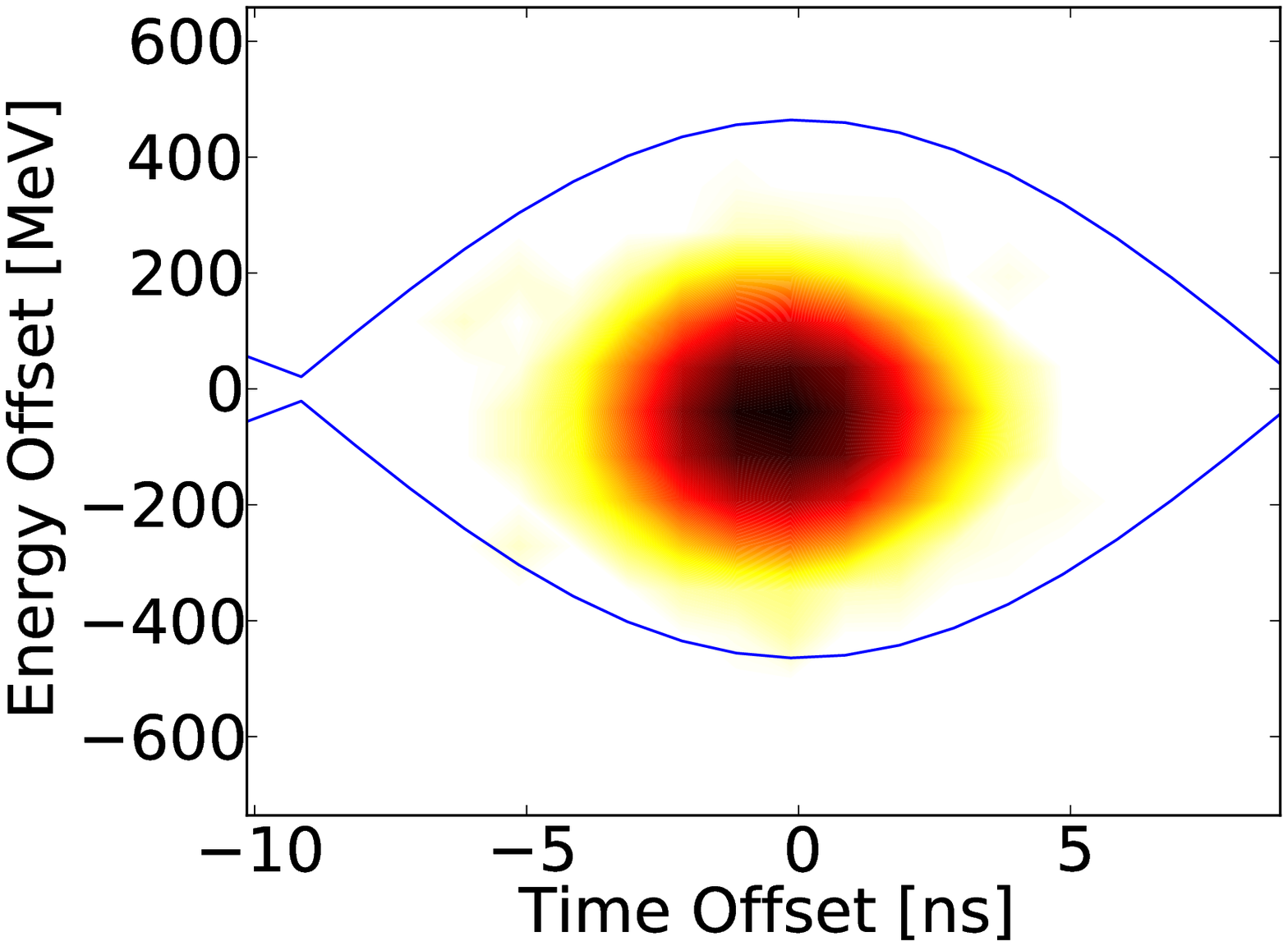}
\caption{Phase space plots of proton bunch 1 in Store 8146 at 4 stages 
during injection (top) and 4 stages at 980 Gev (bottom).}
\label{fig: prot1-8146}
%
\centering
\includegraphics[scale=0.3]{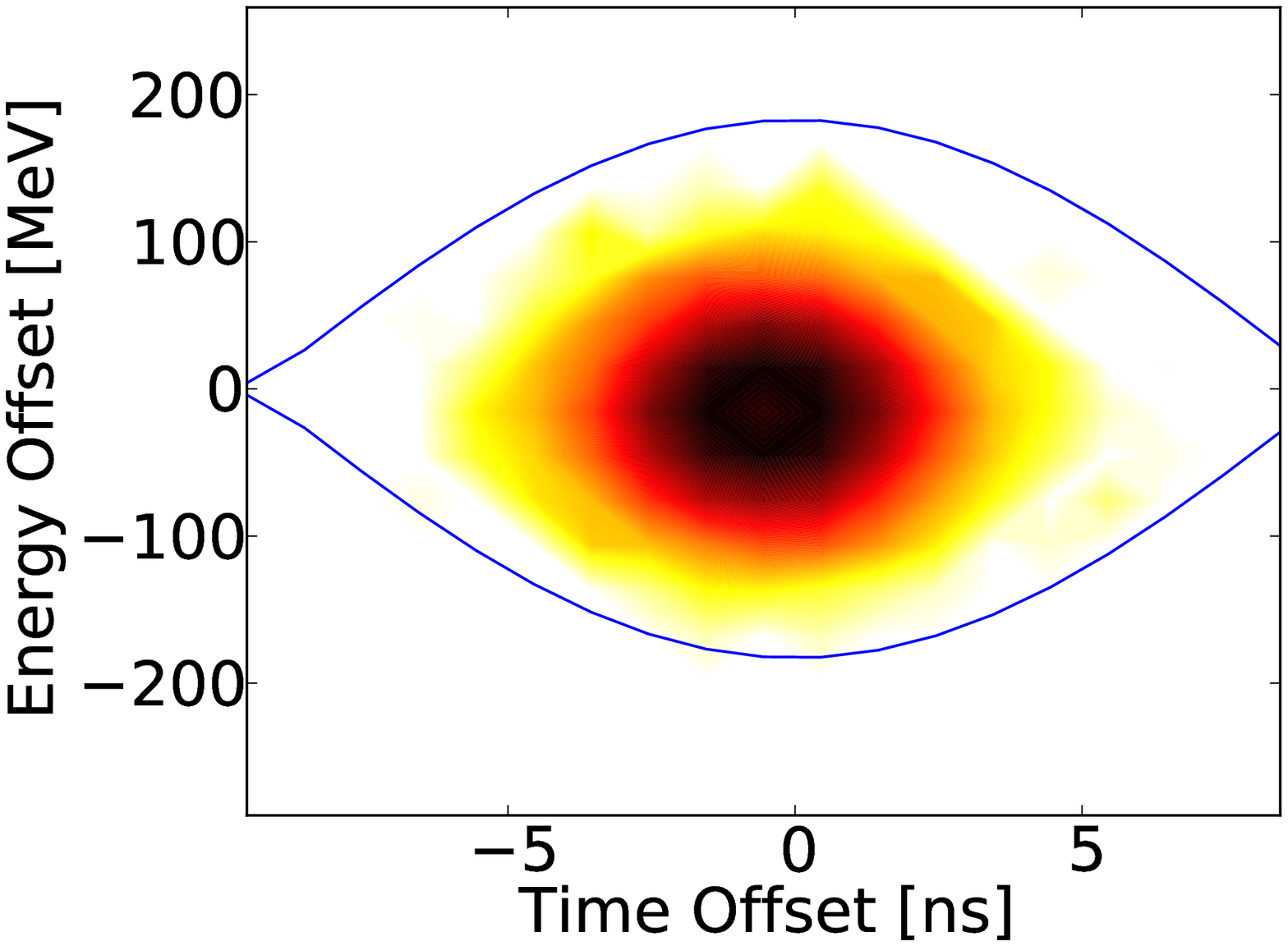}
\newline
\includegraphics[scale=0.15]{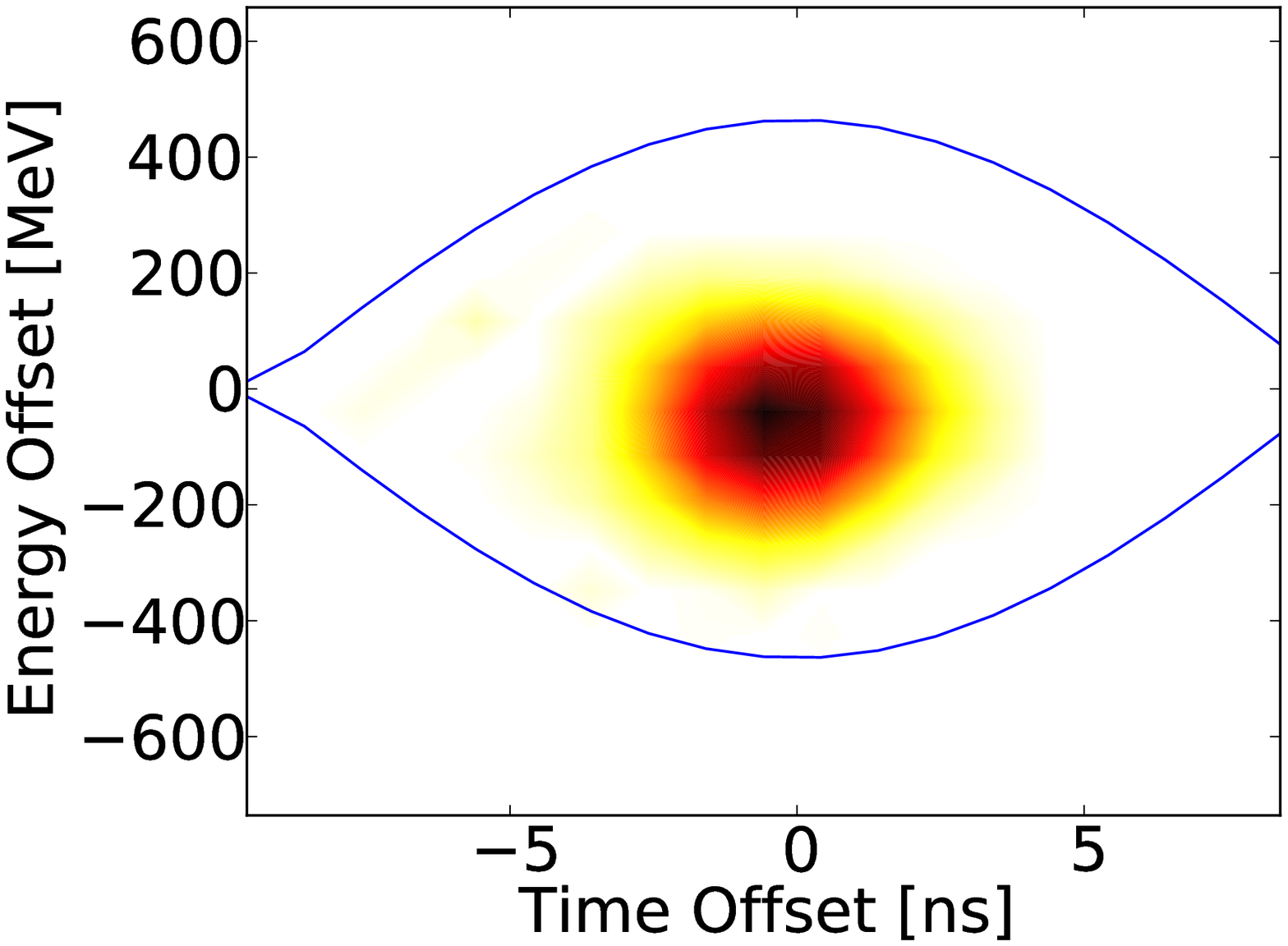}
\includegraphics[scale=0.15]{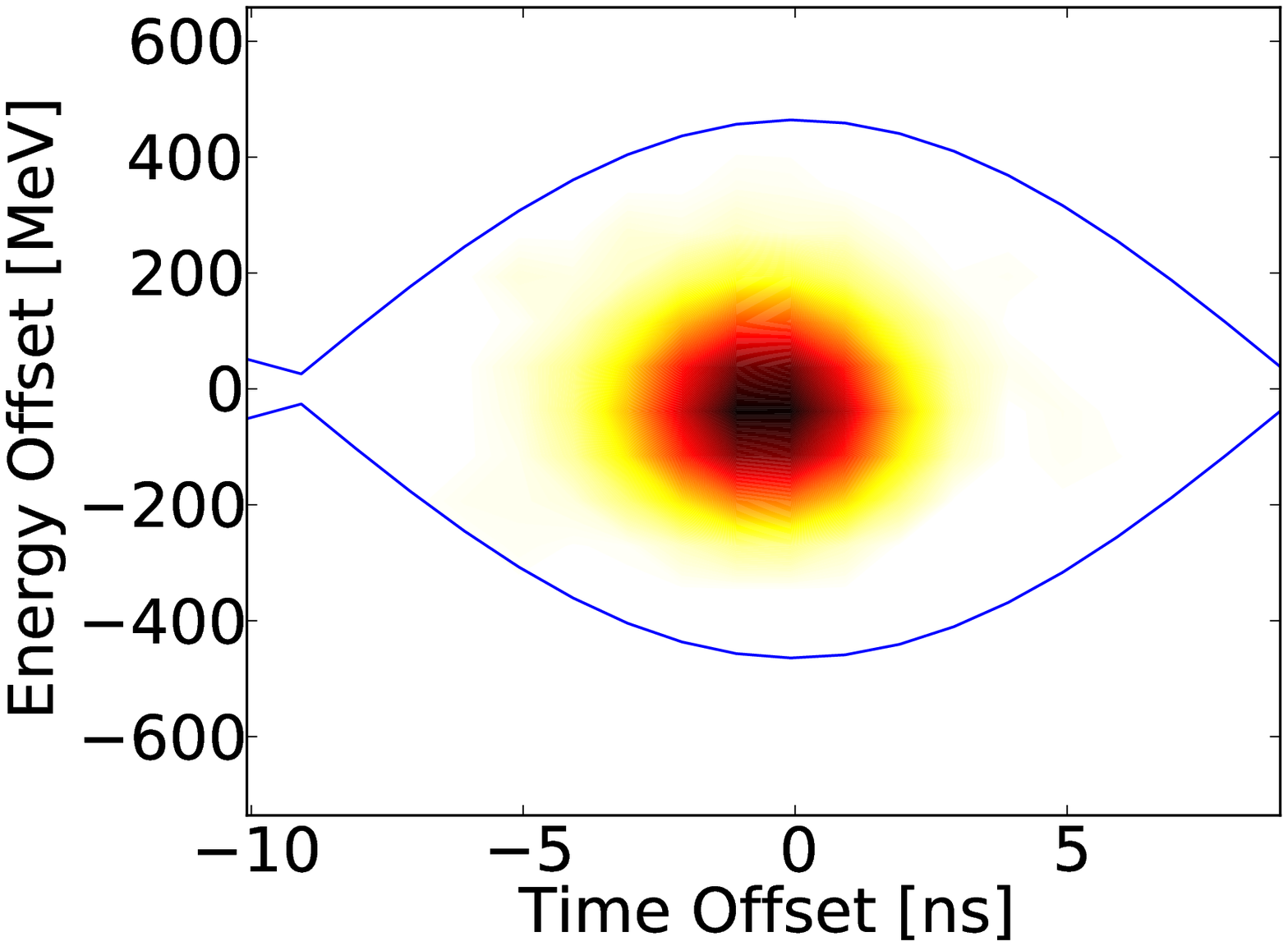}
\includegraphics[scale=0.15]{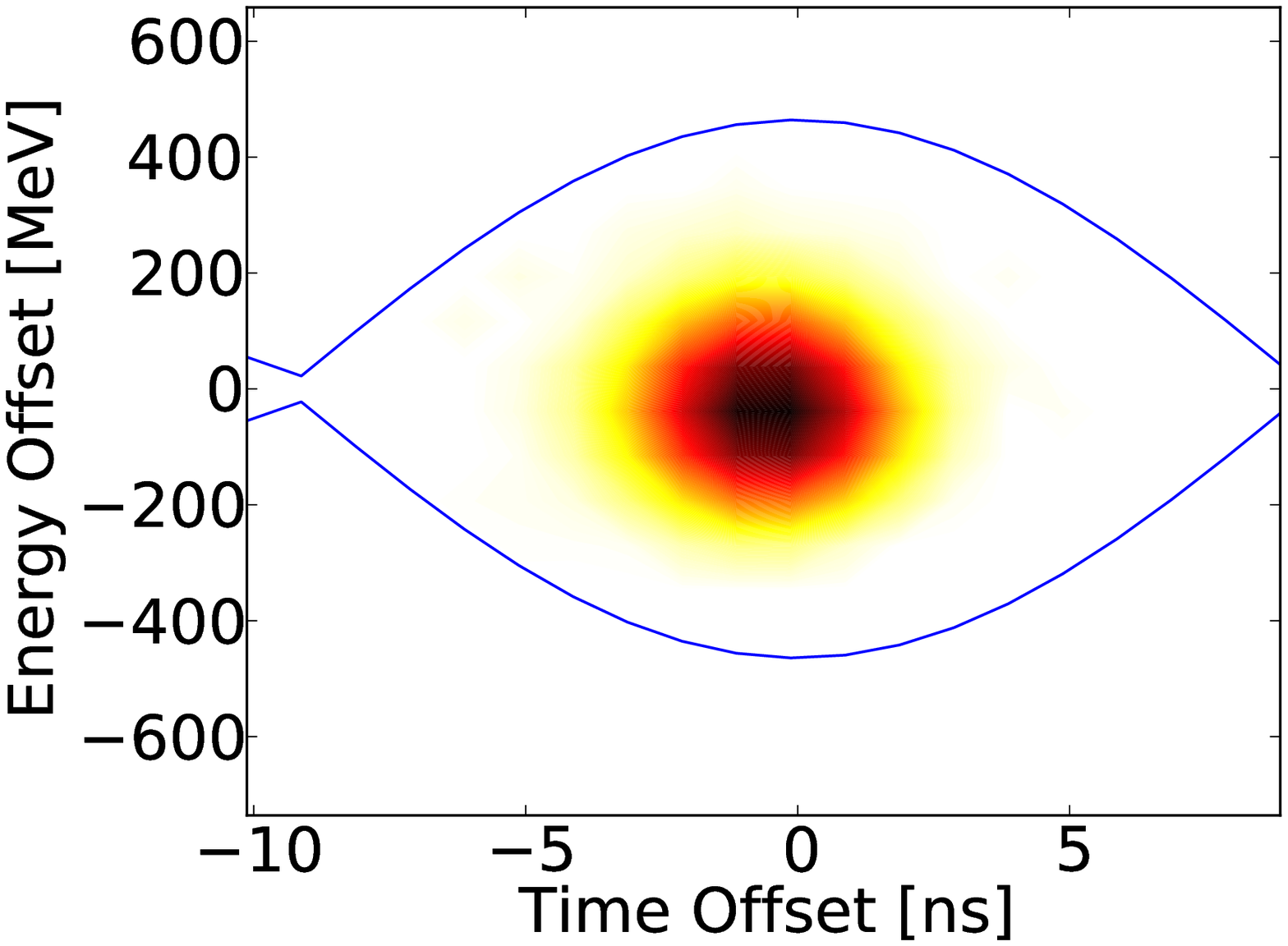}
\includegraphics[scale=0.15]{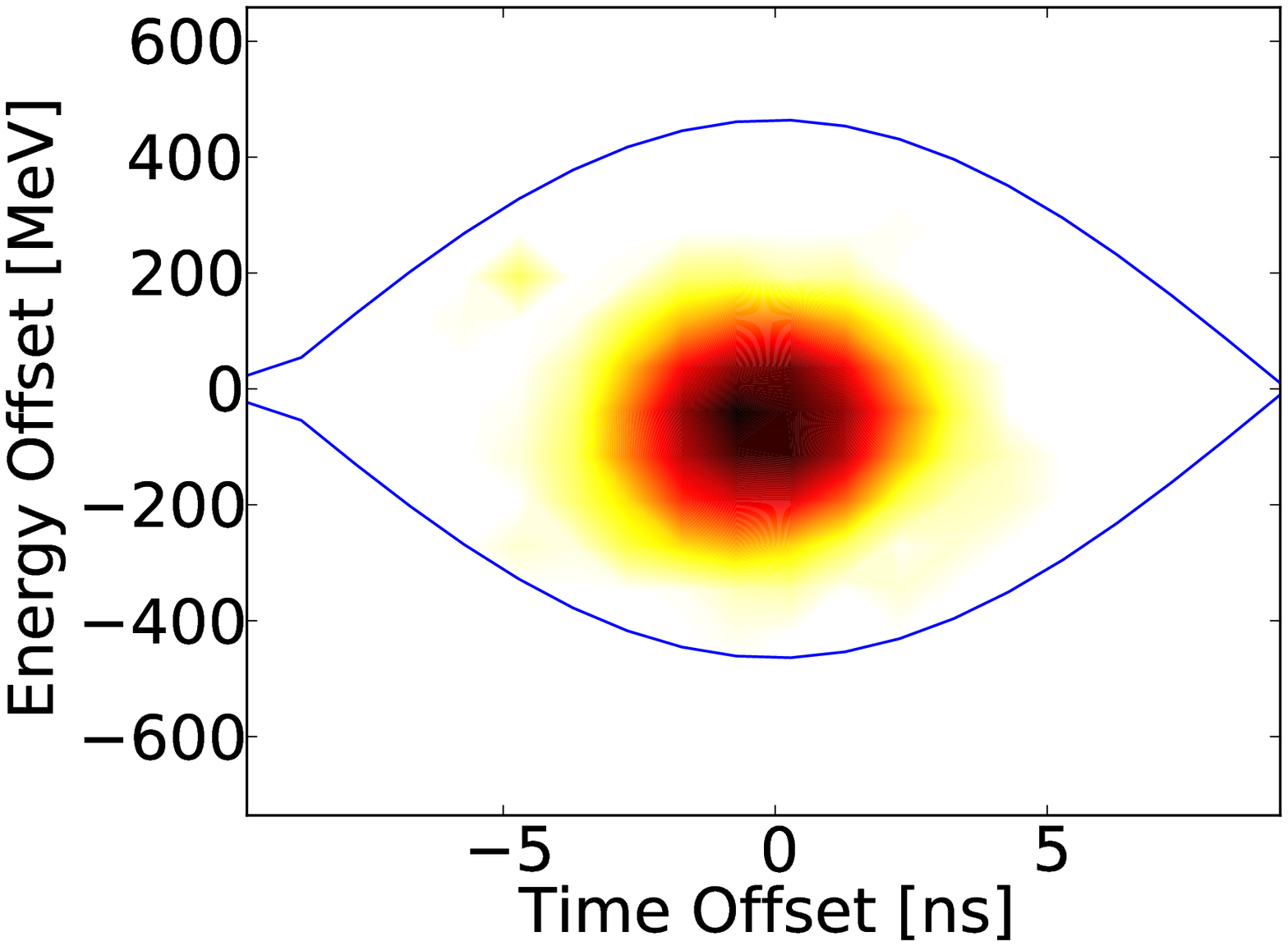}
\caption{Phase space plots of anti-proton bunch 2 in Store 8146  at the 4th stage of injection (top) and at 4 stages at 980 Gev (bottom).}
\label{fig: pbar2-8146}
\end{figure}
Figure \ref{fig: prot1-8146} and \ref{fig: pbar2-8146} show the phase
space of proton bunch 1 and anti-proton bunch 2 during different stages
of store 8146.
As mentioned in Section \ref{sec: measure}, anti-proton bunches numbered
1-4 are injected during the first stage of injection but at locations
closer to the tail of the proton train. They are then moved towards the 
head of the proton train in subsequent coggings. Consequently in
store 8146, we have longitudinal profiles and phase space 
reconstruction at only the third stage at injection energy for 
the profiles of anti-proton bunches A2, A3.

The phase space plots of proton bunches at 150 GeV in both stores 
show small blobs at the edges and these move around in bunch phase space. 
These blobs are formed during coalescing of seven bunches in the Main Injector to form one Tevatron bunch. The existence of these blobs shows that the bunch is not yet in equilibrium at 
150 GeV. We have found these blobs 
to exist in all proton bunches recorded in both stores. The more intense
 proton bunches nearly fill the bucket at injection and we also observe
beam loss and some longitudinal clipping, e.g. in Figure \ref{fig: sigt_kurt_8146}. 

The anti-proton bunches in the Tevatron are formed by the coalescing of five lower 
intensity anti-proton bunches in the Main Injector. The data from store 8146 
allowed a finer resolution of phase space due to the greater number of projection
 angles. Substructure or  blobs are not seen in the phase space of either of the 
two anti-proton bunches captured in this store and there is no observable 
longitudinal clipping on anti-protons at 150 GeV.

At 980 GeV, the bunches are smaller in the bucket and neither protons
or anti-protons appear to have any structure or distortions from the
expected shapes. The differences between stages are not easily
discernible. 

There are, of course, errors in the reconstructed phase space.
Some of the errors are fundamental to the reconstruction process,
others arise in the particle tracking since we assume design values for
machine parameters, only the ideal rf voltage is used, and no perturbations are included in the tracking.

\section{Momentum distributions}

The momentum distributions can be obtained by projecting the phase
space distribution onto the momentum axis. 
\begin{figure}
\centering
\includegraphics[scale=0.5]{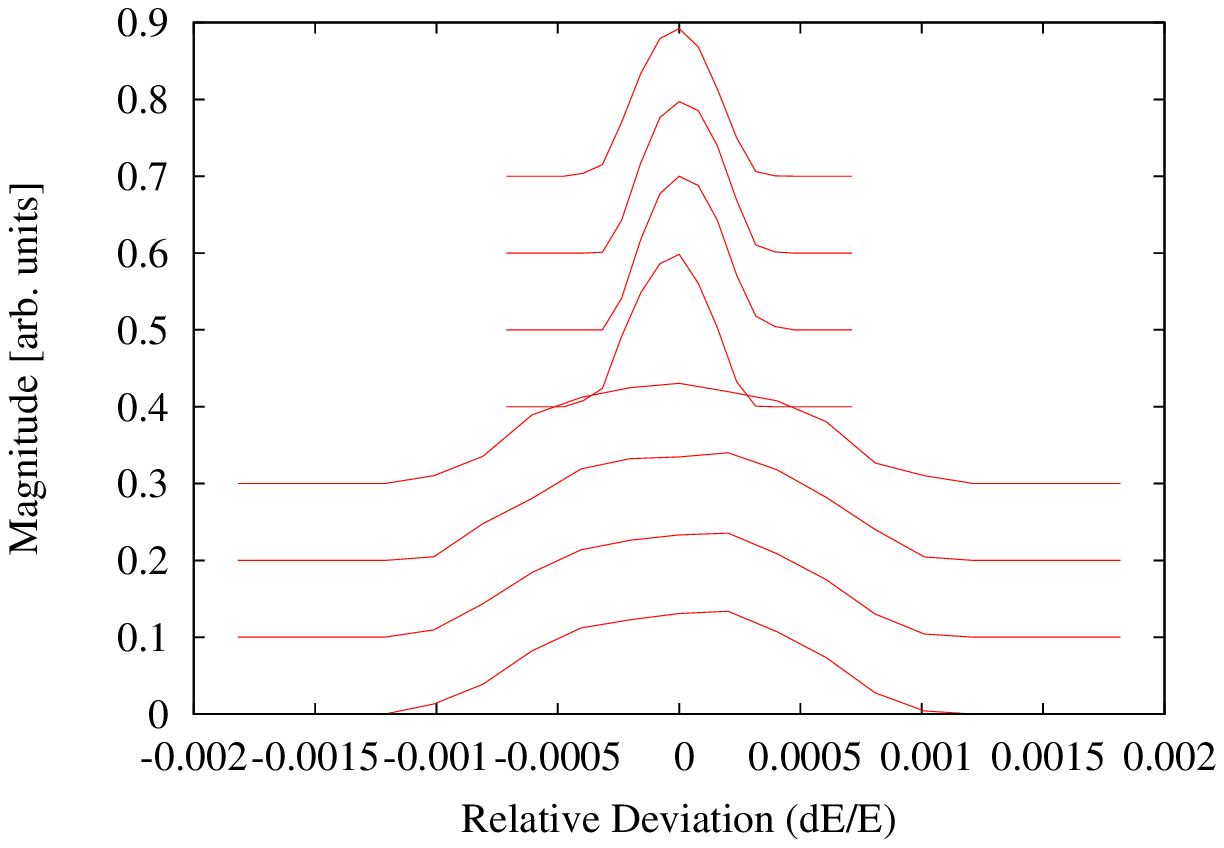}
\includegraphics[scale=0.5]{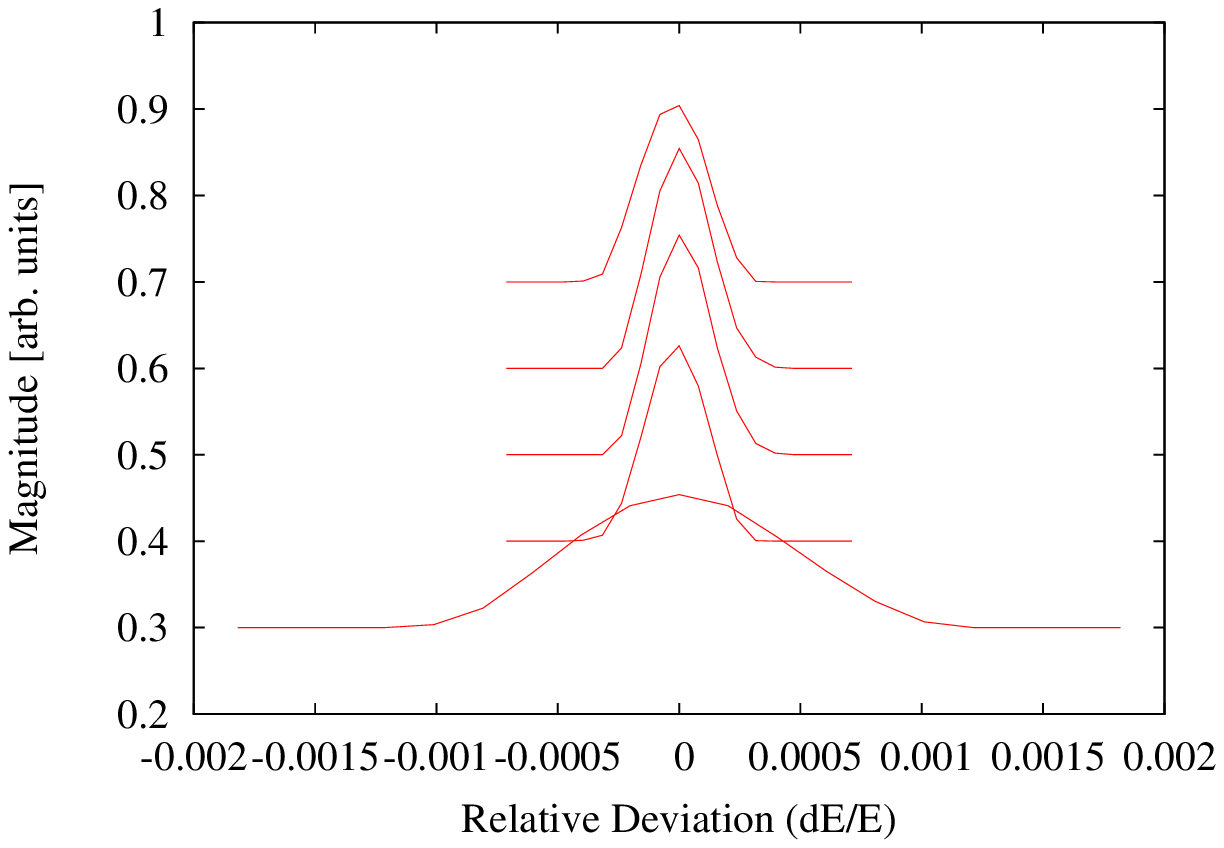}
\caption{Momentum profiles of the first proton bunch (left) and the first
anti-proton bunch (right) during the different stages from injection to 3 hours 
after start of collisions in Store 8146. The bottom four stages for
the proton bunch and the bottom stage for the anti-proton bunch are
at injection energy.}
\label{fig: momdistrib_8146}
\end{figure}
Figure \ref{fig: momdistrib_8146} shows representative momentum profiles
constructed for a proton bunch and an anti-proton bunch over different
stages from 150 GeV to 980 GeV in store 8146. 
As expected, the proton bunches have a larger momentum spread at all stages.

\begin{figure}[h]
\centering
\includegraphics[scale=0.5]{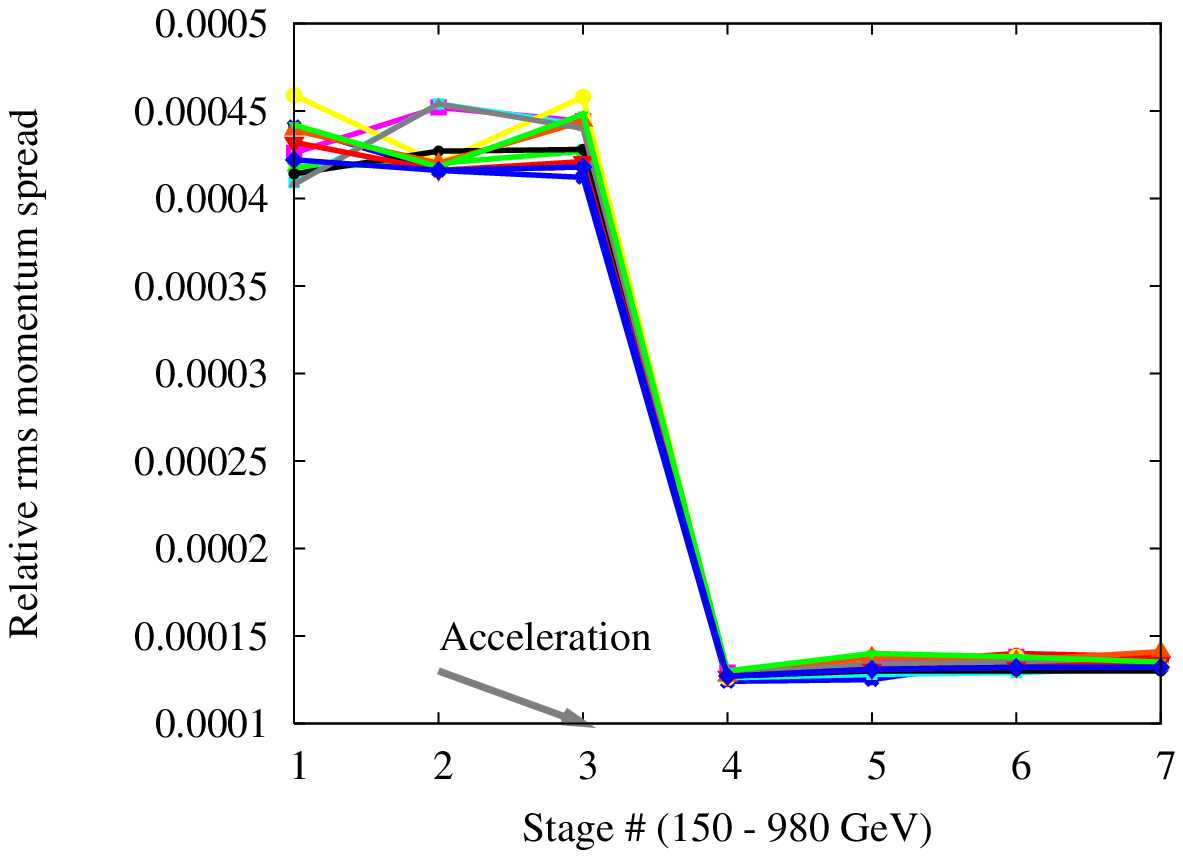}
\includegraphics[scale=0.5]{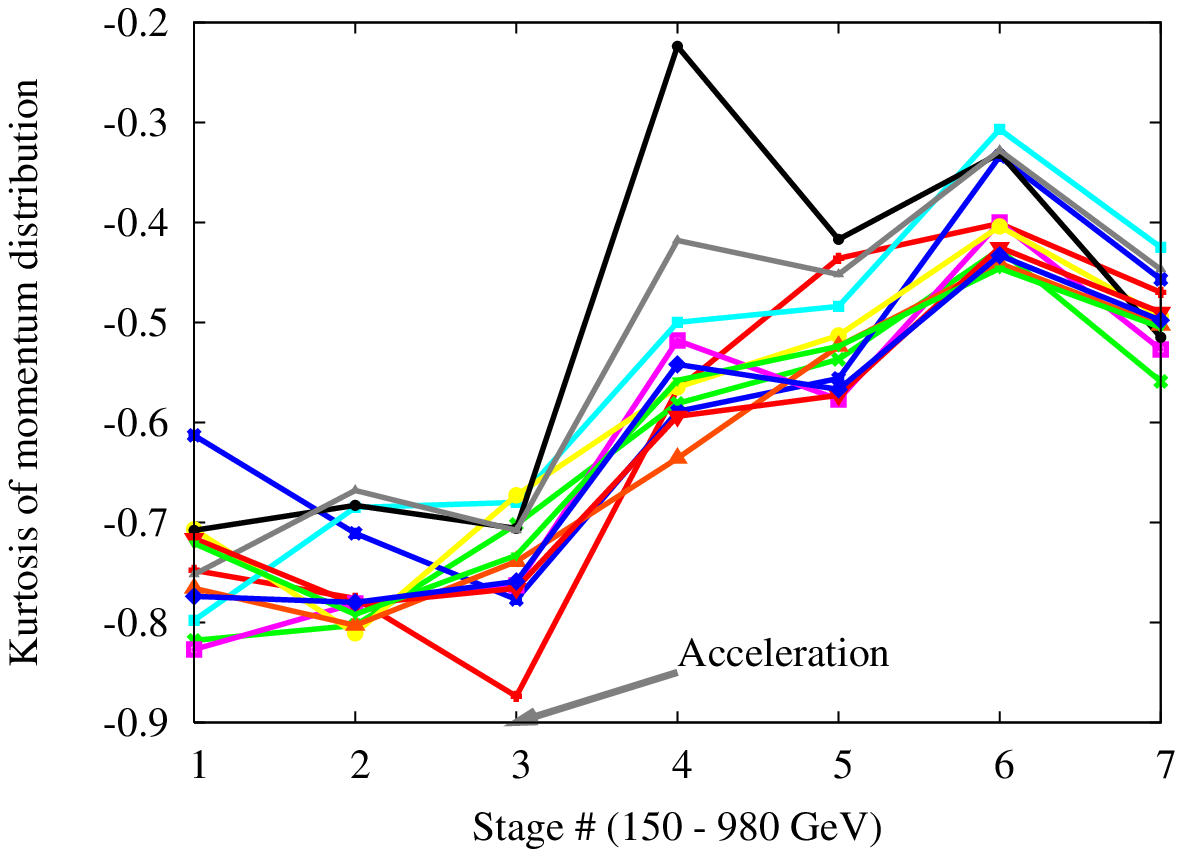}
\caption{Evolution of the proton rms momentum spread (left) and the kurtosis
of the proton momentum distribution (right) in store 7949.}
\label{fig: sigp_kurtosis_7949}
\includegraphics[scale=0.5]{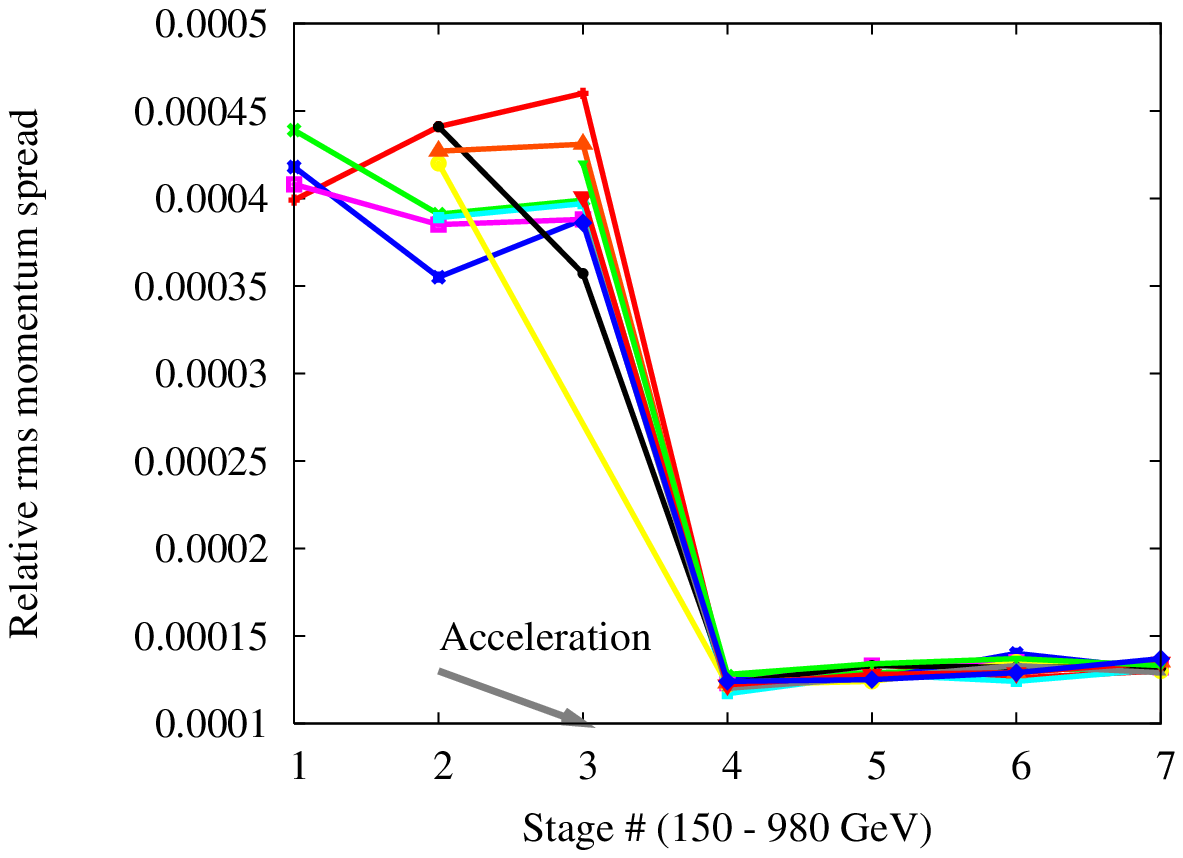}
\includegraphics[scale=0.5]{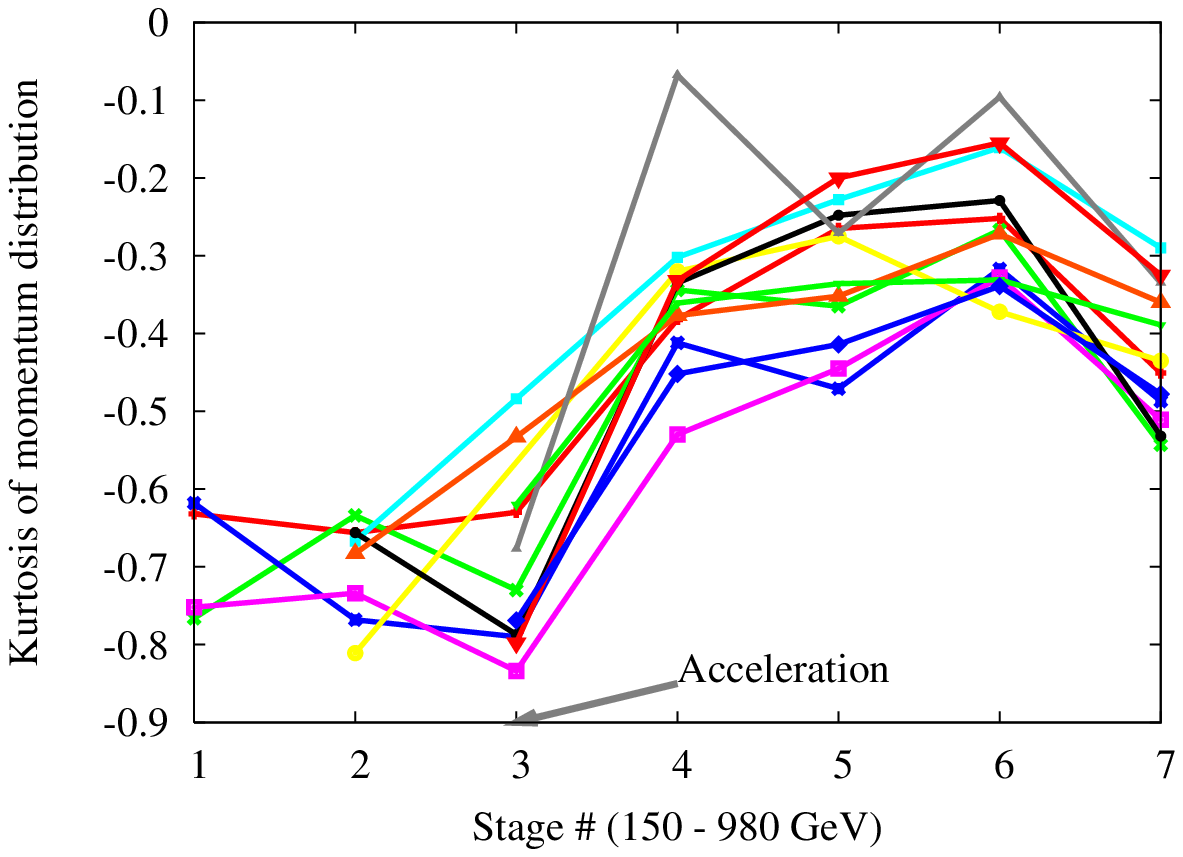}
\caption{Evolution of the anti-proton rms momentum spread (left) and the 
kurtosis
of the anti-proton momentum distribution (right) in store 7949. }
\label{fig: sigp_kurtosis_pbar_7949}
\centering
\includegraphics[scale=0.5]{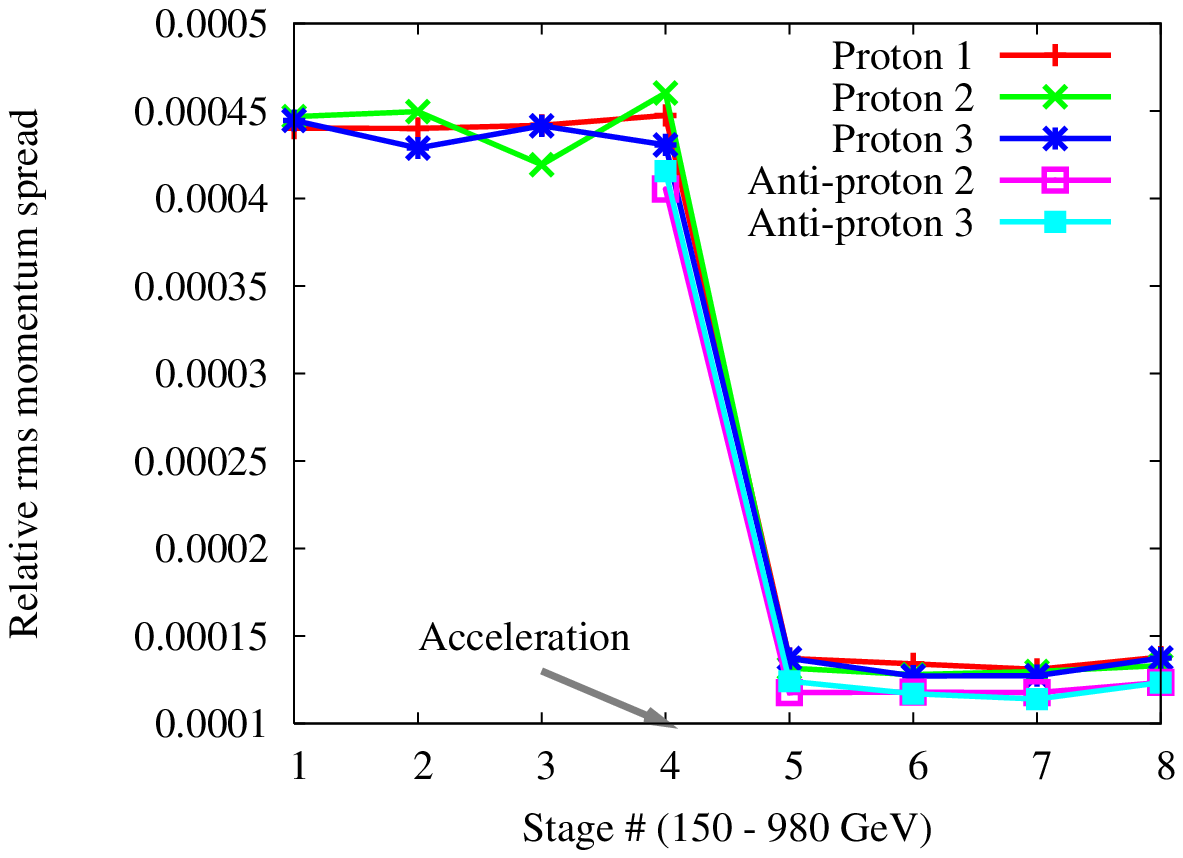}
\includegraphics[scale=0.5]{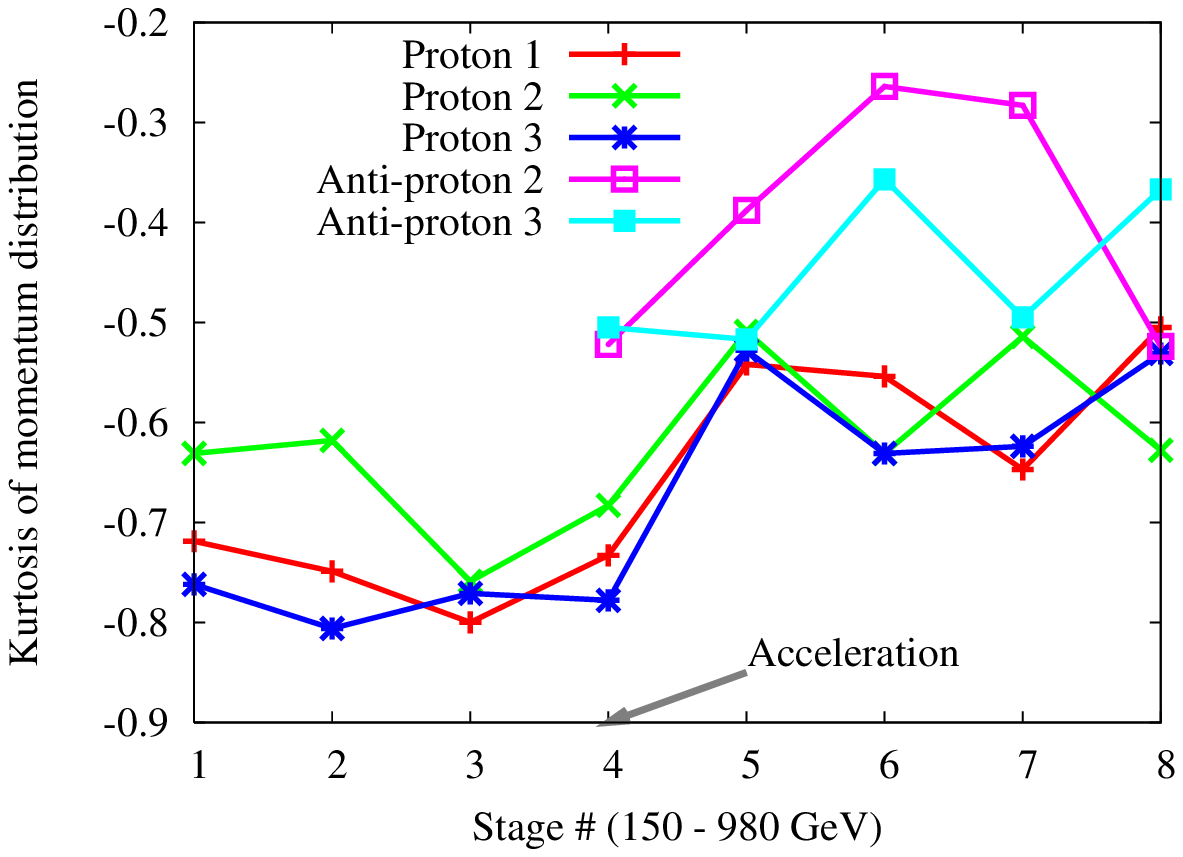}
\caption{Evolution of the rms momentum spread (left) and kurtosis (right)
during the different stages for all 5 bunches monitored in store 8146.}
\label{fig: sigp_kurtosis_8146}
\end{figure}
Figures \ref{fig: sigp_kurtosis_7949} and 
\ref{fig: sigp_kurtosis_pbar_7949} show the evolution of the rms
momentum spread and the kurtosis of twelve proton and twelve anti-proton
 bunches in store 7949.
Since we obtain a single momentum distribution for each bunch from the
phase space, the errors in the calculation of these moments are larger
than those in the calculations of moments of the longitudinal profiles.
The sharp changes in momentum spread seen at injection energy for both
beams are likely a consequence of the reconstruction errors and not
physical. The large drop in the rms spread occurs as the bunch shrinks 
when the bunches are accelerated to 980 GeV. At top energy, the momentum
spread increases gradually for all bunches and both beams.
The kurtosis of protons is negative at injection and keeps decreasing
suggesting shorter tails than for a Gaussian. Acceleration increases the
proton momentum kurtosis as it does for the proton  longitudinal kurtosis.
In general, the kurtosis increases for most bunches until collisions begin
but during the store it decreases, implying
that the momentum distribution does not approach a Gaussian at long times.
The kurtosis of the anti-protons has a similar behaviour. We note 
however that the error in these kurtosis calculations which involve
the fourth moment are larger than in the rms momentum spread calculations.

In store 8146, the data was collected over 1024 turns so the phase space
reconstruction and hence the momentum distributions are expected to be
more accurate. At injection the rms spread of the proton bunches 
stays nearly constant or fluctuates by small amounts. After 
acceleration, the momentum spread does not change much during the beta
squeeze or the onset of collisions but then grows between 3-8\% during the
2 hours of the store. The right plot in Figure 
\ref{fig: sigp_kurtosis_8146}
shows that the momentum distribution of the protons generally develop 
shorter tails at injection energy as the kurtosis decreases. During
acceleration from stage 4 to stage 5 the kurtosis of all bunches increases.
the behaviour after reaching 980 GeV varies from bunch to bunch. 
For protons the kurtosis fluctuates in a range between -0.5:-0.6 even two 
hours into the store.
The momentum distributions of the two anti-proton bunches tend towards a
Gaussian distribution until stage 6 but thereafter the kurtosis drops
sharply for one bunch indicating a shortening of the tails while it
fluctuates for the other bunch. These bunch by bunch differences are seen
in several observables such as intensity loss and emittance growth and
are often related to the individual tunes of the bunches which differ
due to beam-beam effects \cite{Sen}. 

Analysis of the data at the end of store 8130 shows that the rms
energy spread for the three proton bunches  and two anti-proton bunches
are about the same and have grown to about 325 MeV
since the start of the store.
This shows that the energy spread of the anti-protons has grown much
more rapidly than for protons, since the protons had a larger spread
at the start of the store. 
The excess kurtosis of these five bunches are in the range -1.3 to -1.4
showing that the momentum distribution at long times for both 
species has a much shorter tail compared to a Gaussian.

\section{Intensity loss}

In this section we will consider the beam loss at different stages in 
the cycle. Since the bunch intensity is proportional to the area under the 
longitudinal profile, relative changes in intensity can be calculated.
\begin{figure}[h]
\centering
\includegraphics[scale=0.25,angle=-90]{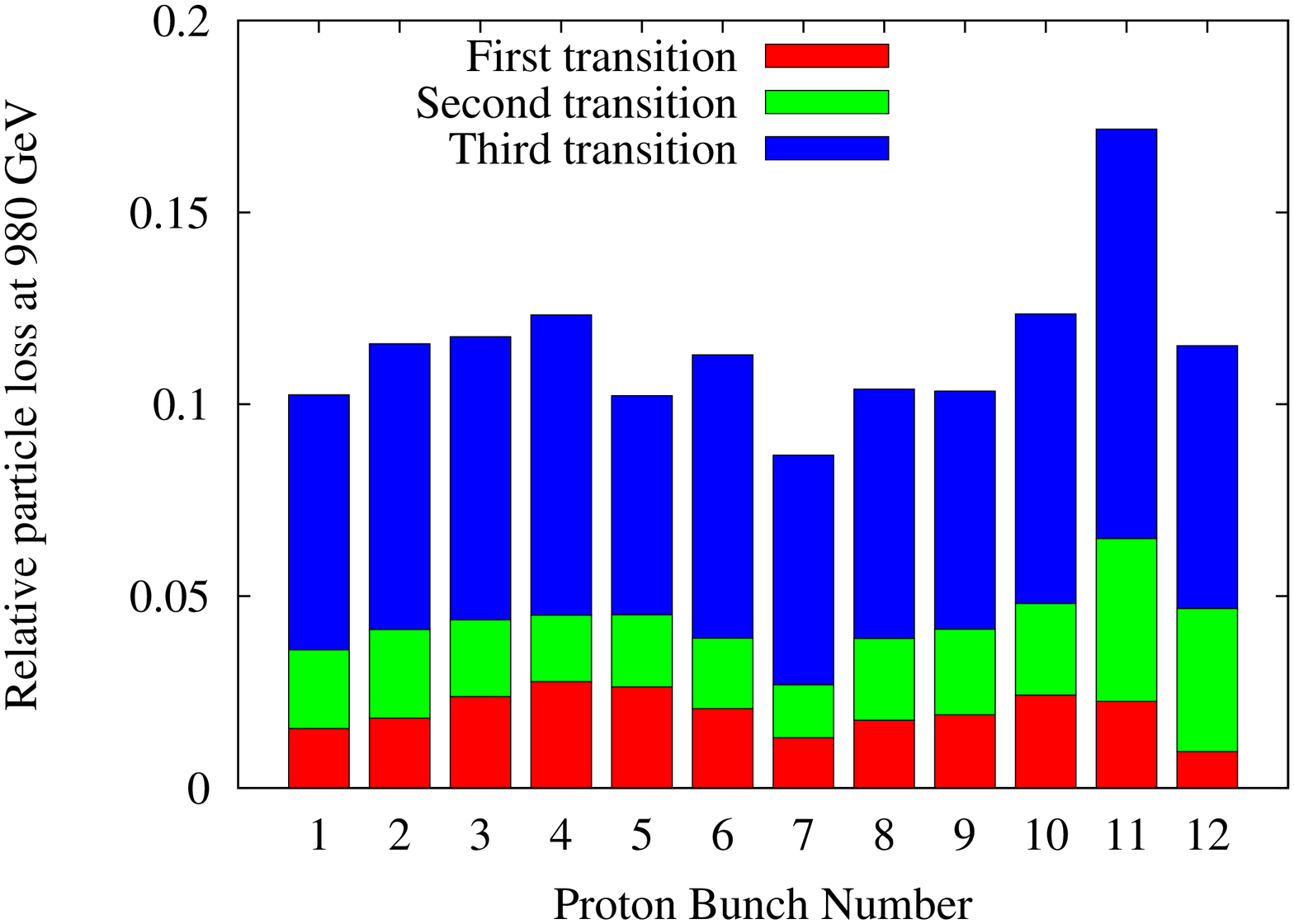}
\includegraphics[scale=0.25,angle=-90]{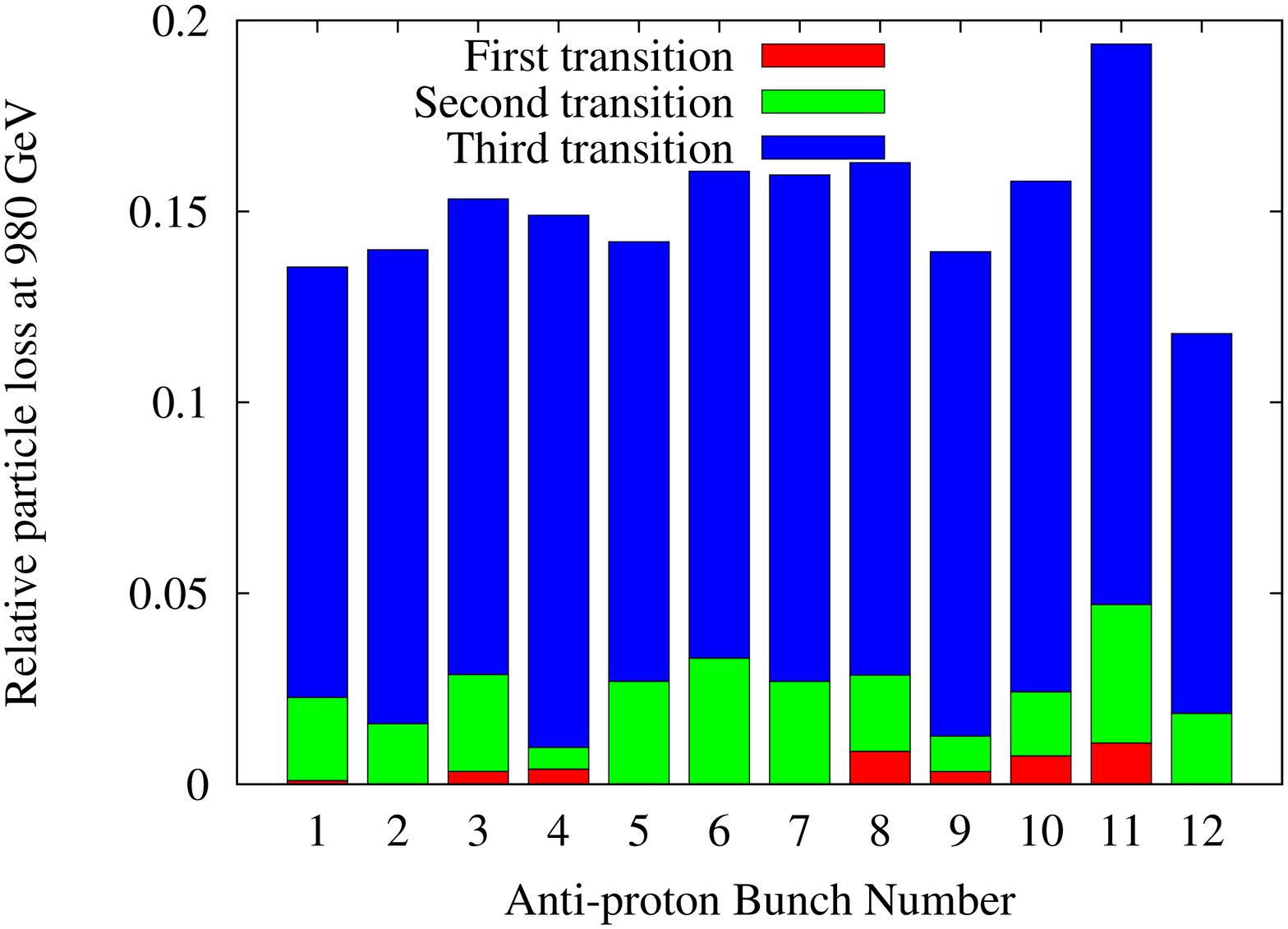}
\caption{Relative particle loss between stages at 980 GeV for proton 
bunches (left) and anti-proton bunches (right) in Store 7949}
\label{fig:particleloss_7949}
\includegraphics[scale=0.5]{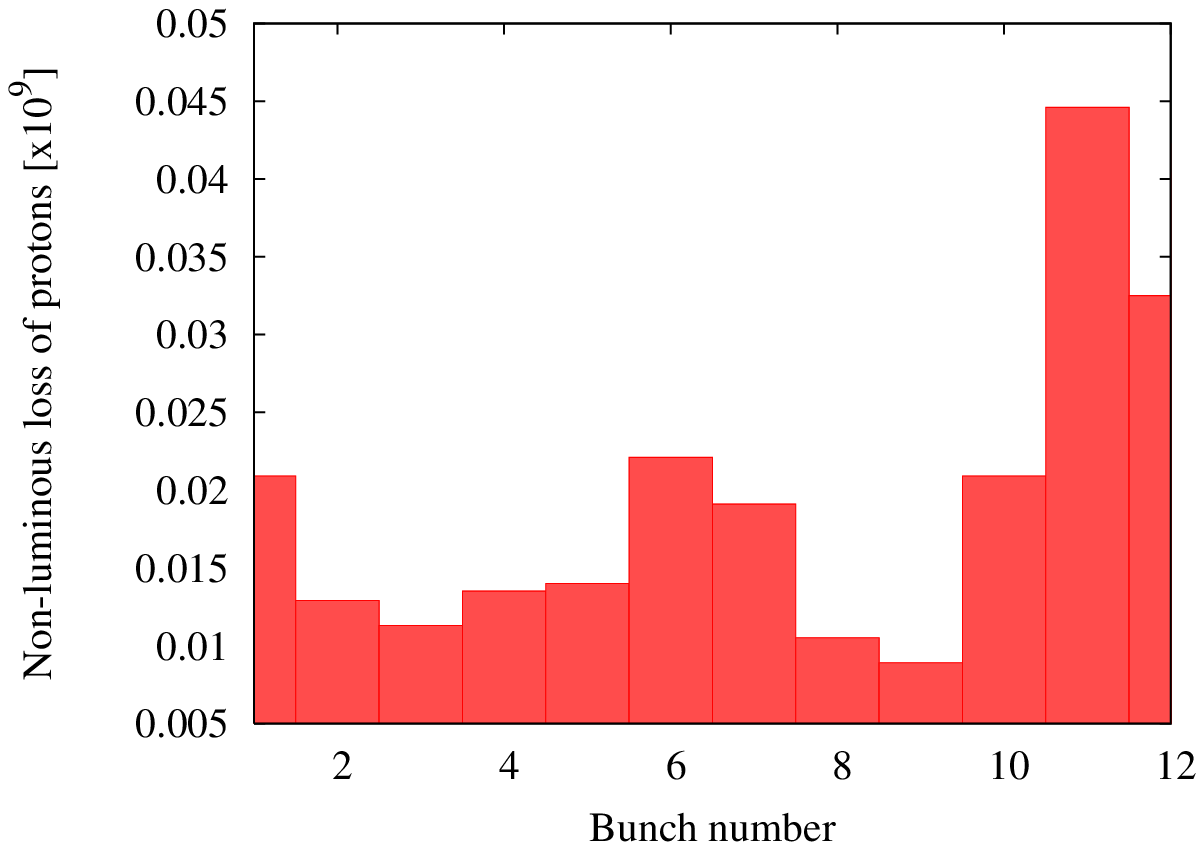}
\includegraphics[scale=0.5]{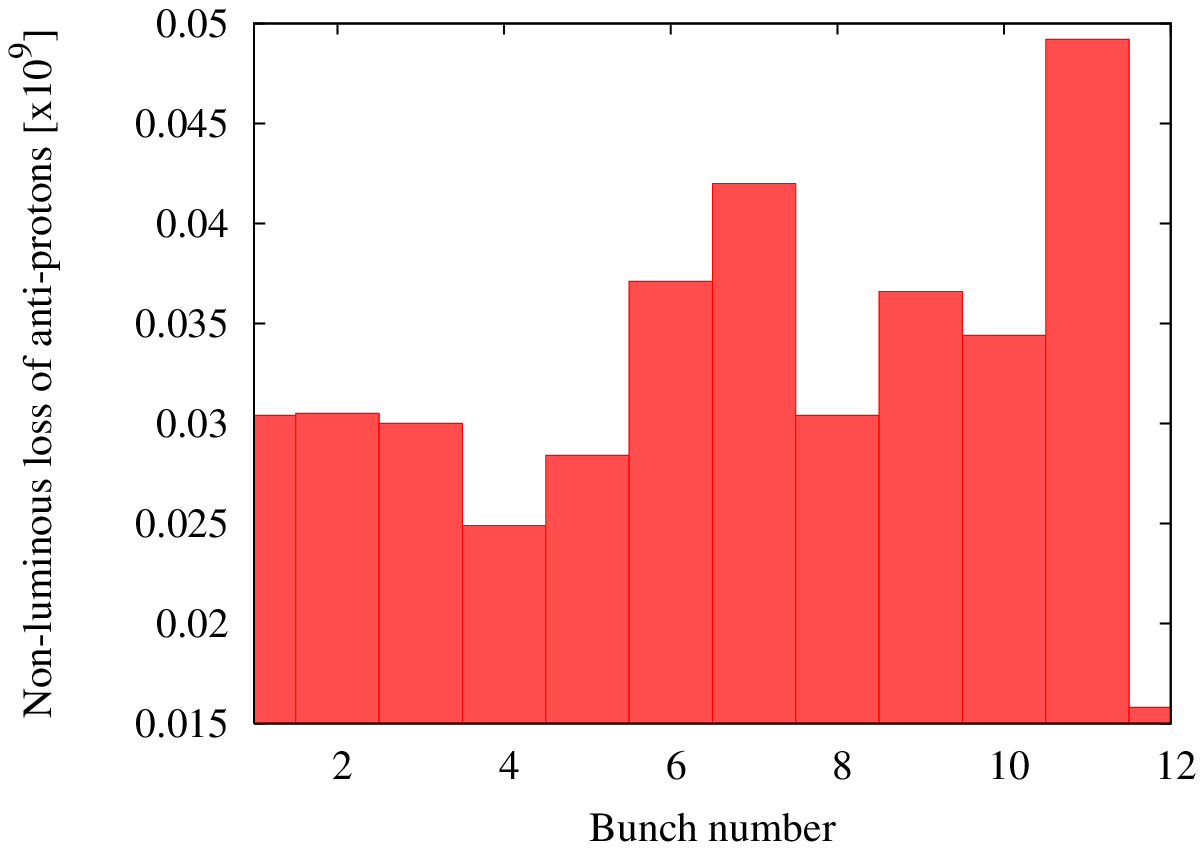}
\caption{Bunch intensity losses not related to luminosity at the end
of store 7949 for the first twelve bunches in each train. the 11th bunch
in each train suffered the largest losses in accordance with the
results in Figure 17 in Store 7949.} 
\label{fig:NLloss_7949}
\end{figure}
Figure \ref{fig:particleloss_7949} shows the relative beam loss computed
from changes in the area for store 7949 during the three transitions
at 980 GeV. The first transition is from before the beta squeeze to
after the squeeze. The second transition occurs to
the start of data taking following removal of the beam halo.
The third transition is to the stage about 2 hours into the store. We
observe that the beta squeeze causes some loss (about 2\%) of protons 
but negligible losses for most anti-proton bunches. The removal of 
beam halo seems to reduce the beam intensity by similar amounts in
both beams. The largest losses occur during the 2 hours into the store
with bunch to bunch variations in both beams. It is interesting that
bunches P11 and A11 have the largest losses during the store and also
during the removal of beam halo. This could be due to a combination of
several factors including tunes closer to the 7/12 resonance, larger
emittances and intensities. 
The major source of beam loss is the inelastic collisions suffered by
the beams at B0 and D0. The beam loss due to beam dynamics, the 
so-called non-luminous loss, can be obtained by subtracting the 
luminosity loss (determined
by the instantaneous luminosity and the inelastic cross-section) from
the total loss. Figure \ref{fig:NLloss_7949} shows the non-luminous loss
at the end of the store for proton bunches P1-P12 and anti-proton 
bunches A1-A12. It is interesting that some of the bunch to bunch
variation seen from the changes in bunch area are also reproduced
in the non-luminous loss. Thus for example, bunches P11 and A11 suffer the
largest loss in bunch area and the largest non-luminous loss. A12 has the lowest beam loss seen among the
anti-protons in both figures but the relative non-luminous loss for A12
is much smaller compared to the loss from the change in bunch area.

\begin{figure}
\centering
\includegraphics[scale=0.5]{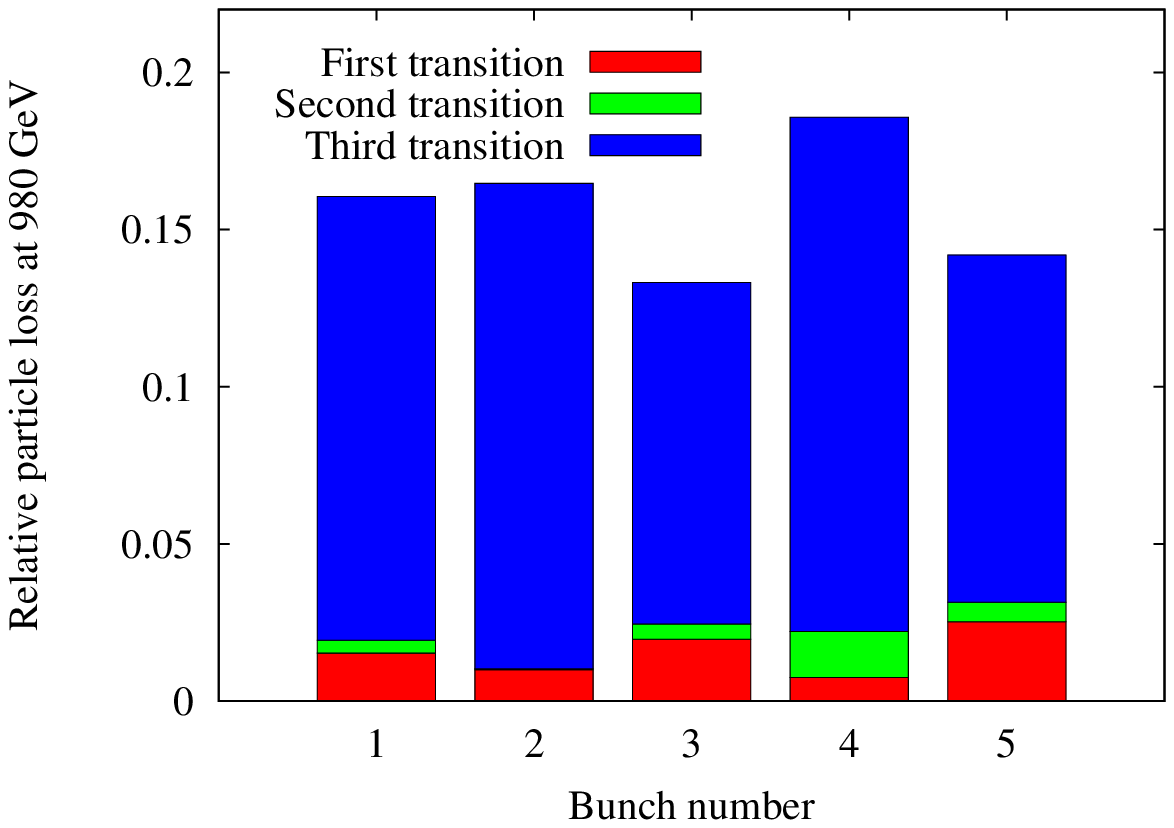}
\caption{Relative particle loss in store 8146. Proton bunches P1-P3 are
numbered 1, 3, 5 while the anti-proton bunches A2-A3 are numbered 2, 4.}
\label{fig: partloss_8146}
%
\includegraphics[scale=0.5]{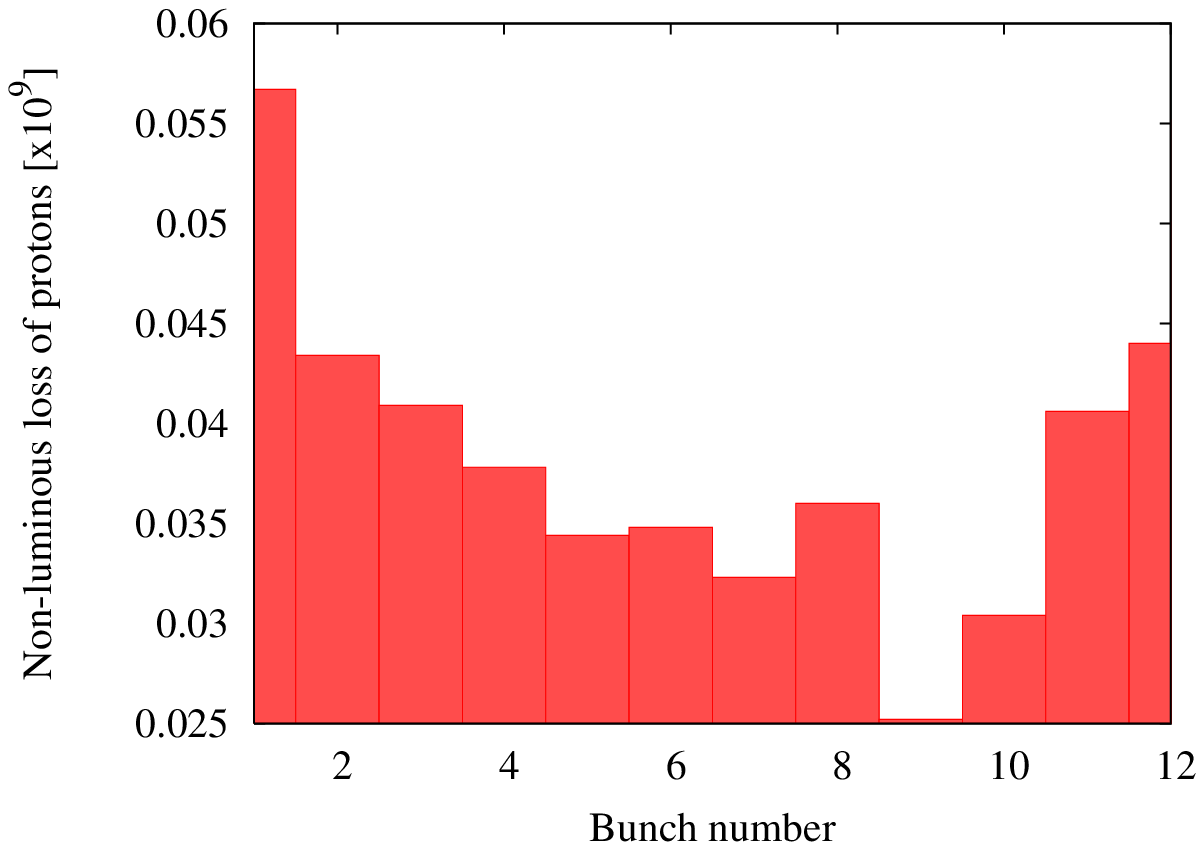}
\includegraphics[scale=0.5]{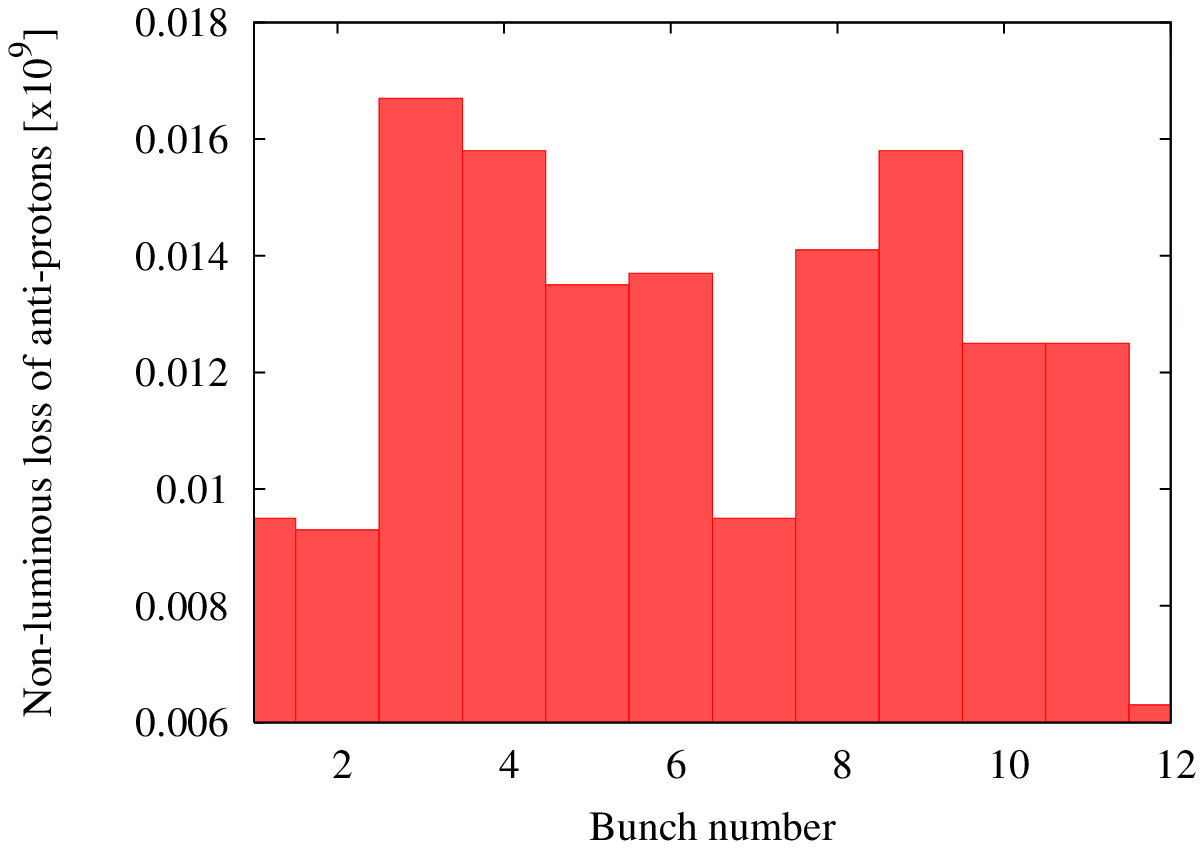}
\caption{Bunch intensity losses not related to luminosity at the end
of store 8146 for the first twelve bunches in each train. Note that the
proton non-luminous losses were significantly higher than the anti-proton
losses.}
\label{fig:NLloss_8146}
\end{figure}
Figure \ref{fig: partloss_8146} shows the particle loss from the change
in bunch area for store 8146. Here the first transition at 980 GeV is
from before the beta squeeze to after the beta squeeze. The second
transition is to a stage just before collisions are initiated. The
third transition is to a stage about 2 hours after collisions started.
Consistent with the data in store 7949, the beta squeeze causes about
a 2\% loss in the proton bunches but much less in the anti-protons. 
The second transition expectedly does not cause much loss in any but
one anti-proton bunch A3 for unknown reasons. Again, the dominant loss
occurs during the 2 hours into the store. Comparison with the 
non-luminous losses in Figure \ref{fig:NLloss_8146} reproduces some
of the bunch to bunch variations. For example, losses are the highest 
in P1 among protons and in A3 among anti-protons in both figures.

\section{Conclusions}

We have used the longitudinal profile of protons and anti-protons
captured over many turns to study the dynamics of the beams during
different stages in the Tevatron cycle. These multiple turn data sets
were subsequently used for tomographic reconstructions of the longitudinal
phase space which allowed us to also study the momentum distribution.

At injection energy we find that the proton bunches continue to 
shorten over
time with some beam loss. This longitudinal clipping and beam loss is due to
the larger proton emittance which nearly fills the bucket and internal motion
which is also seen in the tomographic phase space reconstruction. The 
longitudinal beam-beam effects which should be increasing as more anti-proton
bunches are injected, likely have only a minor impact. The shape
of the proton distribution, as indicated by the excess kurtosis,
becomes 
more Gaussian during injection. After acceleration, the bunch lengths decrease 
but the kurtosis of the proton bunches rises sharply to a maximum 
value of 2 for many bunches, suggesting that the tails have increased 
relative to the core. 
After circulating for several hours during a luminosity store, the
proton longitudinal distribution again approaches a Gaussian. The anti-proton
bunches which have a smaller intensity and smaller longitudinal emittance 
behave rather differently. At injection energy, their bunch length
stays nearly constant, suggesting that the long-range beam-beam effects have
negligible impact. Unlike protons, their kurtosis does not increase 
during acceleration but instead keeps decreasing from negative values in the range
(-0.25:-0.4) at injection to about -0.6 several hours into the store 
implying that the core keeps growing relative to the tails. 

Analysis of the Fourier spectra revealed the presence of a high frequency
line in both beams in both stores. This frequency which changes with 
energy was found to be the same for all bunches analyzed. In store 7949,
this high frequency line was at 3.3 kHz at 150 GeV and it dropped to
1.7 kHz at 980 GeV while in store 8146, the line was at 3.0 kHz at 150 GeV
and dropped to 1.3 kHz at 980 GeV. The amplitudes of these lines grow
during injection and only start to drop a few hours after collisions 
start. The fact that all bunches had the same
frequency and that the line persists over several hours suggests that an
external source such as the rf cavity may be responsible. Since these lines
are at high harmonics of the coherent synchrotron frequency, they appear
to not have much impact on the beam.

The reconstructed phase space profiles show that at injection there is
some substructure in the proton bunches but less so in the anti-proton 
bunches. Proton bunches are more intense and larger, and there may be
coherent motion of smaller bunchlets within the main bunch. These bunchlets
are known to be created during coalescing in the Main Injector. 
This motion 
may also be responsible for the proton beam loss observed during injection.
At top energy, the phase space structure of both beams is smooth and
unremarkable, at least on the resolution of our data.

The phase space distributions are used to construct a single momentum
distribution for a bunch at each stage. Based on these projections, we 
find that the momentum distributions of both beams 
have shorter tails compared to Gaussian distributions and the 
distributions become more non-Gaussian over time. The momentum spread
of the anti-protons grows at a faster rate than that of the protons.
These conclusions are tentative, since the statistical error of the
results obtained for the momentum distributions is significant. 

Intensity losses were computed from the change in area under the 
longitudinal bunch profile. Stage by stage comparisons showed that
the beta squeeze induces about 2\% loss in protons and less so in the
anti-protons. The losses during a store vary from bunch to bunch
and these variations are qualitatively similar to the variations in the
non-luminous losses due to beam dynamics.

 We have shown that longitudinal profiles gathered over a few
synchrotron periods can be used to reveal the richness of beam 
dynamics and they
have the potential to be equally useful in other accelerators.

\vspace{2em}

\noi {\bf \large Acknowledgment} 

This study was begun when the first author was an undergraduate intern
in the Lee Teng summer internship program of 2010 at Fermilab. We
thank the program for its support.


\begin{thebibliography}{99}
\bibitem{Sen}T. Sen et al, Phys. Rev. ST-AB, {\bf 7 }, 041001 (2004)

\bibitem{Shiltsev}V. Shiltsev et al, Phys. Rev. ST-AB, {\bf 8}, 101001
(2005)

\bibitem{Sen_ICFA} T. Sen, ICFA Beam Dynamics Newsletter, Aug 2010, pg 14

\bibitem{Danilov}V.V. Danilov et al, Proc of 1991 Part. Acc. Conf., 
pg 526 (1991)

\bibitem{Hogan}M. Hogan and J. Rosenzweig, Proc of 1993 Part. Acc. 
Conf., pg 3494 (1993)

\bibitem{Moore}R. Moore et al, Proc of 2003 Part. Acc. Conf., pg 1751 (2003)

\bibitem{Sen_coh} T. Sen et al, Fermilab preprint FERMILAB-TM-2431-APC 
(2009)

\bibitem{Hancock}S. Hancock et al,Phys. Rev. ST-AB, {\bf 3}, 124202 (2000)


\end{thebibliography}
\end{document}